%% file: 12p5-altpol-main.tex
\documentclass[twocolumn,appendixfloats,twocolappendix]{aastex63}

\usepackage{natbib}
\usepackage{amsmath,amssymb}
\usepackage{xspace}
\usepackage{rotating}
\usepackage{multirow}
\usepackage{enumitem}

\definecolor{mpl_blue}{HTML}{1F77B4}
\definecolor{mpl_orange}{HTML}{FF7F0E}
\definecolor{mpl_green}{HTML}{2CA02C}
\definecolor{mpl_red}{HTML}{D62728}

\usepackage[all]{hypcap}
\usepackage[caption=false]{subfig}
\newcommand{\m}[1]{\mathbf{#1}}

\def\be{\begin{equation}}
\def\ee{\end{equation}}
\newcommand{\bb}{\begin{bmatrix}}
\newcommand{\eb}{\end{bmatrix}}
\def\bea{\begin{eqnarray}}
\def\eea{\end{eqnarray}}

\def\R{\color{red}}

\defcitealias{abb+15}{NG9}
\defcitealias{abb+16}{NG9gwb}
\defcitealias{abb+18a}{NG11}
\defcitealias{abb+18b}{NG11gwb}
\defcitealias{aab+20}{NG12}
\defcitealias{NG}{NG12.5}

\usepackage{graphicx}
\usepackage{dcolumn}
\usepackage{bm}
\usepackage{xcolor}
\usepackage{aas_macros}
\usepackage{tikz,xcolor,hyperref}
\begin{document}
\title{The NANOGrav 12.5-year data set: Search for Non-Einsteinian Polarization Modes in the Gravitational-Wave Background }

\shorttitle{NANOGrav 12.5-year Gravitational-Wave Background}
\shortauthors{The NANOGrav Collaboration}

\input{12p5-altpol-authors}

\collaboration{1000}{The NANOGrav Collaboration}

\noaffiliation

\correspondingauthor{Nima Laal}
\email{laaln@oregonstate.edu}

\begin{abstract}
We search NANOGrav's 12.5-year data set for evidence of a gravitational wave background (GWB) with all the spatial correlations allowed by general metric theories of gravity. We find no substantial evidence in favor of the existence of such correlations in our data. 
We find that scalar-transverse (ST) correlations yield signal-to-noise ratios and Bayes factors that are higher than quadrupolar (tensor transverse, TT) correlations. Specifically,
we find ST correlations with a signal-to-noise ratio of 2.8 that are preferred over TT correlations (Hellings and Downs correlations) with Bayesian odds of about 20:1. However, the significance of ST correlations is reduced dramatically when we include modeling of the Solar System ephemeris systematics and/or remove pulsar J0030$+$0451 entirely from consideration. Even taking the nominal signal-to-noise ratios at face value, analyses of simulated data sets show that such values are not extremely unlikely to be observed in cases where only the usual TT modes are present in the GWB. 
In the absence of a detection of any polarization mode of gravity, we place upper limits on their amplitudes for a spectral index of $\gamma = 5$ and a reference frequency of $f_\text{yr} = 1 \text{yr}^{-1}$. Among the upper limits for eight general families of metric theories of gravity, we find the values of $A^{95\%}_{TT} = (9.7 \pm 0.4)\times 10^{-16}$ and $A^{95\%}_{ST} = (1.4 \pm 0.03)\times 10^{-15}$ for the family of metric spacetime theories that contain both TT and ST modes.

\end{abstract}
\keywords{
Gravitational waves --
Alternative theories of gravity --
Alternative polarization modes of gravity --
Methods:~data analysis --
Pulsars:~general
}
\section{\label{sec:level1}Introduction}

Pulsar timing experiments~\citep{Shazin,Detw} allow us to explore the low-frequency ($\sim$ 1-100~nHz) part of the gravitational-wave (GW) spectrum. 
By measuring deviations from the expected arrival times of radio 
pulses from an array of millisecond pulsars, we can search 
for a variety of GW signals and their sources. The most promising sources in the
nHz part of the GW spectrum are super-massive binary black holes (SMBHBs)
that form via the mergers of massive galaxies. Orbiting SMBHBs
produce a stochastic GW background
(GWB)~\citep{lb01,jb03,Wyithe:2002ep,vhm03,ein+04,svc08,s13,McWilliams:2012jj,Ravi_2015,Rosado_2015,Sesana:2016yky,Kelley_2016,2017MNRAS.471.4508K,2017MNRAS.470.4547D,Ryu:2018yhv,2018MNRAS.477.2599B},
individual periodic signals or continuous waves
(CWs)~\citep{svv09,sv10,rs11,rwh+12,Mingarelli:2012hh,Ravi_2015,Rosado_2015,2016MNRAS.459.1737S,2017NatAs...1..886M,2018MNRAS.477..964K},
and transient GW bursts~\citep{Becsy:2020utk,vl10,Cordes:2012zz,Ravi_2015,2017PhRvD..96l3016M,Islo:2019qht}.
We expect to detect the GWB first, followed by detection of individual
SMBHBs~\citep{Rosado_2015, Siemens:2013zla,2016ApJ...819L...6T,2017NatAs...1..886M} that stand out above the GWB. 
Detection of GWs from SMBHBs will yield insights into galaxy mergers
and evolution not possible through any other means. Other potential sources in the 
nanohertz band include cosmic strings~\citep{Berezinsky:2000vn,Damour:2000wa,Damour:2001bk,Damour:2004kw,Siemens:2006vk,Siemens:2006yp,Olmez:2010bi,Sanidas:2012tf,2018PhLB..778..392B, CS1, CS2, CS3, CS4, CS5, CS6, CS7, CS8, CS9, CS10}, phase transitions in
the early universe~\citep{PhysRevD.30.272,ccd+10, PT1, PT2, PT3, PT4, PT5, PT6, PT7}, and relic GWs from
inflation~\citep{sa79,PhysRevD.37.2078, INF1, INF2,INF3,INF4,INF5,INF6,INF7}, all of which would provide
unique insights into high-energy and early-universe physics. 

NANOGrav, the North American Nanohertz Observatory for Gravitational Waves, has been taking pulsar timing data since 2004, and currently monitors over 70 pulsars~\citep{Brazier:2019mmu}. NANOGrav is one of several pulsar timing arrays (PTAs) around the world, which include the European PTA (EPTA;~\citealt{{dcl+16}}), the Parkes PTA (PPTA;~\citealt{krh+20}), the Indian PTA (InPTA;~\citealt{InPTA}), and the Chinese PTA~\citep{CPTA}. Two additional telescope-centered pulsar timing programs are ongoing which use the MeerKAT telescope in South Africa~\citep{MeerTime} and the CHIME telescope in Canada~\citep{CHIMEPulsar}. These collaborations form the International Pulsar Timing Array (IPTA;~\citealt{pdd+19}).
In recent years, PTAs have produced increasingly longer and more sensitive data sets, resulting in upper limits on the GWB that have continued to improve~\citep{vhj+11,dfg+13,src+13,ltm+15,srl+15,vlh+16,abb+16,abb+18b}. Very recently, NANOGrav detected a common red noise process in our 12.5-year data set~\citep{NG}. This common process could be the first hints of a stochastic background of GWs; but unfortunately, the data were not sufficiently sensitive to show statistically significant evidence for quadrupolar correlations~\citep{hd83}, the tell-tale sign of a GWB. 

PTAs provide an important test bed for theories of gravity~\citep{Yunes:2013dva}. By modifying Einstein's theory of General Relativity, alternative theories of gravity are often invoked to explain the origin of cosmic acceleration, provide an alternative to dark matter, and reconcile quantum mechanics and gravity, some of the 
most profound challenges facing fundamental physics today~\citep{Yunes:2013dva}. 
General Relativity predicts the existence of GWs
which travel at the speed of light, are transverse, and have two polarizations.
Other theories of gravity generically predict the existence of GWs with
different properties: additional polarization modes and modified dispersion relations.
For instance, metric theories of gravity can have up to six possible GW
polarization modes~\citep{Eardley:1973br,Eardley:1974nw}. PTA searches for alternative polarization modes of gravity can therefore shed light on important foundational questions by exploring the different types of correlations that these additional modes produce.

LIGO has already made possible a number of GW tests of General Relativity~\citep{LIGO1, LIGO2, LIGO3}. Until very recently~\citep{Chen:2021wdo,Wu:2021kmd,Chen:2021ncc}, PTA data had not been used to perform GW tests of gravity due to the absence of a strong signal that can be attributed to GWs. However, as we mentioned, this situation has 
changed~\citepalias{NG}, \citep{Goncharov:2021oub}. Even though NANOGrav's 12.5-year dataset did not contain strong evidence for quadrupolar correlations, the detection of a common red noise process brings PTAs to a regime where the exploration of non-Einstenian theories could prove to be fruitful.

Due to the nature of pulsar timing experiments, PTAs offer advantages over interferometers for detecting new polarizations or constraining the polarization content of GWs. For instance, each line of sight to a pulsar can be used to construct an independent projection of the various GW polarizations, and since PTAs typically observe tens of pulsars, linear combinations of the data can be formed to measure or constrain each of the six polarization modes many times over~\citep{Yunes:2013dva,ljp08,ss12,2015PhRvD..92j2003G}. Additionally, PTAs have an enhanced response to the longitudinal
polarization modes~\citep{ss12,2018PhRvL.120r1101C,2019PhRvD..99l4039O}.
Indeed, the constraint on the energy density of longitudinal modes inferred
from recent NANOGrav data is about three orders of magnitude better than
the constraint for the transverse modes~\citep{2018PhRvL.120r1101C}.

In this paper, we complement our work in \citetalias{NG} by searching for evidence of non-Einsteinian modes of gravity. We start our analyses by studying simulated PTA datasets similar to NANOGrav's 12.5-year dataset~\citep{Astroforcast} and show that for current datasets (with tens of pulsars having observational baselines less than 15 years and for typical amplitudes of the GWB signal $\sim 2 \times 10^{-15}$), the correlations induced by transverse modes of GWs can be hard to distinguish from one another. 
These results are shown first to set our expectations for our analyses of the dataset in hand as well as future datasets. 

We then report on the results of detection analyses on our 12.5-year data set. We analyze the data assuming that the observed stochastic common red noise process across pulsars is due to various combinations of the possible modes available in metric theories of gravity and perform a suite of Bayesian and frequentist searches on our data. 

We find that a model with a phenomenological correlation pattern, the GW-like monopole \footnote{GW-like monopole is a phenomenological correlation pattern (introduced first in this paper) that we have found to be the most preferred among all other tested correlation patterns by our 12.5 year data set. This correlation patterns follows the equation $\Gamma_{ab}^{\text{mono}}=\frac{\delta_{ab}}{2} + \frac{1}{2}$, in which $\delta_{ab}$ is the Kronecker delta function, and $a$ and $b$ are two pulsars. Refer to \S\ref{sec:ORF} for more information. }, is the most favored model (being preferred by an odds ratio of over 100 to 1 compared to a model without correlations), followed by a model with correlations induced entirely by the scalar-transverse mode of gravity (the breathing mode). The latter finding was first reported by~\cite{Chen}, though we disagree with some aspects of their methodology and conclusions. 
Note that on theoretical grounds, we expect the presence of these types of correlations to be accompanied by the standard quadrupolar $+$- and $\times$-modes of General Relativity: metric theories of gravity have at least the $+$- and $\times$-modes and possibly additional modes. In addition, our simulations show that at short observational baselines, for weak correlations, it is hard to distinguish between the different polarization modes; specifically, we show that when only the $+$- and $\times$-modes of General Relativity are present, one can nevertheless, by chance, find evidence in favor of scalar-transverse (breathing modes) modes. We also find that the significance of non-quadrupolar correlations is reduced significantly (the Bayes factor drops to about 20) when the pulsar J0030$+$0451 is removed from our analyses. This pulsar has a history of being problematic in detection searches~\citep{11YEAR_SLICES}, and our results point to the possibility of noise modeling issues involving this MSP. We conclude that the apparent (and weak) presence of non-Einsteinian modes of gravity is likely un-physical, though worth following up in analyses of future data sets.

Finally, since we do not find statistically significant evidence in favor of any correlations, we place upper limits on the amplitudes of all possible subsets of polarization modes of gravity predicted by metric spacetime theories.

The structure of this paper is as follows. In \S\ref{sec:background}, we summarize alternative theories of gravity in the context of pulsar timing experiments. 
We begin the section with a discussion of the most general form the polarization tensor of gravitational waves can have in a general metric theory of gravity and show the effects these generalized gravitational waves have on PTA data. In \S\ref{sec:actualsearches}, we apply these results to a series of simulated data sets and to NANOGrav's 12.5 year data set. 
In \S\ref{sec:summary}, we present our conclusions.
\section{Background}\label{sec:background}
In this section, we review some of the concepts related to pulsar timing and GWs in general metric theories of gravity necessary to lay the foundations for the stochastic GWB detection pipeline.  
We begin with the form of the most general gravitational wave GW polarization tensor and discuss the signature of a GWB in PTA data. We then present a way to integrate pulsar timing and non-Einstenian modes of gravity into a single framework that we can use to search a PTA data set for the GWB. 
\subsection{Polarization Modes in Metric Theories of Gravity}%
In a general metric theory of gravity, GWs can have up to six independent polarization modes~\citep{Eard}. Using the notation of Newman and Penrose~\citep{NP-not} and adapting a coordinate system in which the GW travels along the +z axis, these modes can be written in terms of the electric components of the Riemann tensor through the following relations~\citep{Eard,Will}
\begin{eqnarray}
\psi_{2}(u) &&\equiv -\frac{1}{6} R_{0303}, \\ 
\psi_{3}(u) &&\equiv -\frac{1}{2} R_{0103} + \frac{1}{2}i R_{0203}, \\ 
 \psi_{4}(u) &&\equiv - R_{0101} + R_{0202} + 2iR_{0102}, \\ 
 \phi_{22}(u) &&\equiv - R_{0101} - R_{0202}, \\
 A_{spatial} &&= \begin {bmatrix}  A_{B} + A_{+} & A_{\times} & A_{V1} \\ A_{\times} & A_{B} - A_{+} & A_{V2} & \\ A_{V1} & A_{V2} & A_{L} \end{bmatrix} \label{gpt}, 
\end{eqnarray}
where $u = t-z$ is the retarded time, $ \operatorname{Re}(\psi_{4})=A_{+}$, $ \operatorname{Im}(\psi_{4})=A_{\times} $, $ \phi_{22}=A_{B} $, $ \operatorname{Re}(\psi_{3})=A_{V1} $, $ \operatorname{Im}(\psi_{3})=A_{V2} $, and $ \psi_{2}=A_{L} $ are the plus, cross, breathing, $x$-vector, $y$-vector, and longitudinal modes of gravity, respectively.
This particular choice of the six independent components has the advantage of yielding the standard result of General Relativity in the transverse-traceless gauge when all modes except cross and plus are set to zero:
\begin{eqnarray}
A^E_{spatial} = \begin {bmatrix}  A_{+} & A_{\times} & 0 \\ A_{\times} & - A_{+} & 0 & \\ 0 & 0 & 0 \end{bmatrix}.
\end{eqnarray}
Eq.~(\ref{gpt}) is sufficient to search for all six modes of gravity in pulsar timing data in a fully general way, i.e, without constraining ourselves to a particular metric theory of gravity. 

\subsection{Isotropic Gravitational Wave Background and Pulsar Timing}%
GWs perturb the geodesics of photons traveling from a pulsar to our radio telescopes on
earth. In the late 1970s,~\citet{Shazin} and ~\citet{Detw} first calculated this effect and expressed it in terms of the red-shifting and blue-shifting induced by a continuous gravitational wave propagating through the earth-pulsar system. Setting the speed of light as well as Newton's constant to unity ($c = G = 1$), the GW-induced redshifts for signals from pulsar $a$ are of the form
\begin{eqnarray}
z_a= \frac{\hat{n}_a^{ i} \hat{n}_a^{ j} }{2 (1 + \hat{\Omega} \cdot \hat{n}_{a})} [h^e_{ij} - h^p_{ij} ],
\end{eqnarray}
where $h^e_{ij}=h_{ij} (t,\vec{X} = 0)$ is the metric perturbation at the earth when the pulse is received, $h^p_{ij}=h_{ij} (t-d_a,\vec{X} = d_a\hat{n}_{a})$ is the metric perturbation at the pulsar when the pulse is emitted,
$\hat{n}_a$ is a unit vector pointing from the earth to the pulsar $a$, $\hat{\Omega}$ is a unit vector in the direction of propagation of the gravitational wave, and $d_a$ is the distance to pulsar $a$. The terms proportional to $h^e_{ij}$ and $h^p_{ij}$ are usually referred to as the earth and pulsar terms. 
The metric perturbation can be written in terms of a plane wave expansion as
\begin{eqnarray}
{{h}_{ij}}(x^\mu)=&&\sum\limits_{A}{\int_{-\infty }^{\infty }{df\int{d\hat{\Omega}\, \tilde{h}_A(f,\hat{\Omega} )\varepsilon _{ij}^{A}( {\hat{\Omega}} ){{e}^{-2\pi i f( t-\vec{X}\cdot \hat{n} )}}}}},\nonumber\\
&&
\end{eqnarray}
where $A$ denotes the polarization mode, $\varepsilon^A_{ij}(\hat{\Omega})$ is the polarization tensor of the GW coming from $\hat{\Omega}$ direction, and $f$ is the frequency of GW. Using this expansion, we can re-express the total redshift induced by GWs in the form
\begin{eqnarray}
  z_a(t)=&&\sum\limits_{A}{\int_{-\infty }^{\infty }{df\int{d\hat{\Omega}\, \tilde{h}_A( f,\hat{\Omega } )F_{a}^{A}( {\hat{\Omega }} )e^{-2\pi ift}U_a(f,\hat{\Omega})}}},\nonumber \label{rsh}\\
  \end{eqnarray}
with
\begin{eqnarray}
 F_{a}^{A}( {\hat{n}} )=&&\frac{\hat{n}_{a}^{i}\hat{n}_{a}^{j}\varepsilon _{ij}^{A}( {\hat{\Omega }})}{2( 1+\hat{\Omega }\cdot {{{\hat{n}}}_{a}})},\\ 
 U_a(f,\hat{\Omega})=&&\left[ 1-{{e}^{2\pi if{{d}_{a}}( 1+\hat{\Omega }\cdot {{{\hat{n}}}_{a}} )}} \right] \label{U},
\end{eqnarray}
where the $F_{a}^{A}$ are the so-called \emph{antenna pattern} functions. 

In pulsar timing, we measure the pulsar timing $R(t)$ residuals rather than the redshifts. The GW contribution to the residuals are simply the integral of the GW-induced redshifts, i.e,
\begin{eqnarray}
R^{GW}_a(t) = \int_{0}^{t}{dt'\,z_a(t')}. \label{r_t}
\end{eqnarray}

Taking the stochastic gravitational wave background to be isotropic, unpolarized, and stationary, the correlation function for the strain can be written as
\begin{eqnarray}
\left\langle {{{\tilde{h}}}_{A}}^{*}f,\hat{\Omega} ){{{\tilde{h}}}_{{{A}'}}}( {f}',{\hat{\Omega}}') \right\rangle =&&\delta ( f-{f}')\frac{{{\delta }^{2}}( \hat{\Omega},{\hat{\Omega}}')}{4\pi }{{\delta }_{A{A}'}}\nonumber\\&&\times\frac{1}{2}{{H}( f )},\label{gpe}\nonumber\\
\end{eqnarray}
where ${H}(f)$ is the one-sided \emph{power spectral density} of the GWB. This quantity is related to the fractional energy density spectrum in GWs, $\Omega_{GW}(f)$, through the equation
\begin{eqnarray}
{{H}}\left( f \right)=\frac{3{{H}_{0}}^{2}}{2{{\pi }^{2}}}\frac{{{\Omega }_{GW}}\left( f \right)}{{{f}^{3}}},
\end{eqnarray}
where $H_0$ is the present value of the Hubble parameter and 
\begin{eqnarray}
{{\Omega }_{GW}}\left( f \right)=\frac{1}{{{\rho }_{c}}}\frac{d{{\rho }_{GW}}}{d\ln \left( f \right)},
\end{eqnarray}
for critical density $\rho_c$ and GW energy density $\rho_{GW}$.
Combining Eq.~(\ref{rsh}), Eq.~(\ref{r_t}), and Eq.~(\ref{gpe}) results in
\begin{eqnarray}
  & \left\langle {{R}^{GW}_{a}}{{R}^{GW}_{b}} \right\rangle =\displaystyle \frac{1}{2}\sum\limits_{A}{\int_{{{f}_{L}}}^{{{f}_{H}}}{df\frac{{{H}({f})}}{\left(2\pi f\right)^2}\Gamma _{ab}^{A}\left( \xi_{ab} ,f \right)}}\nonumber\\ \label{2pc}
  \end{eqnarray}
for
\begin{eqnarray}
 \Gamma _{ab}^{A}\left( \xi_{ab} ,f \right)=&&\int^{2\pi}_{0}d\phi\int^\pi_0{d\theta \gamma_{ab}^A}\nonumber\\
 \gamma_{ab}^A =&& \sin(\theta){{{U}_{a}}\left( f,\Omega  \right)U_{b}^{*}\left( f,\Omega  \right)F_{a}^{A}\left( {\hat{n}} \right)F_{b}^{A}\left( {\hat{n}} \right)},\nonumber\\ \label{spher_integ}
 \end{eqnarray}
where $f_H$ and $f_L$ are the upper and lower bounds of frequency, and $\Gamma _{ab}^{A}$ is the so called \emph{overlap reduction function} (ORF). The ORF is a function of the angular separation $\xi_{ab}$ between two pulsars and the GW frequency $f$. This function plays a key role in GW stochastic background searches in a PTA data set.
\subsection{Explicit Form of the GWB Signal in a PTA Data Set}%
Here we discuss i) the properties of the ORFs for each of the polarization modes, and ii) the characterization of the power spectral density of GWs. 
These provide the final set of tools for creating the framework that enables us to search our 12.5-year data set for evidence of existence of non-Einstenian polarization modes of gravity.

\subsubsection{Overlap Reduction Functions} \label{sec:ORF}
The ORFs for all polarization modes of gravity have been  studied extensively in the literature (see e.g,~\citealt{SX} and~\citealt{2015PhRvD..92j2003G}). In the following, we summarize the most important results of these studies.

For the tensor transverse mode of gravity (TT mode), the ORF is found to be
\begin{eqnarray}
 \Gamma _{ab}^{TT} = \Gamma _{ab}^{\times } + \Gamma _{ab}^{+}\simeq&&\frac{\delta_{ab}}{2}+C(\xi_{ab}) \label{TTORF},
\end{eqnarray}
where $\delta_{ab}$ is the Kronecker delta function and $C(\xi_{ab})$ is best known as the Hellings and Downs (HD) correlations \citep{HD}:
 \begin{eqnarray}
  C(\xi_{ab})=&&\frac{3}{2}\left\{ \frac{1}{3}+k_{ab}\left[ \ln \left( k_{ab}  \right)-\frac{1}{6} \right] \right\},
\end{eqnarray}
for
 \begin{eqnarray}
  k_{ab} &&= \frac{1-\cos \left( {{\xi }_{ab}} \right)}{2}.
 \end{eqnarray}
To an excellent approximation, the HD correlation curve is frequency- and pulsar-distance-independent for all angular separations over the range of $fd_a$ values relevant to pulsar timing experiments~\citep{abc+09}. This can be understood by noting that the ratio of pulsar distances to the GW wavelengths at nHz frequencies is large (typically larger than 100); hence, the exponential terms of Eq.~(\ref{spher_integ}) oscillate rapidly while making a negligible contribution to the overall integral. Also, in case of $\xi_{aa} = 0$ (i.e, $a = b$), the product of $U_a(f,\hat{\Omega})$ and  $U^*_a(f,\hat{\Omega})$ doubles the ORF to relative to what is predicted by $C(\xi_{ab})$ alone, hence the need for $\delta_{ab}$ in Eq.~(\ref{TTORF}).
\begin{figure}
    \includegraphics[width=\linewidth]{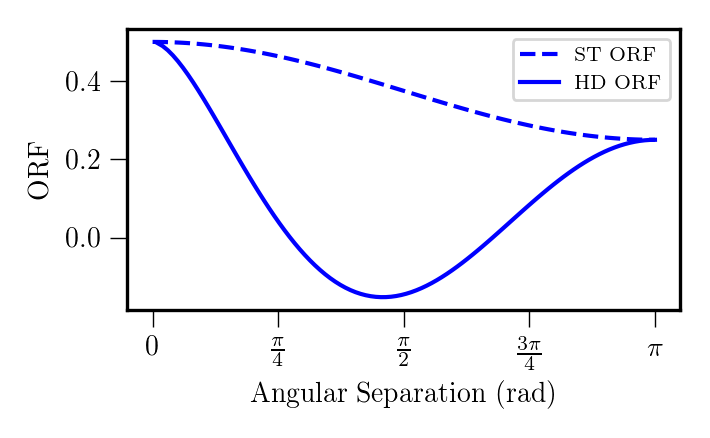}
    \caption{Overlap reduction functions ($a \neq b$) for the transverse modes of gravity normalized to $1/2$ at zero angular separation. The solid line is the tensor transverse mode ORF and the dashed line is the scalar-transverse mode ORF.}
    \label{TTST}
\end{figure}

For the scalar-transverse (ST) mode, also known in literature as the breathing mode, the ORF is found to be~\citep{SX}
\begin{eqnarray}
 \Gamma _{ab}^{ST}\approx&& \frac{\delta_{ab}}{2}+\frac{1}{8}(3+\cos(\xi_{ab})).
\end{eqnarray}
Similar to the case of the TT modes, the ST mode ORF is frequency  and  pulsar-distance-independent to an excellent approximation. Fig.~\ref{TTST} shows the transverse ORFs as a function of angular separation for the case of $a \neq b$.  

\begin{figure}
    \includegraphics[width=\linewidth]{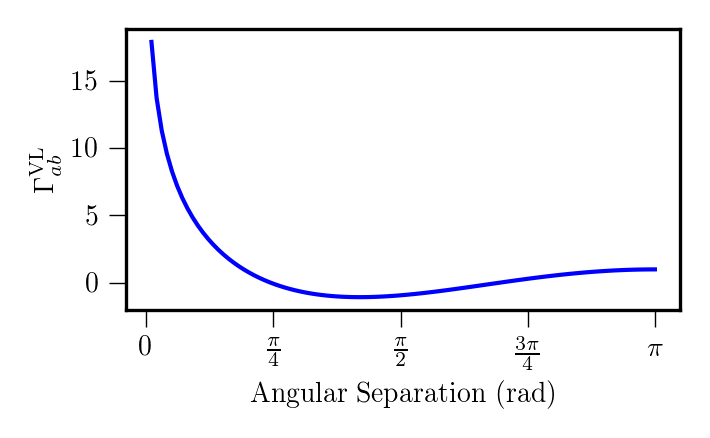}
    \caption{Overlap reduction function for the vector longitudinal modes of gravity. Here we have taken $a \neq b$. Note the values on the y-axis at low angular separations are significantly larger than for the 
    TT and ST modes (see Fig.~\ref{TTST}). PTAs are more sensitive to VL modes than transverse modes.}
    \label{VL}
\end{figure}

For the vector longitudinal (VL) modes, the ORF is found to be~\citep{VLORF}
\begin{eqnarray}
 \Gamma_{ab}^{VL} =&& \Gamma_{ab}^{(VL)_y} + \Gamma_{ab}^{(VL)_x}\nonumber\\ \approx&&3 \log{\left(\frac{2}{1 - \cos{\left(\xi_{ab} \right)}} \right)} - 4 \cos{\left(\xi_{ab} \right)} - 3 \label{VLORF},
\end{eqnarray}
where a normalization factor of $3/(4\pi)$ has been applied for consistency with the transverse ORFs (see Fig.~\ref{VL}) and $a \neq b$. These modes are also  frequency independent in the limit of large $fd_a$ values relevant to pulsar timing, albeit to a lesser extent than the transverse modes. The approximation fails at zero angular separations requiring the inclusion of the pulsar (exponential) terms in the calculation of ORF to cancel the divergence. For the case of $a = b$ (i.e, the case of a pulsar correlated with itself), $\Gamma_{aa}^{VL}$ is~\citep{SX}
\begin{eqnarray}
 \Gamma _{aa}^{VL}=2\ln \left( 4\pi {{d}_{a}}f \right)-\frac{14}{3}+2{{\gamma }_{E}},
\end{eqnarray}
where ${{\gamma }_{E}}$ is Euler's constant, and $fd_a \gg 1$.

\begin{figure}
    \includegraphics[width=\linewidth]{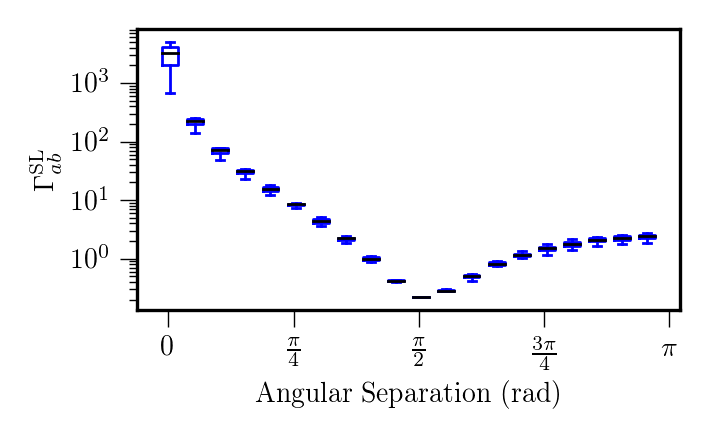}
    \caption{Box plots for the ORF for the scalar longitudinal mode of gravity. The box plots depict the variance of SL ORF over $fd_a$ values ranging from 100 to 1000.  Pulsars $a$ and $b$ are assumed to be different. Note the much larger values on the y-axis compared to the transverse (Fig.~\ref{TTST}) and VL (Fig.~\ref{VL}) modes showing that the SL mode will produce the largest signal in a PTA.}
    \label{SL}
\end{figure}

Finally, for the scalar longitudinal (SL) mode, the ORF cannot be evaluated analytically for all angular separations. Hence, the integral in Eq.~(\ref{spher_integ}) needs to be evaluated numerically given a set of pulsar distances, frequencies and angular separations. Fig.~\ref{SL} shows the strong dependence of $\Gamma _{ab}^{SL}$ to $fd_a$ values. However, similar to the case of vector longitudinal modes, for the same pulsar and large $fd_a$ values relevant to pulsar timing experiments, an estimate of the $\Gamma^{SL}$ can be found~\citep{SX}
\begin{eqnarray}
 \Gamma _{aa}^{SL}=\frac{{{\pi }^{2}}}{4}f{{d}_{a}}-\ln \left( 4\pi f{{d}_{a}} \right)+\frac{37}{24}-{{\gamma }_{E}} \label{SL_aa}.
\end{eqnarray}

So far, we have only discussed ORFs that result from generic metric theories of gravity. In light of recent observations of ST correlations in NANOGrav's 12.5 year data set \citep{Chen}, it will be useful to define other ORFs that are more phenomenological in nature and are not necessarily due to any single metric theory of gravity. Two of such ORFs are the \emph{GW-like monopole} and \emph{GW-like dipole} with explicit forms
\begin{eqnarray}
\Gamma_{ab}^{\text{GW-mono}}=&&\frac{\delta_{ab}}{2} + \frac{1}{2}, \label{GW_MONO}\\
\Gamma_{ab}^{\text{GW-dipole}}=&&\frac{\delta_{ab}}{2} + \frac{\cos{\xi_{ab}}}{2} \label{GW_DIPOLE}.
\end{eqnarray}
We will use these ORFs when searching NANOGrav's 12.5 year data set and compare them to the different polarization modes of gravity, specifically the results of ST mode searches. These ORFs should not be confused with systematic monopole (e.g, clock-error-induced) or dipole (e.g, ephemerides-induced) uncertainties, as those produce correlations that do not distinguish between co-aligned pulsars (i.e, $\Gamma(\xi_{ab} = 0)$) and a pulsar paired with itself (i.e, $\Gamma(\xi_{aa} = 0)$). Namely, the type of correlations in Eqs.~(\ref{GW_MONO}) and (\ref{GW_DIPOLE}) only affect half of the signal, the earth term, whereas clock and ephemerides errors affect the entire signal. This is the reason we introduce the terminology \emph{GW-like} to these phenomenological ORFs. 

It is worth noting that even though a pure monopole of the form of Eq.~(\ref{GW_MONO}) is not predicted by any metric theory of gravity, massive GWs originating from a scalar-tensor metric theory of gravity could alter the form of ST ORF into more of a monopolar looking correlation pattern~\citep{MG}. For example, a metric theory of gravity can have two types of contributions to the GWB, a massive scalar-transverse wave contribution of the form  \begin{eqnarray}
{{\sigma}_{ij}}(x^\mu)=&&{\int_{-\infty }^{\infty }{df\int{{{d}{}}\hat{\Omega}\tilde{\sigma}( f,\hat{\Omega} )\varepsilon _{ij}^{\text{ST}}( {\hat{\Omega}} ){{e}^{ Y_\sigma(f,m)}}}}},\nonumber\\ \\
Y_\sigma(f,m) = && { -2\pi i f \left(t -d\sqrt{1-\frac{m^2}{4\pi^2 f^2}}\right)},
\end{eqnarray}
and a massless transverse tensor contribution of the form
\begin{eqnarray}
{{h}_{ij}}(x^\mu)=&&{\int_{-\infty }^{\infty }{df\int{{{d}{}}\hat{\Omega}\tilde{h}_{\text{TT}}( f,\hat{\Omega} )\varepsilon _{ij}^{\text{TT}}( {\hat{\Omega}} ){{e}^{Y_h}}}}},\\
Y_h=&& -2\pi i f(t-d),
\end{eqnarray}
where $d$ is the distance to a pulsar.
Depending on the values of the mass and the frequency, the resulting ORF due to the scalar-transverse mode could approach a monopolar form (see~\citealt{MG} for a detailed discussion).

\subsubsection{Spectral Density of Gravitational Waves and Correlations in Timing Residuals} \label{sec: the signal}
In PTA analyses, the spectral density $H$ is often written in terms of the dimensionless characteristic strain $h_c$ defined by
\begin{eqnarray}
h_c(f)\equiv\sqrt{fH(f)}.
\end{eqnarray}
NANOGrav analyses have included various models for the characterization of $h_c$ including a power-law model, free-spectral model, and broken power-law model depending on the nature of the analysis (see e.g, the 12.5-yr GWB analysis \citealt{NG}). In this paper we will restrict ourselves to the power-law model as it is the simplest model to implement and interpret. Namely, for each polarization mode of gravity we will use
\begin{eqnarray}
h_c\left(f\right) = A\left(\frac{f}{f_{\text{yr}}}\right)^{\alpha},
\end{eqnarray}
where $A$ is a dimensionless amplitude, $f_{\text{yr}}$ is a reference frequency chosen to be $1/1(\text{yr})$, and $\alpha$ is the spectral index.  The values of the amplitude and spectral index depend on the sources that produce the GWs and the polarization content of the metric theory under consideration. 
The expected correlation in the timing residual time-series for two pulsars can be written as follows
\begin{eqnarray}
\left\langle {{R}_{a}}{{R}_{b}} \right\rangle = \int_{f_L}^{f_H}{dfS_{ab}(f)},
\end{eqnarray}
where 
\begin{eqnarray}
{{S}_{ab}}=&&  \sum\limits_{m}{\Gamma _{ab}^{m}A_{m}^{2}{{\left( \frac{f}{{{f}_{yr}}} \right)}^{3-\gamma_m}}}\frac{1}{8{{\pi }^{2}}f^3}, \label{GWSignal}
\end{eqnarray}
for the sum ranging over all six polarization modes and 
\begin{eqnarray}
\gamma_m \equiv {3 - {2\alpha }_{m}}.
\end{eqnarray}
Hereon, the term \emph{spectral index} will refer to the value of $\gamma_m$ rather than $\alpha_m$.

Before proceeding with analyzing PTA data sets using the framework presented in this section, it is worth mentioning that Eq.~(\ref{GWSignal}) can be written in a more general form encapsulating the frequency-dependent effects of differing emission rates of binary sources of GWs. One such effect relevant to the study of alternative theories of gravity is caused by dipole radiation. To leading order in the post-Newtonian approximation and in $\vartheta$ (i.e, the difference in the self-gravitational binding
energy per unit mass), the rate of change of orbital energy of a binary source is \citep{GeneralEQ}
\begin{eqnarray}
\frac{dE}{dt}=\frac{2}{3}\lambda \mu {{\vartheta }^{2}}{{\left( \frac{2\pi }{P} \right)}^{2}}K\left( e \right)E \label{dipole_rate}
\end{eqnarray}
where $\mu$ is the reduced mass, $\lambda$ is the dipole parameter, $P$ is the orbital period of the binary system, $K(e)$ is a function of the binary's eccentricity $e$, and $E$ is system's instantaneous energy. Applying the Newtonian approximation, $E= -\frac{1}{2}{{M}^{\frac{2}{3}}}{{(2\pi f_s) }^{\frac{2}{3}}}\mu$, where $M$ is the total mass and $f_s$ is the orbital frequency, Eq.~(\ref{dipole_rate}) can be used to calculate the rate of change of orbital frequency:
\begin{eqnarray}
\frac{d\ln{f_s}}{dt}=8{{\pi }^{3}}{{\vartheta }^{2}}\lambda K\left( e \right)\mu {{f_s}^{2}} \label{df/dt}.
\end{eqnarray}
Assuming a Keplerian rest frame, the instantaneous GW characteristic strain radiated by a circular binary system  is \citep{kappa_param}
\begin{eqnarray}
{h_c}\left( f \right)=2{{\left( 4\pi  \right)}^{\frac{1}{3}}}\frac{{{f}^{\frac{2}{3}}}{{M}^{\frac{5}{3}}}}{{{D}_{L}}},
\end{eqnarray}
where $D_L$ is the luminosity distance to the source. This alongside Eq.~(\ref{df/dt}) and the fact that $f_s = f/2$ can be used to yield 
\begin{eqnarray}
h^D_{c}={{C}_{D}}{{f}^{-\frac{2}{3}}}
\end{eqnarray}
in which $h^D_c$ is the characteristic amplitude of the GWB due to dipole radiation and $C_D$ is a constant related to parameters introduced in Eq.~(\ref{dipole_rate}) such that $\lambda = 0$ results in $C_D = 0$. Adding the quadrupolar contribution to the characteristic amplitude and treating it as more dominant than the dipole contribution results in \citep{NanoFirstAlt}
\begin{eqnarray}
{{S}_{ab}}=&&  \left( \frac{1+{{\kappa}^{2}}}{1+{{\kappa}^{2}}{{\left( \frac{f}{{{f}_{\text{yr}}}} \right)}^{\frac{2}{3}}}} \right)\sum\limits_{m}{\Gamma _{ab}^{m}A_{m}^{2}{{\left( \frac{f}{{{f}_{yr}}} \right)}^{3-\gamma_m}}}\frac{1}{8{{\pi }^{2}}f^3}, \label{GWSignal with kappa}\nonumber\\
\end{eqnarray}
for a constant parameter $\kappa$ denoting the relative value of the amplitude of dipole radiation over the amplitude of quadrupolar radiation driving the binary system to a merger. 
In this paper, we set $\lambda$, and consequently $\kappa$, to zero and use Eq.~(\ref{GWSignal}) instead. See \S\ref{sec: Bayesian} for a justification of this choice.

\section{Searches for non-Einstenian Modes in the Gravitational-Wave Background}\label{sec:actualsearches}
Real pulsar timing data sets require significantly more complex modeling than what Eq.~(\ref{GWSignal}) might suggest. Eq.~(\ref{GWSignal}) only includes the GWB content of the pulsar timing residuals; other chromatic (radio-frequency-dependent) and achromatic noise contributions to the timing residuals need to be included in a robust detection analysis. To accomplish this goal, we add the stochastic GW signal modeling presented here to our already existing detection pipeline and pulsar inference tool, ENTERPRISE \citep{ENTP}, and search for various modes of gravity using NANOGrav's 12.5-year data set.

\begin{figure}
    \includegraphics[width=\linewidth]{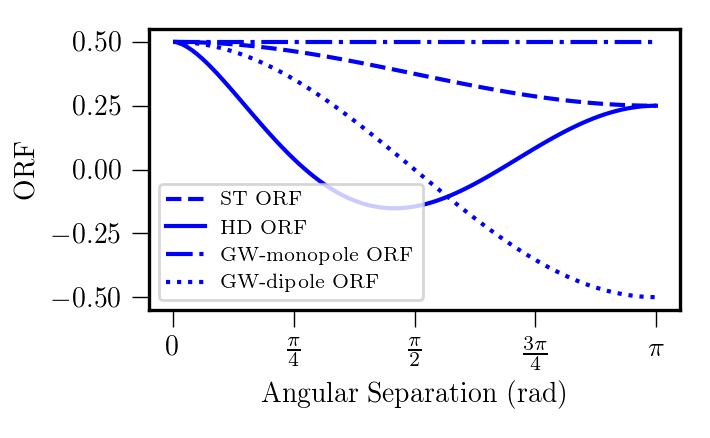}
    \caption{The blue solid line is the HD correlation curve, the blue dashed line is the ST ORF, the solid dotted line is the monopole ORF, and the fully dotted line is the dipole ORF. These four ORFs can be hard to distinguish if the uncertainties in the timing residual cross-correlations are sufficiently large.}
    \label{STTTDipoleMonopole}
\end{figure}

The detection procedure for the ST and VL modes does not require significant modifications to the already existing tools for searches for isotropic gravitational-wave backgrounds in ENTERPRISE. This is due to the fact that the cross-correlation curves are a function of angular separation only, and not frequency. However, the similarities between some of the tensorial ORFs such as the GW-like dipole, GW-like monopole, HD, and ST correlations can pose a significant detection challenge: distinguishing between these ORFs requires high-significance measurements of the cross-correlated power as a function of the angular separation. Fig.~\ref{STTTDipoleMonopole} shows the ORFs for the TT and ST modes, as well as the GW-like monopole and dipole. It is easy to see that given large enough uncertainties in the cross-correlations, the detection of a data set's actual correlation pattern can become problematic.

To address the challenge of reduction of the uncertainties of the cross-correlations, improvements in four key areas can be pursued: i) increasing the observation time; ii) improving the observing instrumentation used at our radio telescopes; iii) increasing the number of pulsars being observed; and iv) improving noise modeling of individual pulsars. All these avenues are actively being pursued by NANOGrav. 

In this section, we will use our detection pipeline to search for, and set upper limits on, the polarization modes present in general metric theories of gravity. We start by performing our analyses on simulated data sets and then proceed to perform similar analyses on NANOGrav's 12.5 year data set. In this paper, we only perform upper limit analyses (not detection analyses; see \S\ref{sec:uls}) for the vector and scalar-longitudinal modes of gravity. This is for three reasons: i) large correlations at small angular separations predicted for the longitudinal (VL, SL) polarization modes, are absent in the current data set, ii) as shown in Fig.~\ref{SL}, the values of the ORF for the SL mode are very sensitive to pulsar distances (which are not well known) for most angular separations, and, iii) the addition of frequency-dependent terms to our current detection pipeline required for the SL mode demands significant modifications, testing, and simulations which are outside the scope of this work. These additions to our detection pipeline are currently under development and will be deployed in analyses of future data sets. 

\subsection{Detection of Additional Polarization Modes of Gravity in Simulated Pulsar Timing Data}%
\label{sec:sims}
\begin{figure}
    \includegraphics[width=\linewidth]{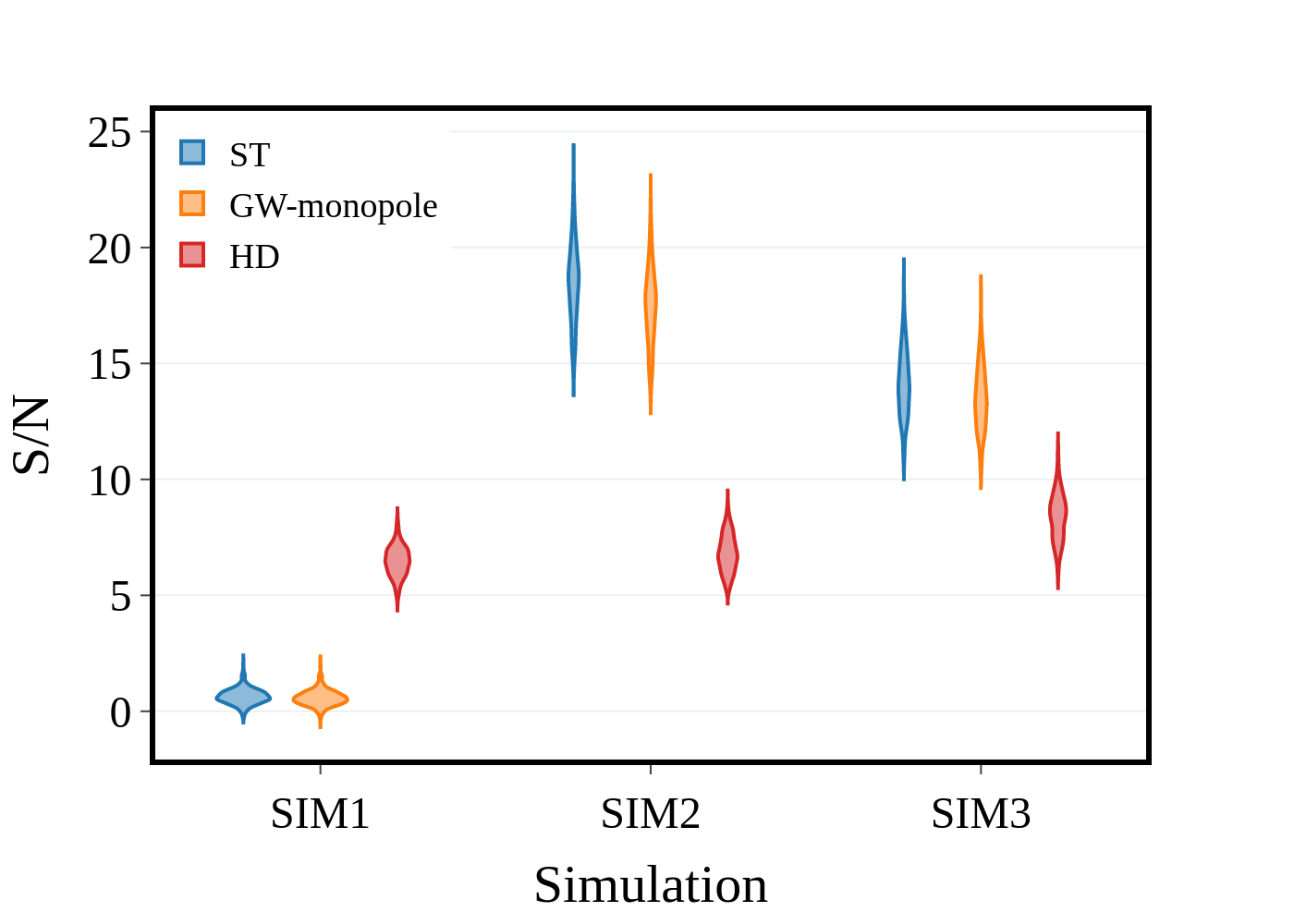}
    \caption{Violin plots depicting the S/N distribution of the noise marginalized optimal statistic for one realization of simulated data sets SIM1, SIM2, and SIM3 (see the main text for a description). The data are searched for three different correlation patterns: ST (blue), HD (red), and GW-like monopole (orange). The S/N distribution for each simulated data set is obtained from the calculation of the noise-marginalized optimal statistic evaluated 1000 times. Even in the case of a strong injection of ST correlations (SIM2), GW-like monopole and ST correlations yield similar S/Ns.}
    \label{SIM_S/N}
\end{figure}

\begin{figure}
    \includegraphics[width=\linewidth]{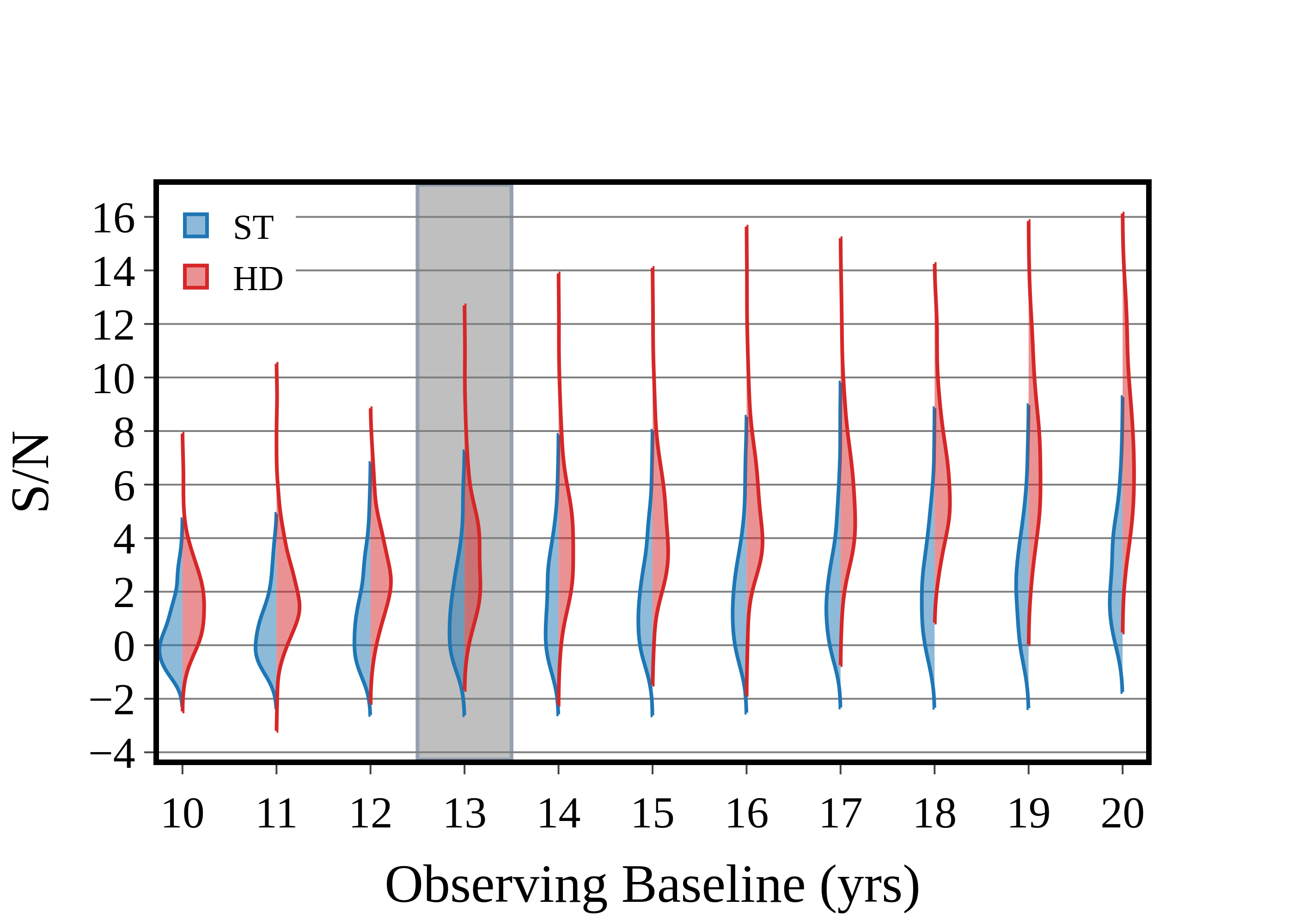}
    \caption{S/N evolution for the optimal statistic using the SIM1 data set as a function of observation time. The split violin plots show the distribution of S/N over 100 different realization of SIM1 for observing baselines of 10 to 20 years for HD (red) and ST (blue) correlations. To obtain the S/N value for each realization, the noise-marginalized optimal statistic is performed 1000 times and the median of the calculated S/Ns is plotted as the given S/N of that realization. The shaded region highlights the approximate region where NANOGrav's 12.5-year data sets resides in, which is a regime where the correlated signal is weak and the correlations cannot be distinguished from one other easily. As the baseline increases, the distinction between HD and ST S/N becomes more manifest. Specifically, 11\%  of realizations yield higher S/N for ST than HD correlations at 13 years whereas 4\% of realizations yield higher S/N for ST than HD correlations at 20 years. See the description of Fig.\ref{SN_SCATT}.}
    \label{SN_EVO}
\end{figure}
\begin{figure*}
    \includegraphics[width=\linewidth]{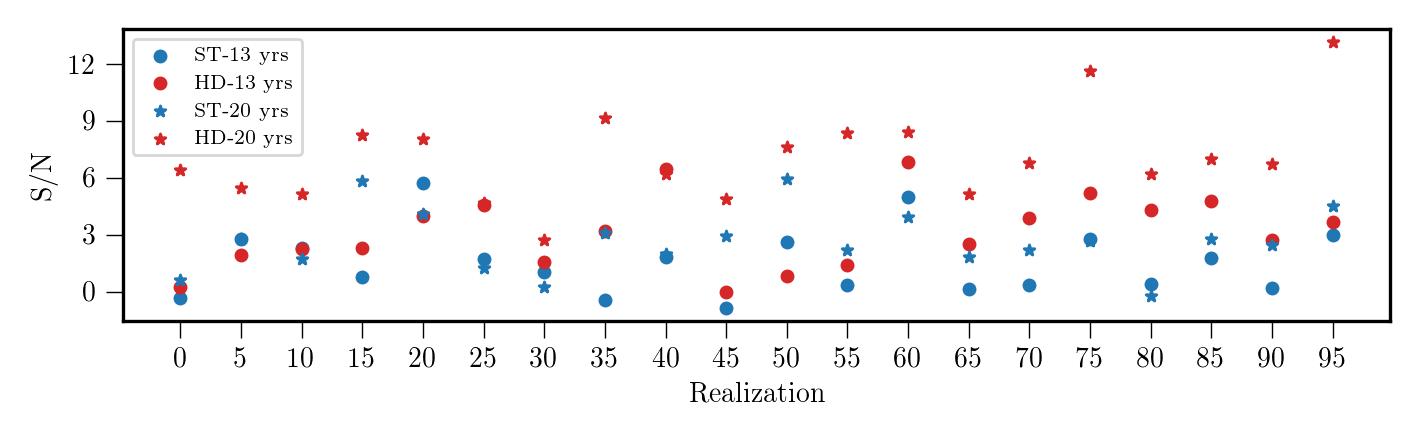}
    \caption{S/N estimation using the noise marginalized optimal statistic technique for 100 different realizations of SIM1 at observation baseline of 13 (solid circle) and 20 years (stars). For clarity, only 20 realizations out of 100 are show. In blue we show the S/N values of ST and in red the S/N values of HD correlations. 
    Over the 100 realizations, 11 yield higher S/N for ST than HD correlations at 13 years whereas 4 out of 100 yield higher S/N for ST than HD correlations at 20 years.}
    \label{SN_SCATT}
\end{figure*}
It is useful to test our detection techniques on simulated data sets in order to set our expectations for the analysis presented in this paper and future projects.

The first simulated PTA data set we have analyzed is obtained from NANOGrav's Astro4cast project~\citep{Astroforcast}. The
data set is made out of simulated pulsar time of arrivals (TOAs) for the same 45 pulsars as used in \citetalias{NG}, with similar noise characteristics as are present in our real data set along with an injected GWB signal of amplitude of $A_{TT} = 2 \times 10^{-15}$ and spectral index of $\gamma = 13/3$. The observational baseline for this simulated data set is 20 years.  
Hereon we refer to this data set as SIM1. 

The second (SIM2) and the third (SIM3) simulated data sets are identical to SIM1 except for the polarization-mode content and spectral indices of the injected GW signals. SIM2 has a GWB of ST GWs with $A_{ST} = 2 \times 10^{-15}$ and $\gamma_{ST} = 5$, and SIM3 has both ST and TT type GWB with amplitudes of $A_{ST} = A_{TT} = 2 \times 10^{-15}$ and spectral indices of $\gamma_{ST} = 5$, and $\gamma_{TT} = 13/3$ respectively. All of the simulated data sets have been analyzed using NANOGrav's ENTERPRISE to search for a common correlated red noise process.

One of the most powerful and computationally inexpensive analyses is the noise-marginalized optimal statistic technique~\citep{OPT11}. Fig.~\ref{SIM_S/N} shows the distributions for S/N of the optimal statistic with HD, monopole, and ST correlations and all three simulated data sets. We conclude the following as a result of these S/N calculations:

1. The relative high value of S/N of HD correlations relative to monopole and ST correlations observed in SIM1 at late observational times gives us confidence that if significant HD correlations are present in our data, our current techniques are capable of detection without mistaking HD correlations for ST or GW-like monopole correlations.

2. The large value of the S/N of the TT mode observed in SIM2 suggests that a ST GWB signal could be mistaken for a TT GWB signal if the ST mode is excluded from a noise-marginalized optimal statistic analysis. Given that the optimal statistic and Bayesian analyses used by \citetalias{NG} yield consistent results, ST mode and monopolar correlations of Eq.~(\ref{GW_MONO}) need to be included in searches for a GWB signal to ensure an unbiased determination of the type of correlations present in a particular data set.

3. ST and GW-like monopolar correlations yield broadly similar S/Ns. In fact, in the absence of a ST mode (as in SIM1), the two correlations give nearly identical S/Ns. Thus, distinguishing ST correlations from GW-like monopole correlations is challenging. This result requires further investigation and will be the subject of a future publication. See appendix \S\ref{app:simultsearch} for a brief discussion of one technique to distinguish ST from GW-like monopole correlations in the noise marginalized optimal static.

To conclude this subsection, we discuss the evolution of the distributions for the S/N in SIM1 as a function of observational baseline, shown in Fig.~\ref{SN_EVO} and Fig.~\ref{SN_SCATT}. One hundred different realizations of 10 to 20 year slices of SIM1 are treated as independent data sets in which we find the S/N for HD and ST correlations (GW-like monopolar correlations are almost identical to ST correlations and hence are not shown) through the noise-marginalized optimal statistic technique. Each slice of SIM1 has 100 different realizations making the total number of data sets to be 1100. These results confirm our earlier expectation for the degeneracy of ST and HD correlations (and GW-like monopole correlations) at short observational baselines. If a GWB signal with HD correlations is weak (in the case of SIM1, ``weak'' can be defined as having amplitude of $A_{TT} = 2 \times 10^{-15}$ and baseline of less than 15 years), a GWB with HD correlations can be easily mistaken for a ST and GW-like monpolar GWB. This is due to the stochastic nature of the gravitational-wave background and the non-isotropic sky distribution of pulsars used in the analyses. As can be seen in Fig.~\ref{SN_EVO}, there is a significant overlap between S/Ns calculated for ST and TT modes suggesting that we should not be surprised to observe a high relative S/N value of ST or GW-like monopole over the TT mode when only HD correlations are present but weak.

\subsection{Searching for Non-Einstenian Polarization Modes of Gravity in NANOGrav's 12.5 Year data set}%
\begin{figure*}[hbt!]
\includegraphics[width=\textwidth] {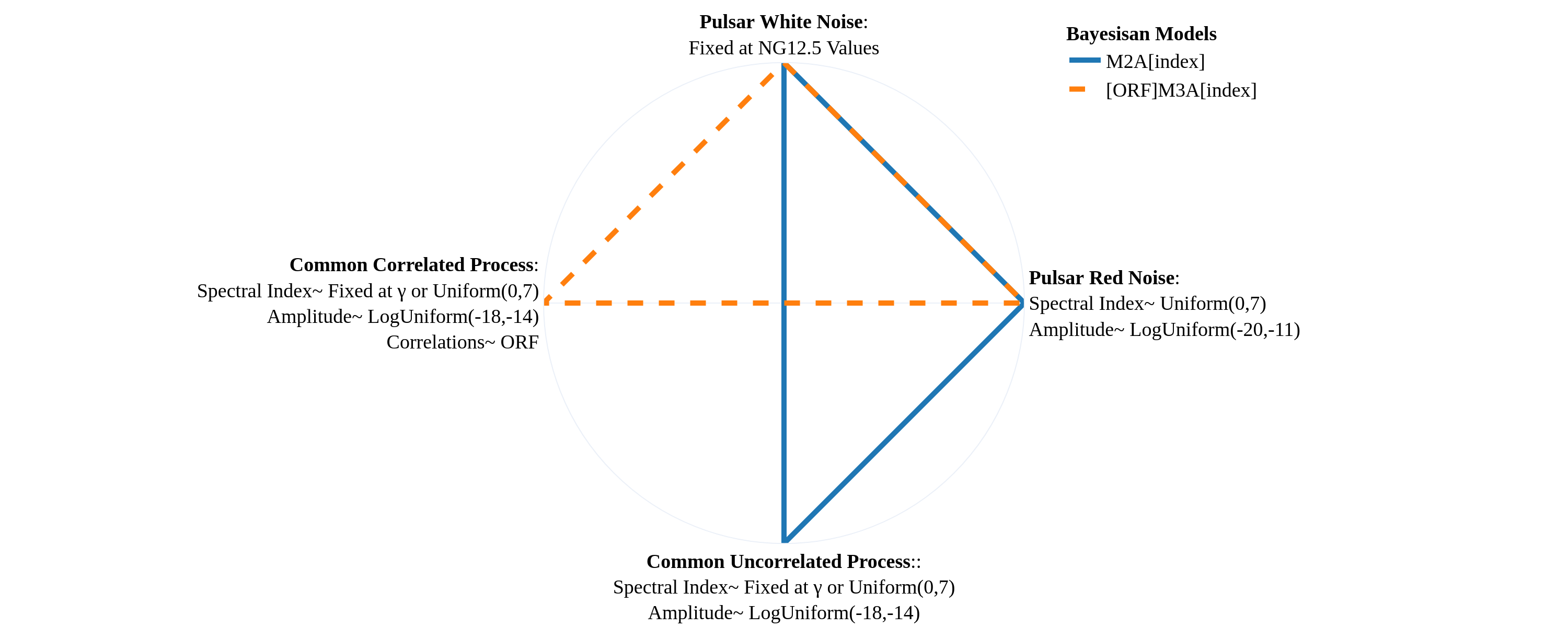}
\caption{A chart depicting the structure of the Bayesian models used. The blue line connects the pieces of a M2A[index] model while the orange dashed line connects the pieces of a [ORF]M3A[index] model. As can be seen from the figure, a M2A model consists of pulsar intrinsic red noise, white noise, and a common uncorrelated process with a given spectral index, while a M3A replaces the common uncorrelated process with a common correlated process of the type ORF and a given spectral index. More technical details of each component of a M2A or a M3A is also included in this illustration.}
\label{BModels}
\end{figure*}
\begin{figure}
\includegraphics[width=\linewidth] {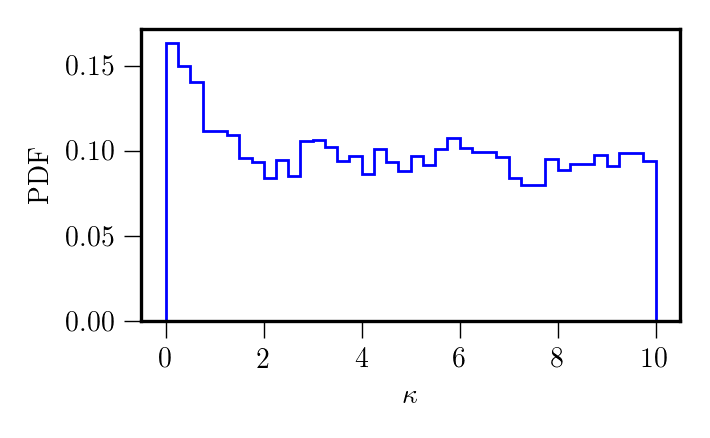}
\caption{The Bayesian posterior for the $\kappa$ parameter obtained from NANOGrav's 12.5 year data set using Eq.~(\ref{GWSignal with kappa}). We take $\gamma_m$ to be $13/3$ for the TT mode and $\gamma_m$ to be $5$ for the ST, VL, and SL modes which is the appropriate choice for binary systems. We take log-uniform priors between $-18$ and $-14$ for the TT and ST modes, a log-uniform prior between $-18$ and $-15$ for the VL mode, a log-uniform prior between $-18$ and $-16$ for the SL mode, and a uniform prior between $0$ and $10$ for the $\kappa$ 
parameter~\citep{NanoFirstAlt}. The posterior curve is uninformative for large $\kappa$ and shows a slight preference for small values. Based on this result, for all the runs featured in this paper, the $\kappa$ parameter is set to zero.}
\label{kappa}
\end{figure}

The NANOGrav 12.5-year data set was searched for an isotropic gravitational wave background consistent with Einstein's gravity in \citetalias{NG}. 
In this subsection, we extend the analyses presented in \citetalias{NG} by including searches for common red noise processes with ST, HD, and GW-like monopole correlations and their expected spectral indices. A few issues are worth keeping in mind while interpreting the results of our searches:

1. When the correlations are weak, the transverse modes of gravity can be easily mistaken for one another as seen in the S/N evolution analysis of SIM1. It is possible to obtain S/Ns as high as 3 for the ST (or GW-like monopole) mode even in the case of absence of such a mode in a PTA data set so long as the TT mode is present.

2. Though a large optimal statistic S/N value for a particular mode of gravity can be significant, 
the amplitude of that mode as seen in the correlations needs to be consistent with the amplitude of the common red noise process. For instance, in \citetalias{NG} we showed that a process with monopolar cross-correlations has a S/N distribution with a peak around 2.8 for spectral index of $\gamma = 13/3$. However, the amplitude of this monopolar process was shown to be significantly smaller than the amplitude of the uncorrelated common red noise process indicating that the majority of the common signal did not have monopolar correlations.  This is because the optimal statistic estimate of the amplitude does not include the auto-correlation terms in the covariance matrix, only the cross-correlation terms. We show further examples of this below. In \citetalias{NG}, a monopolar process was disfavored in the full Bayesian analysis which includes both auto- and cross-terms of the covariance matrix, due to the inconsistency of the amplitude of the common process with the best-fit cross-correlation-based estimate of the monopole amplitude.

3. The threshold for detection has to be large enough that it is robust to the modeling of uncertainties in the Solar System Ephemeris, \emph{BayesEphem} \citep{BayesEphem}. Long term, this will not be a problem for detection of the TT mode; the  impact of BayesEphem has been shown to be minimal as the observation time increases (see \citealt{BayesEphem}). This is likely true for the other modes, but the impact of BayesEphem on other polarization modes has not been fully explored to date.

4. Bayes factors, S/Ns, and upper-limits are all model dependent. Extreme care must be taken when interpreting Bayes factors, S/N values, or upper-limit estimates:  different choices for spectral indices, priors, and competing models can significantly affect the results of these calculations. 
\subsubsection{Bayesian Analyses} \label{sec: Bayesian}

In \S\ref{sec: the signal}, we stated that we approximate Eq.~(\ref{GWSignal with kappa}) with Eq.~(\ref{GWSignal}) by setting the dipole parameter $\lambda$ (and consequently $\kappa$) to zero. This approximation follows from our analysis of the 12.5-year data set using Eq.~(\ref{GWSignal with kappa}), with the choice of $\gamma_m$ of 13/3 for the TT mode and $\gamma_m$ of 5 for the ST, VL, and SL modes
which is appropriate for binary sources~\citep{NanoFirstAlt}.
The result of such modeling is shown in Fig.~\ref{kappa}. The posterior for $\kappa$ is uninformative for large values, and shows a slight preference for values close to zero. Hence, for simplicity, we set the $\kappa$ parameter to zero for all analyses in this paper. 

\begin{figure*}[hbt!]
\includegraphics[width=\textwidth] {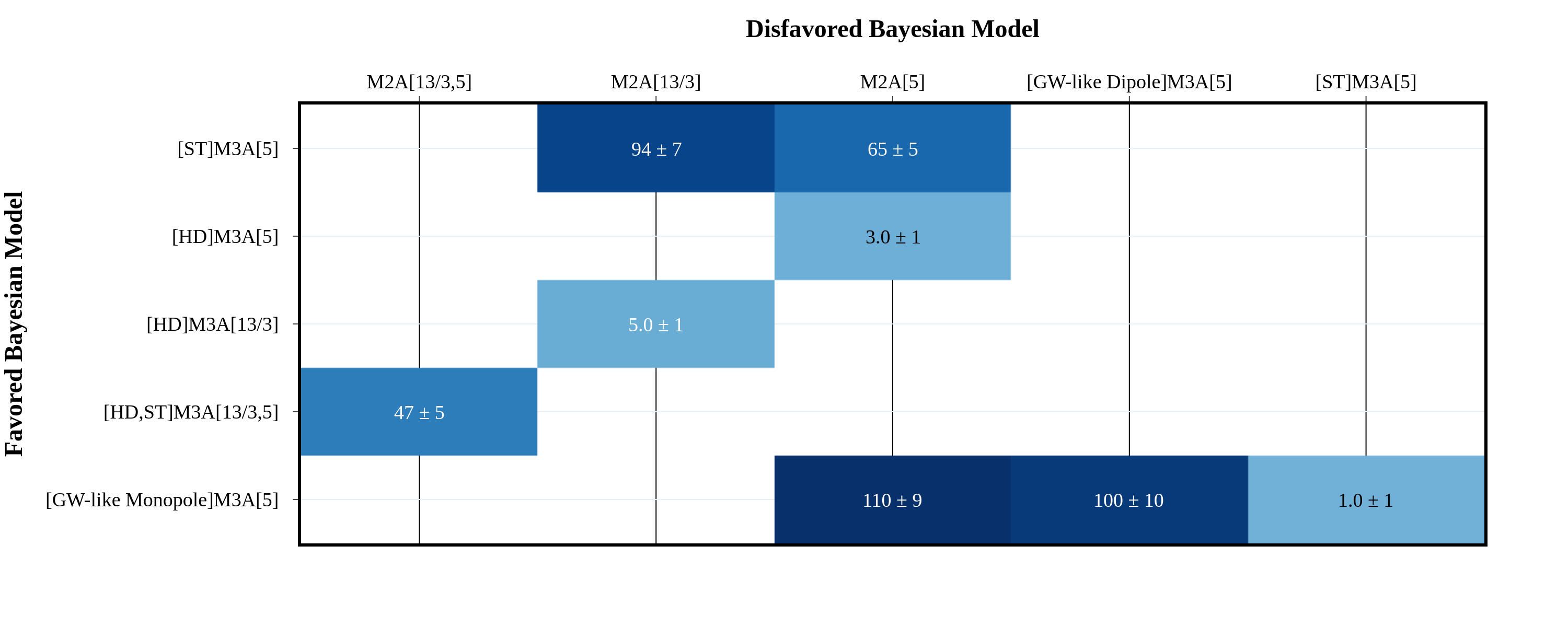}
\caption{A table illustrating estimated Bayes factors from comparison of various Bayesian models. The choice of ephemeris model is fixed at DE438 for all of the comparison in this figure. The darker the color of the blocks, the higher the value of the Bayes factor. The most favored model in all of the comparisons is a GW-like monopole. The naming convention of the models follows the structure defined in Fig.~\ref{BModels}. One can take advantage of the transitive nature of Bayes factors to compute Bayes factors for model comparisons that are not explicitly featured in this table. For instance, Bayes factor obtained from comparing [ST]M3A[5] to [HD]M3A[5] can be estimated by dividing the Bayes factor obtained from [ST]M3A[5] over M2A[5] by the Bayes factor obtained from [HD]M3A[5] over M2A[5]. The result is about $65/3 \approx 21$.}
\label{BFHeatmap}
\end{figure*}
\begin{figure}
\includegraphics[width=\linewidth] {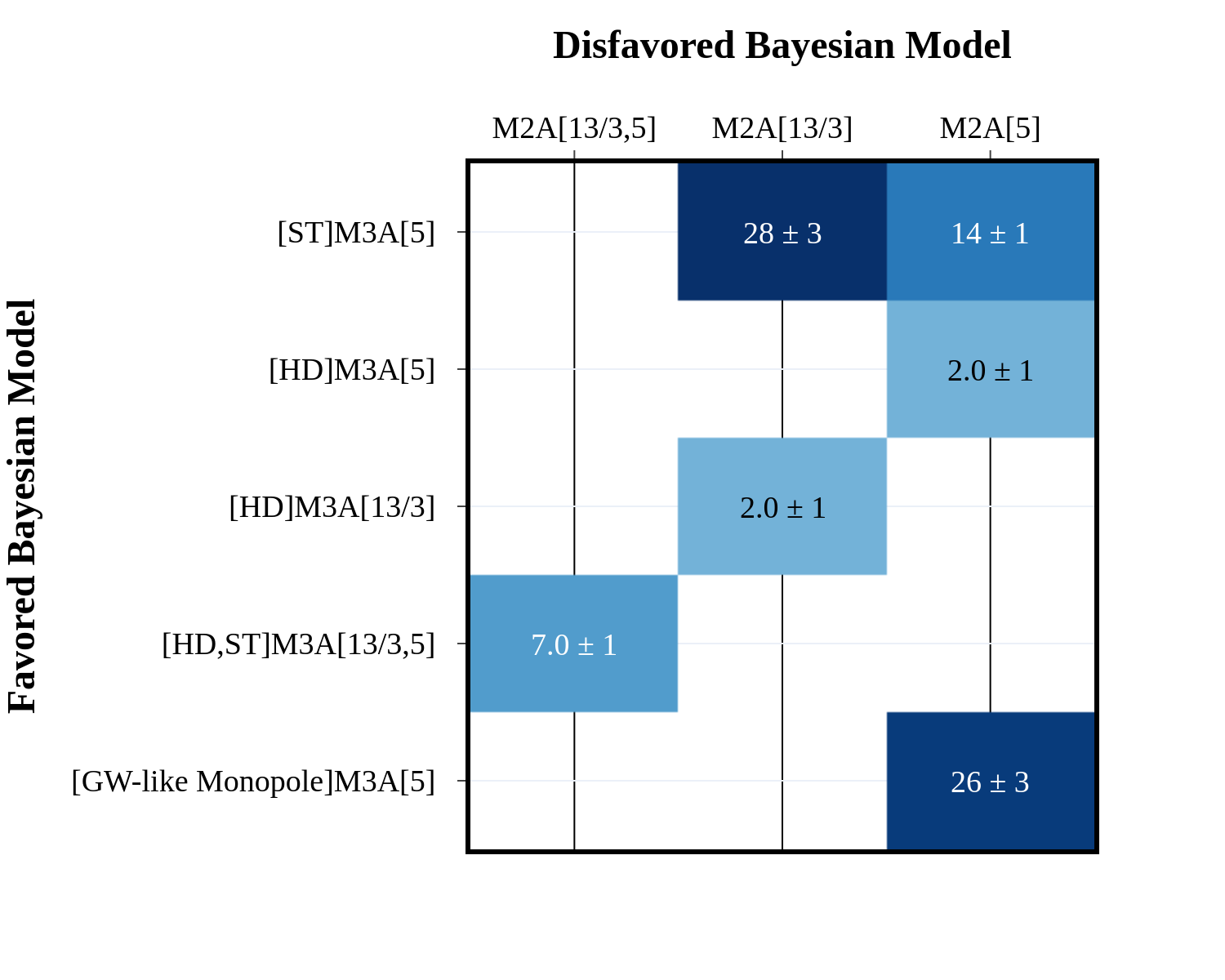}
\caption{A table illustrating estimated Bayes factors for comparison of various Bayesian models. The choice with DE438 including BayesEphem corrections for all of the comparison in this figure. The darker the color of the blocks, the higher the value of the Bayes factor. BayesEphem removes almost all of the significance from the ST and GW-like correlations. The naming convention of the models follows the structure defined in Fig.~\ref{BModels}.}
\label{BEBFHeatmap}
\end{figure}
Before describing the results of the rest of the Bayesian analyses, it is worth defining our Bayesian modeling terminology clearly. Following the naming convention of \citetalias{NG}, two general types of Bayesian models have been used in this paper: \emph{M2A} and \emph{M3A}. M2A includes a \emph{common red noise process}, pulsar intrinsic red noises, plus various backend-dependent white noise terms such as EFAC, ECORR, and EQUAD\footnote{The white noise components are EQUAD, which adds white noise in quadrature; ECORR, which describes white noise that is correlated within the same observing epoch but uncorrelated between different observing epochs; and EFAC, which scales the total template fitting TOA uncertainty after the inclusion of the previous two white noise terms. For all of these components, we used separate parameters for every combination of pulsar, backend, and receiver.}. M2A does not include correlations between pulsars so the full PTA covariance matrix is block-diagonal. 
M3A includes the same noise processes as M2A with the addition of \emph{correlations} of the common red noise process, i.e, the off-diagonal terms in the full PTA covariance matrix are populated. The type of  correlations considered for a M3A model are specified in square brackets preceding the term ``M3A''. Furthermore, for both M2A and M3A, the choice of the spectral index of the common process is specified inside square brackets following ``M2A'' or ``M3A''. For example, [HD]M3A[5] refers to a M3A model in which the type of the correlations considered for the common process is Hellings-Downs (quadrupolar) and the spectral index of this common correlated process is fixed at 5. Some M3A models may include more than one type of common correlated red noise process. For these models, we include more than one type of ORF in the square bracket preceding the term ``M3A''. For instance, [HD,ST]M3A[13/3,5] means that the M3A contains two different correlated common signals: the first being a red noise process with spectral index of 13/3 following HD type correlations, and second being a red noise process with spectral index of 5 following ST type correlations. Fig.~\ref{BModels} shows a visual illustration of our used terminology.

Extending upon the work presented in \citetalias{NG}, we show the results of 14 different Bayesian analyses that allow us to compare several models of interest. These models follow the structure outlined in Fig.~\ref{BModels}, and the resulting Bayes factors are presented in Figs.~\ref{BFHeatmap} and \ref{BEBFHeatmap} for the choices of ephemeris model DE438 and DE438 with BayesEphem corrections, respectively. Note that like \citetalias{NG}, for computational convenience, we have fixed all the pulsar intrinsic white noise values for the analyses in this section. 

As shown in Fig.~\ref{BFHeatmap}, the most favored Bayesian model is a GWB with GW-like monopolar correlations of Eq.~(\ref{GW_MONO}) with a Bayes factor greater than 100. Additionally, as a cross-check, we have reproduced the results of~\citet{Chen}, where a model with ST correlations with a spectral index of 5, [ST]M3A[5], was compared to a model without correlations and a spectral index of 13/3, M2A[13/3]. We obtain a Bayes factor of around 94 in favour of [ST]M3A[5], which is consistent with their results.

We note, however, that the calculation of the [ST]M3A[5] to M2A[13/3] Bayes factor is not the right one to make to answer the question of whether or not the data prefer ST correlations to no correlations. This is because the difference in spectral indices between the two models accounts for a significant fraction of the Bayes factor. A more appropriate comparison is obtained by calculating the Bayes factor for [ST]M3A[5] vs M2A[5], where both models have the same spectral index. For this model comparison, we obtain a Bayes factor of around 65 in favour of [ST]M3A[5]. Though tantalizing, this Bayes factor is not sufficient to claim the detection of ST modes in the NANOGrav 12.5-yr data set.  There are several reasons for this. Firstly, the Laplace approximation (see \citealt{Laplace}) gives a S/N of around 2.9 for a Bayes factor of 65, which we do not deem sufficient for a detection claim. Furthermore, given the degeneracy between TT and ST modes when correlations are present but weak (see Fig.~\ref{SN_EVO} and the discussion in \S\ref{sec:sims}), a S/N $\sim 2.9$ in favour of ST correlations is not surprising even when only TT modes are present in our data. Additionally, accounting for uncertainties in the Solar System ephemeris, we show that BayesEphem significantly reduces the Bayes factors to 14 as shown in Fig.~\ref{BEBFHeatmap}\footnote{See \citet{BayesEphem} and \citet{NG11} for a discussion of how BayesEphem changes our sensitivity to the detection of Einsteinian GWs.}. Finally, as we will show below (see \S\ref{sec:j0030}), this result is very sensitive to the inclusion of one MSP, J0030$+$0451.

We note that the data slightly prefers the GW-like monopole to ST correlations; this is again unsurprising given the analyses of simulated data in \S\ref{sec:sims} which show that the ST and GW-monopole to be more or less interchangeable. Though these results are not compelling enough to claim a detection of any mode, they are sufficiently interesting to warrant follow-up analyses in future data sets currently under preparation. 

Taking advantage of the transitive nature of Bayes factors, Fig.~\ref{BFHeatmap} allows us to compute Bayes factors for model pairs that are not featured explicitly in Fig.~\ref{BFHeatmap}. For instance, Bayes factor obtained from comparing [ST]M3A[5] to [HD]M3A[5] can be estimated by dividing the Bayes factor obtained from [ST]M3A[5] over M2A[5] by the Bayes factor obtained from [HD]M3A[5] over M2A[5]. The result is about $65/3 \approx 21$. 

Before we conclude this section, it is worth noting that not all models shown in Fig.~\ref{BFHeatmap} are equally plausible from a theoretical standpoint. All metric theories of gravity must contain, at a minimum, the two Einsteinian $+$- and $\times$-modes. Thus, even though a model with only ST spatial correlations yields a high Bayes factor, ST GWs are not predicted on their own by any metric theory of gravity. On the other hand, a compound model such as [HD,ST]M3A[13/3,5] is theoretically more well-motivated. 
\subsubsection{Frequentist Analyses and S/N Estimation}
\label{sec:freq_analyses}
\begin{figure}
    \centering
    \includegraphics[width=\linewidth]{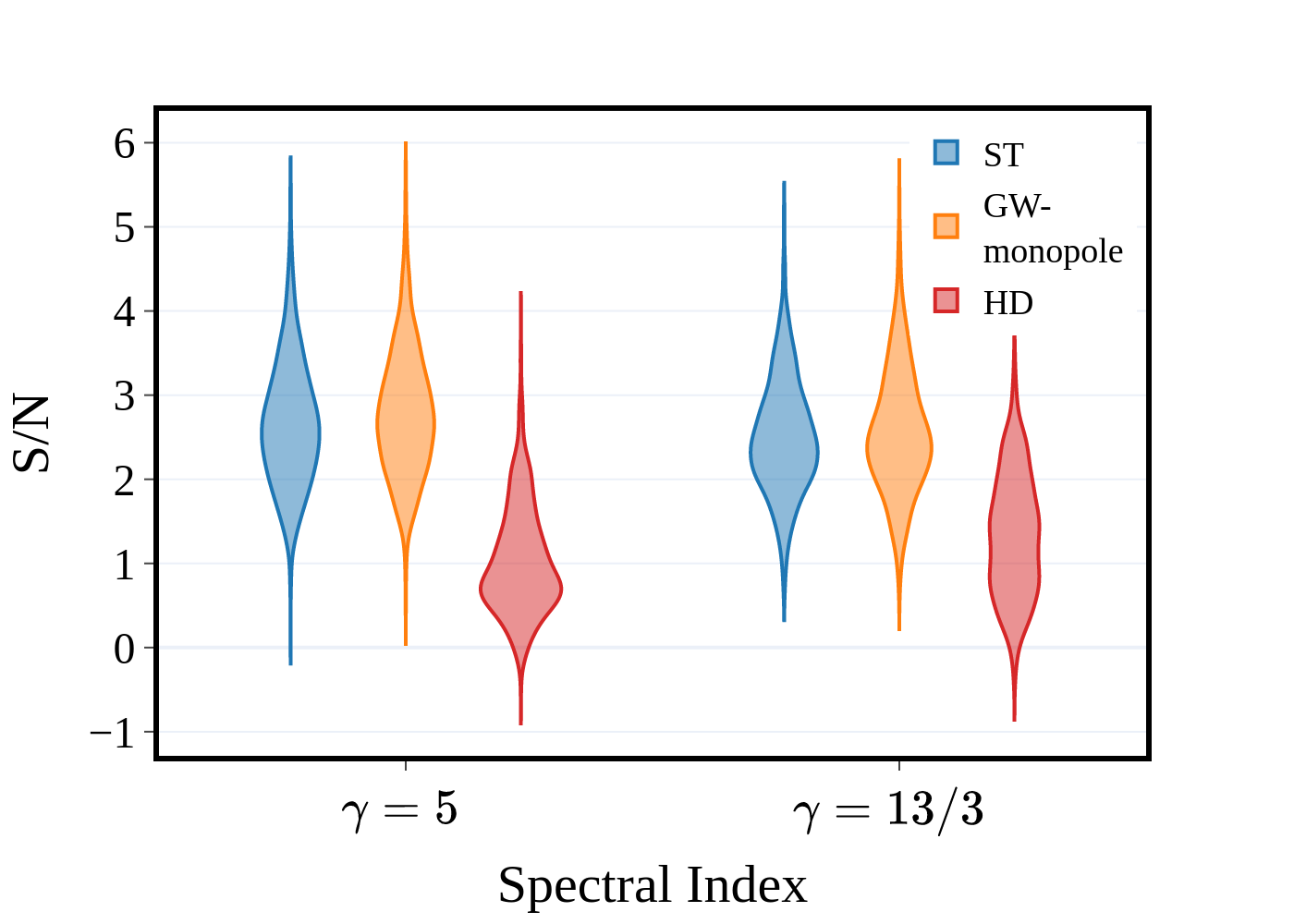}
    \caption{Violin plots depicting the S/N distribution of the 12.5 year data set for ST (blue), HD (red), and GW-like monopole (orange) correlations at different spectral indices $\gamma = 5$ and $\gamma = 13/3$. The choice of the spectral index does not affect the S/N distribution of any of the correlation patterns. ST and GW-like monopole yield similar S/Ns which is higher than S/Ns obtained from HD correlations. This surprising result can be easily understood from S/N evolution of SIM1 shown in Fig.~\ref{SN_EVO}.}
    \label{SN_comp}
\end{figure}
\begin{figure*}
    \centering
    \includegraphics[width=\linewidth]{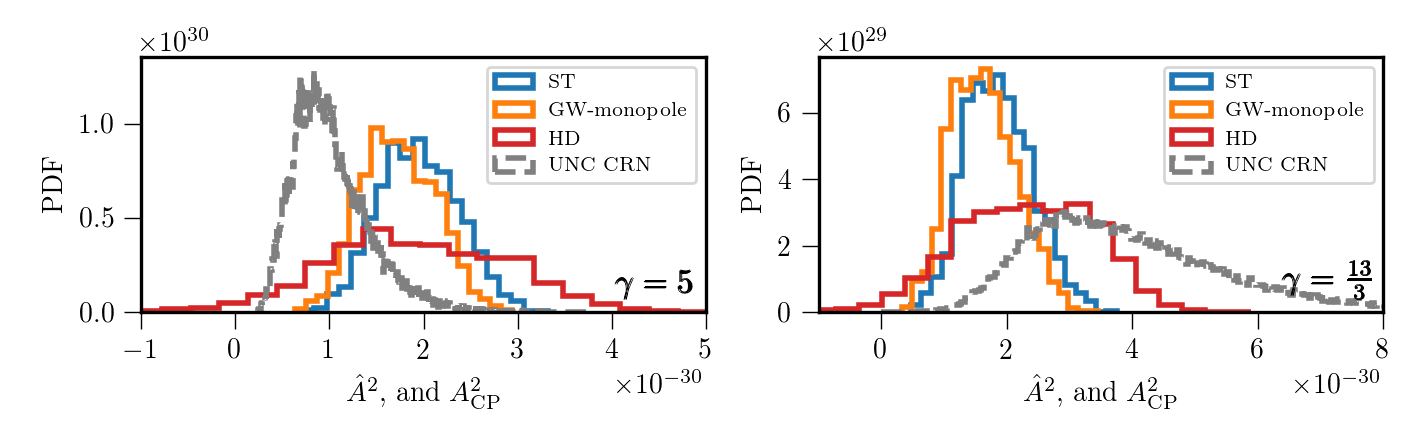}
    \caption{Distributions of the noise marginalized
    optimal statistic for
HD (red), GW-like monopole (orange), and ST (blue) spatial correlations for spectral indices of $5$ (left figure) and $13/3$ (right figure). The addition of the uncorrelated common process from a Bayesian search that only includes the auto-correlation terms, labeled as ``UNC CRN'', (grey) guides us to determine which correlations make what portion of the observed common red noise process regardless of a value of S/N. The correlations result in a range of amplitudes that are mostly not consistent with the amplitude of the common uncorrelated red noise process for spectral index of $5$ while HD correlations' amplitude is somewhat more consistent with the UNC CRN for a spectral index of $13/3$.}
    \label{AMP_comp}
\end{figure*}
As we discussed in \S\ref{sec:sims}, the noise-marginalized optimal statistic offers a very robust and computationally inexpensive alternative to the Bayesian techniques by estimating the S/N. The S/N can be related to the Bayes factor using the Laplace approximation \citep{Laplace}; specifically,
\begin{equation}
    \ln B \approx \rho^2/2
\end{equation}
where, $B$ is the Bayes factor and $\rho$ is the S/N. Later, we will show how our calculated Bayes factors are consistent with our S/N estimates through this Laplace relation.

Fig.~\ref{SN_comp} shows the distribution of S/N for ST, GW-like monopole, and HD correlations obtained by calculating the noise-marginalized optimal statistic for the 12.5 year data set. The S/N calculation is performed for two choices of the spectral index, 13/3 and 5. Even though the choice of spectral index does not affect the results of S/N estimation significantly, the estimates for the amplitude of the red noise process change because of the covariance between amplitudes and spectral indices: the amplitude of a red noise process is lower with spectral index of 5 compared to a spectral index of 13/3. Figure ~\ref{AMP_comp} shows distributions of the amplitudes for spectral indices of 13/3 and 5. In the case of $\gamma = 5$, none of the correlated models match the amplitude of the common red noise process suggesting that despite the high S/N value of ST and  GW-like monopole at this spectral index, these modes do not makeup much of the observed common process. The best match occurs in the case of $\gamma = 13/3$ where the amplitude of a model with HD correlations overlaps somewhat significantly with the amplitude of the uncorrelated common red noise process.
This is noted in \citetalias{NG} as well.

\begin{figure*}
    \centering
    \includegraphics[width=\linewidth]{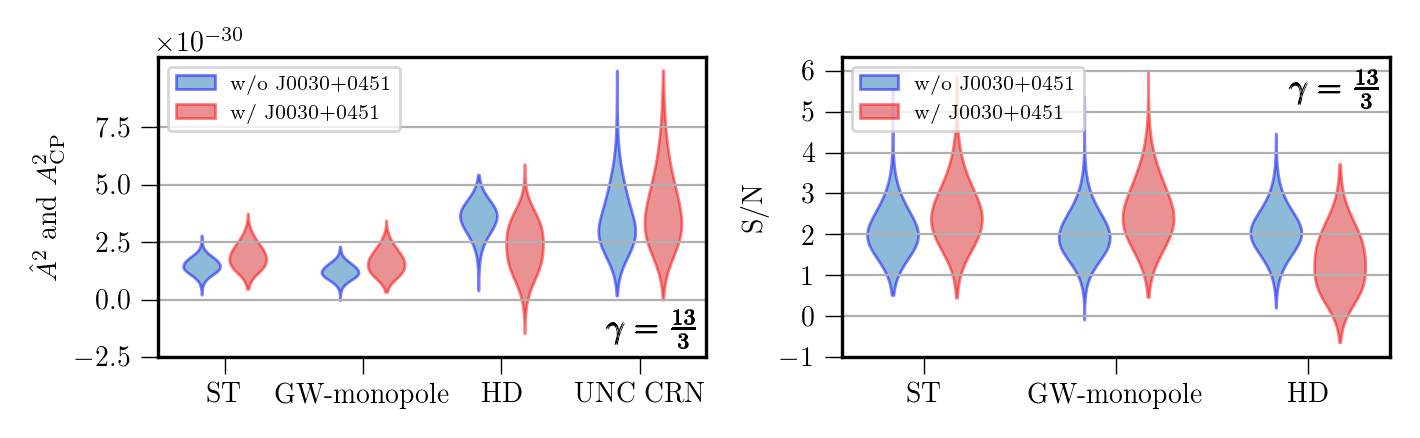}
    \caption{Distributions of the noise marginalized optimal statistic and S/N for
HD, GW-like monopole, and ST spatial correlations for $\gamma=13/3$. The red violin plots show the results of optimal statistic analyses done on the full 12.5-year data set whereas the blue violin plots showcase the results of optimal statistic analyses done on the 12.5-year data set excluding the pulsar J0030$+$0451. The S/N of non-HD correlations is significantly reduced by omitting MSP J0030$+$0451. Additionally, there is a notable improvement in HD correlation's amplitude consistency with the amplitude of the uncorrelated common red noise process (UNC CRN) and an increase in the HD S/N.}
    \label{comp_J0030_4.33}
\end{figure*}
\begin{figure*}
    \centering
    \includegraphics[width=\linewidth]{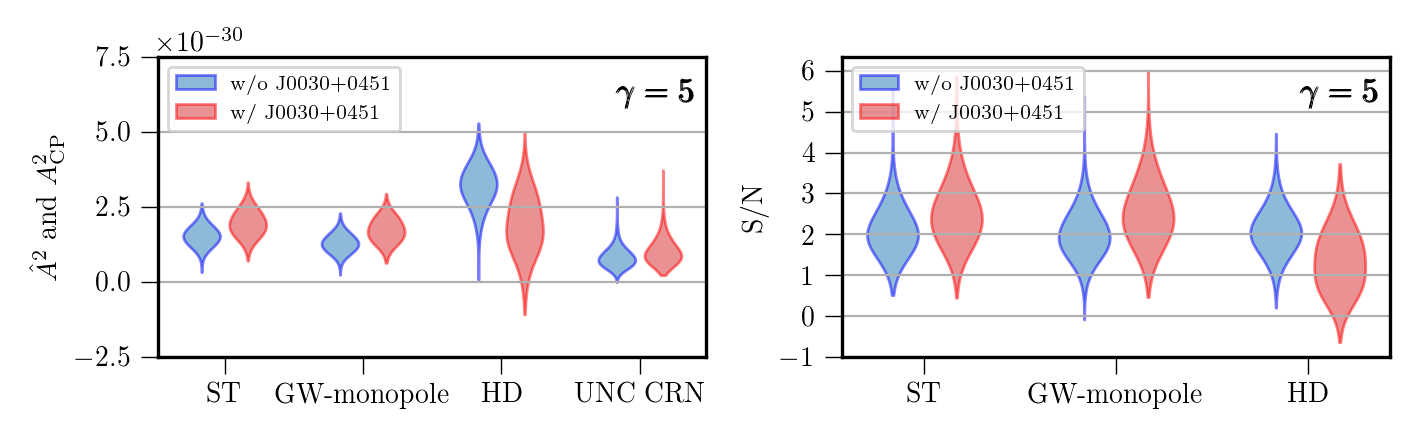}
    \caption{Distributions of the optimal statistic and S/N for
HD, GW-like monopole, and ST spatial correlations for $\gamma=5$. The red violin plots show the results of optimal statistic analyses done on the full 12.5-year data whereas the blue violin plots showcase the results of optimal statistic analyses done on the 12.5-year data set excluding the pulsar J0030$+$0451. The problem of relatively high S/N of non-HD correlations is resolved by omitting MSP J0030$+$0451. However, no noticeable improvement can be seen in consistency of amplitudes of any of the correlations relative to the amplitude of the uncorrelated common red noise process (UNC CRN).}
    \label{comp_J0030_5}
\end{figure*}
\subsubsection{MSP J0030$+$0451 and GW-monopole/ST Correlations} \label{sec:j0030}

\citetalias{NG} identified 10 of the 45 pulsars included in the analysis to be the most significant contributors to the common red noise process that was detected. These pulsars are J1909-3744, J2317$+$1439, J2043$+$1711, J1600$-$3053, J1918$-$0642, J1744$-$1134, J1910$+$1256, J0030$+$0451, J2145$-$0750, and J1640$+$2224. We performed our analyses anew, this time removing each of the above pulsars one at a time, and identified a main culprit for the GW-monopole/ST correlations we find in our data: MSP J0030$+$0451.

Removing this pulsar from our analyses results in the most significant changes to the S/N, recovered amplitude, and Bayes factor estimation. This effect can be seen in Figs.~\ref{comp_J0030_4.33} and \ref{comp_J0030_5} for the choice of spectral indices $13/3$ and $5$, respectively. 

As shown in Fig.~\ref{comp_J0030_4.33} and Fig.~\ref{comp_J0030_5}, the optimal statistic analyses show that the S/N of GW-like monopole (as well as ST) drops from about 2.8 to 2 when MSP J0030$+$0451 is removed. Simultaneously, the S/N of HD increases from 1 to about 2. Furthermore, the amplitude recovery for HD seems to be more consistent with the common red noise process, while the amplitudes of GW-like monopole and ST become less consistent. Our Bayesian analyses agree with the optimal statistic results: when dropping MSP J0030$+$0451 from the analysis, the Bayes factor for [GW-like Monopole]M3A[5] to M2A[5] drops from about 100 to about 15 and the Bayes factor obtained for [HD]M3A[13/3] to M2A[13/3] increases from about 5 to about 10. It is worth pointing out that the changes in the optimal statistic and the Bayes factors are consistent with what we expect from the Laplace approximation.

We do not yet understand why MSP J0030$+$0451 is causing this effect but we suspect incomplete noise modeling of this pulsar as the most plausible cause; MSP J0030$+$0451 has been shown to be problematic in detection analyses in the past (see ~\citealt{11YEAR_SLICES}). However, we do not believe that omitting MSP J0030$+$0451 from our analyses is the right solution to this problem. Rather, a thorough investigation of MSP J0030$+$0451's data, along with improved and more sophisticated noise modeling for this pulsar (and probably others) is the more robust path forward. NANOGrav is actively working on advanced noise modeling of the pulsars used in the 12.5-year data set and the results of these efforts are in preparation.   
\subsection{Upper Limit Estimation}
\label{sec:uls}
\begin{figure*}
    \centering
    \includegraphics[width=\linewidth]{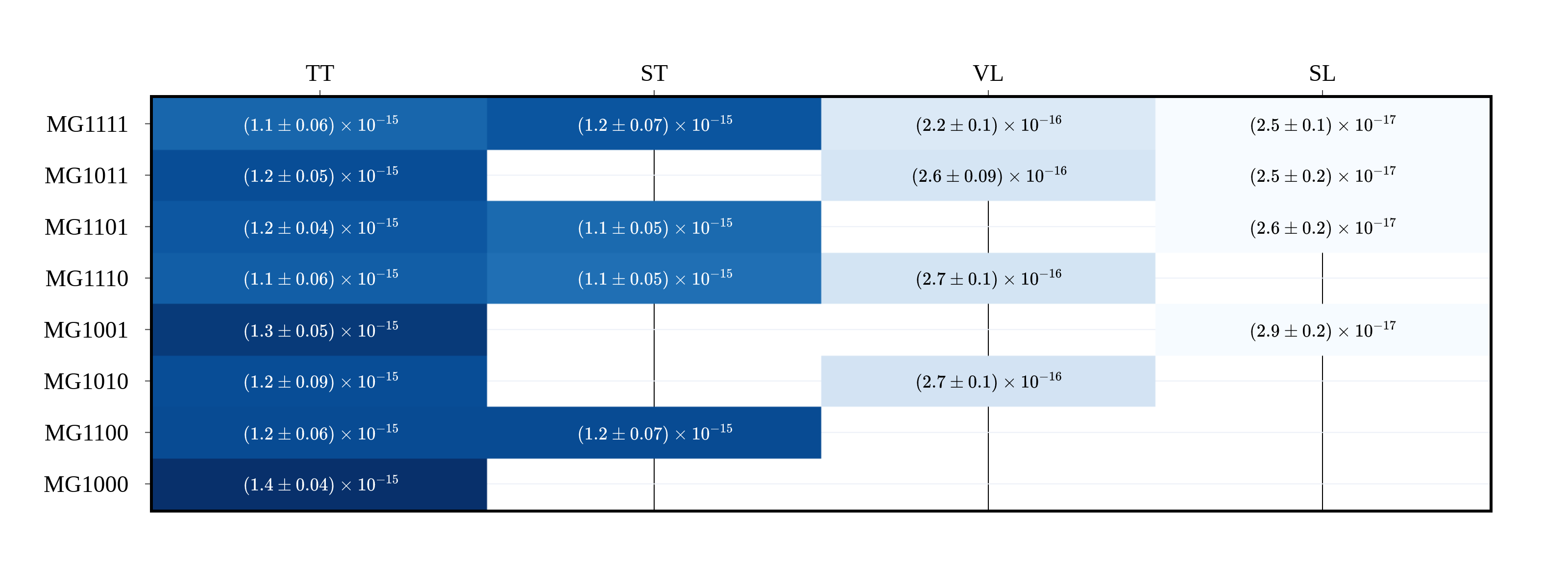}
    \caption{A heat-map illustrating 95\%  upper limit estimated for eight different models labeled based on the naming convention introduced in \S\ref{sec:uls}. The darker the color of a block, the higher the value of the upper limit. The spectral index for all of the polarization modes is fixed at $\gamma = 5$. The low value of SL and VL upper limits attests to the high sensitivity of pulsar timing in detecting these modes.}
    \label{upplim}
\end{figure*}
\begin{figure}
    \centering
    \includegraphics[width=\linewidth]{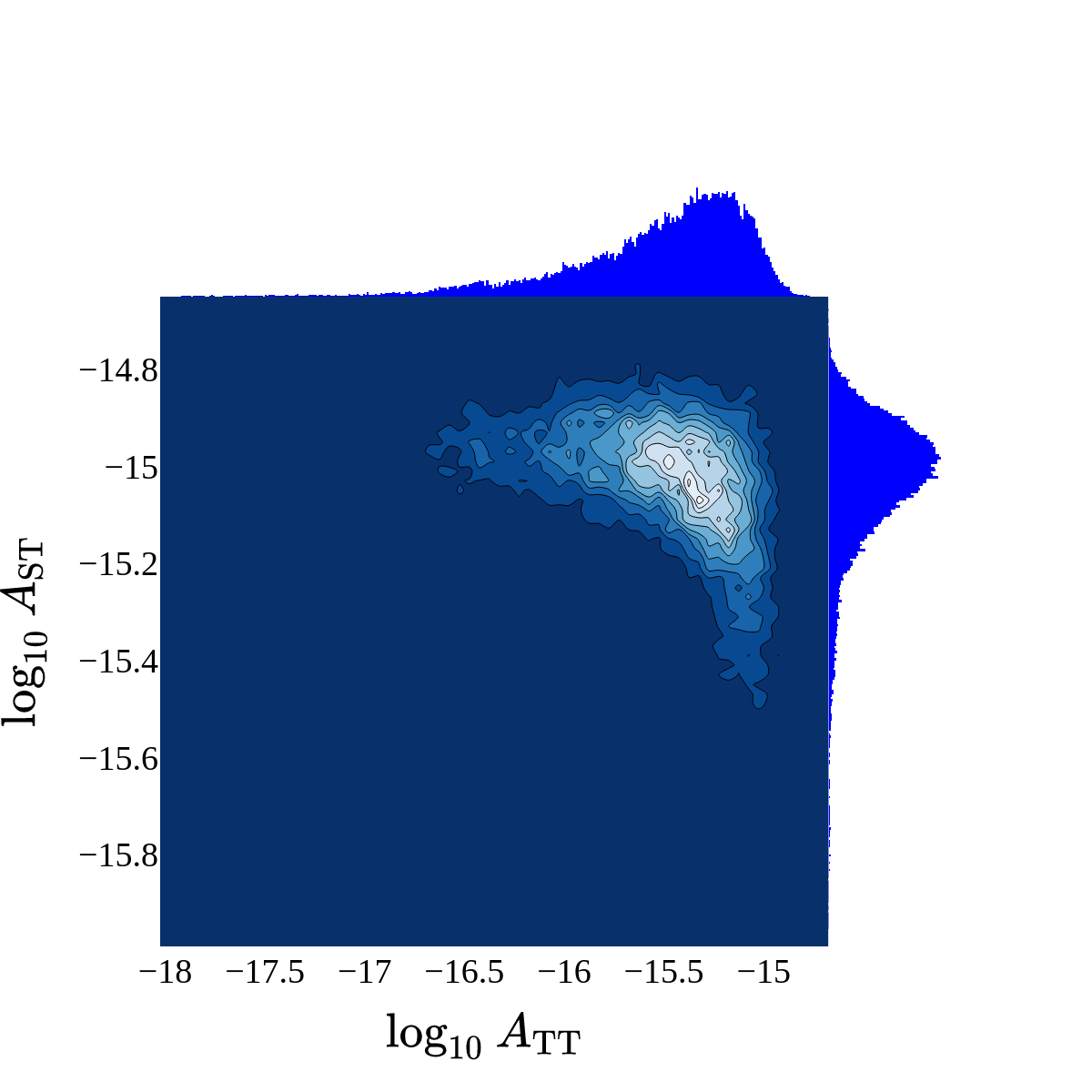}
    \caption{A two dimensional contour plot illustrating the posterior distribution of ST and TT polarization modes obtained from a Bayesian model that contains both polarization modes as its correlated common process for spectral index of $\gamma = 5$. The color scale for this contour plot is inverted: the darker a region is, the less likely it is for an amplitude to belong to those regions. Both mode's amplitude posterior peak around the same value.}
    \label{TTSTContour}
\end{figure}
In the absence of a detection of any polarization mode of gravity, we place constraints on the amplitude of these modes using our Bayesian techniques for specific choices of spectral index and the number of expected polarization modes. As mentioned at the beginning of this section, upper limits are model dependent. Different choices of priors, number of polarization modes considered, and spectral indices can affect the results. We have chosen to report our 95\% upper limits for eight different models. Details of the models are discussed below.

\emph{Common red noise prior}: All modes have uniform priors for their amplitude. More specifically,
\begin{eqnarray}
A_{\text{TT}}&&=\text{Uniform}(10^{-18},10^{-12}),\nonumber\\ A_{\text{ST}}&&=\text{Uniform}(10^{-18},10^{-12}),\nonumber\\ A_{\text{VL}}&&=\text{Uniform}(10^{-18},10^{-15}),\nonumber\\ 
A_{\text{SL}}&&=\text{Uniform}(10^{-18},10^{-16}),\nonumber
\end{eqnarray}
 in all selected models. The VL and SL modes have a narrower and lower range of prior because we expect their amplitudes to be lower than that of transverse modes. For the models in which the pulsar distances are needed (VL, SL), pulsar distances follow a normal distribution with mean at 1 kpc and standard deviation of 0.2 kpc, and we marginalize over the pulsar distance.
 
\emph{Naming convention}: The naming convention adopted for the models considered in this subsection seeks to categorize all metric theories of gravity into eight families based on their predicted polarization content. The prefix ``MG'' is short for  \emph{Metric theory of Gravity} and the succeeding four digits speak to the existence, denoted by 1, or lack of existence, denoted by 0, of the possible four polarization modes $\text{TT}$, $\text{ST}$, $\text{VL}$, and $\text{SL}$. 

For example, MG1000 is Einstein's general theory of relativity, and MG1100 is a theory with TT and ST modes (e.g, Brans-Dicke gravity). Note that all of the eight families of theories possess the TT mode since this is required for all valid metric theories of gravity.

\emph{Spectral index}: For convenience we have taken the power in all modes to have a spectral index of $\gamma = 5$, which corresponds to flat spectrum in $\Omega_{GW}$, the ratio of the density in GWs to the critical density. 

\emph{Intrinsic pulsar noise prior}: All 45 pulsars in our analyses have log-uniform priors on the amplitude from $-20$ to $-11$. The choice of log-uniform priors on the intrinsic red noise is conservative, in the sense that it favours lower amplitudes for the intrinsic red noise relative to the common process (which has uniform priors), and results in larger upper limits on the common red noise process (see \citealt{ModelDep}). The spectral indices of intrinsic pulsar red noises vary uniformly from 0 to 7. 

\emph{Spatial cross-correlations in the models}: For computational convenience we have not included correlations in most of our upper limit analyses. The minor improvements that are possible with the inclusion of cross-correlations do not justify the computational cost of performing such upper-limit analyses. These improvements are particularly small in the case of the non-transverse modes of gravity because the auto-correlation terms dominate the cross terms significantly. 

The upper limit values listed in Fig.~\ref{TTSTContour} can be used to place constraints on the detailed parameters of theories that couple to those modes as well as the astrophysical sources that are capable of producing those modes. Such studies are not within the scope of this work.  

We have also performed an additional analysis that includes correlations for one of our models, MG1100. Unlike the VL and SL modes, the magnitude of the cross-correlations for the TT and ST modes are of the same order as the auto-correlations. We therefore expect the inclusion of the cross terms for theories with TT and ST modes to have the largest effect on their upper limits.  
The contour plot for the amplitude posterior of TT and ST in this model is shown in Fig.~\ref{TTSTContour}. The upper limits obtained from this model ($A^{95\%}_{TT} = (9.7 \pm 0.4)\times 10^{-16}$ and $A^{95\%}_{ST} = (1.4 \pm 0.03)\times 10^{-15}$), are slightly smaller than the ones obtained from the MG1100 model without correlations (see Fig.~\ref{upplim}). Although we expect the corrections to our upper limits to be small, for future data sets we will develop and implement correlations for the VL and SL polarization modes in our pipeline.

\section{Summary}\label{sec:summary}

NANOGrav's 12.5-year data set shows strong evidence for a common stochastic process, a red noise process with the same amplitude and spectral index across all pulsars. This common process, however, does not show strong evidence in favor of any spatial correlations that we have considered. 
The slight preferences for ST and GW-like monopolar correlations are not robust to the modeling of uncertainties in the solar system ephemeris, and seem to be associated with one particular pulsar, J0030$+$0451. A thorough investigation  of MSP J0030$+$0451 data set, along with improved and more sophisticated noise modeling for this and other pulsars is likely to shed more light on this issue. Additionally, as our simulations show, given the baseline, amplitude of the common process, and the levels of white and other noise present in the 12.5 year data, it is possible to misconstrue a weak GWB with HD correlations as a GWB with ST or GW-like monopolar correlations. 

Thus, we disagree with~\citet{Chen} on the existence of strong evidence in favor of a GWB with ST correlations in the NANOGrav 12.5-year data set. Strong evidence for such correlations would require greater Bayes factors and S/N estimations, a higher consistency between the amplitude of the uncorrelated common red noise process and the process with ST correlations, robustness to the modeling of ephemerides uncertainties, and robustness to the removal of individual pulsars. As mentioned, we found MSP J0030$+$0451 to be a significant contributor to the existence of the observed GW-like monopole (or ST correlations): removing this pulsar results in significant reduction of S/N (from 2.8 to 2) and Bayes factor (from 100 to 10) in the case of GW-like monopole. This issue will be followed up in detail in analyses of the upcoming more sensitive 15-year data set.

In the absence of a detection, we place upper limits on the amplitudes of the various modes present in metric theories gravity. Each of the models in this paper have their own set of upper limits which varies from model to model. For sources of GWs that can produce a GWB background signal with spectral index of 5, the estimated upper limits are reported in Fig.~\ref{upplim}. The reported upper limits can be used to place constraints on the parameters of theories that lead to such GW polarization content and the sources that are capable of producing GWs with the various polarization modes.  We do not attempt to make such connections in this paper, but they should be useful in studies of alternative theories of gravity. 

With the release of a new data set on the horizon, NANOGrav's 15 year data set, we will continue to search for evidence of additional polarization modes of gravity. We anticipate that more pulsars, longer observation times, and improved noise modeling of pulsars will aid us greatly in finding and distinguishing the spatial correlation patterns in our data.
\acknowledgments
\emph{Author contributions.}
\input{12p5-altpol-contrib}

\emph{Acknowledgments.}
\input{12p5-altpol-acknowledge}
\facilities{Arecibo, GBT}

\software{\texttt{ENTERPRISE} \citep{enterprise}, \texttt{enterprise\_extensions} \citep{enterprise_extensions}, \texttt{libstempo} \citep{libstempo}, \texttt{matplotlib} \citep{matplotlib}, \texttt{PTMCMC} \citep{ptmcmc}, \texttt{tempo2} \citep{tempo2}, \texttt{plotly} \citep{plotly}}
\clearpage
\appendix
\section{Distinguishing Scalar-tensor from GW-like Monopole Correlations in the Noise Marginalized optimal statistic} 
\label{app:simultsearch}
\begin{figure}
    \centering
    \includegraphics[width=\linewidth]{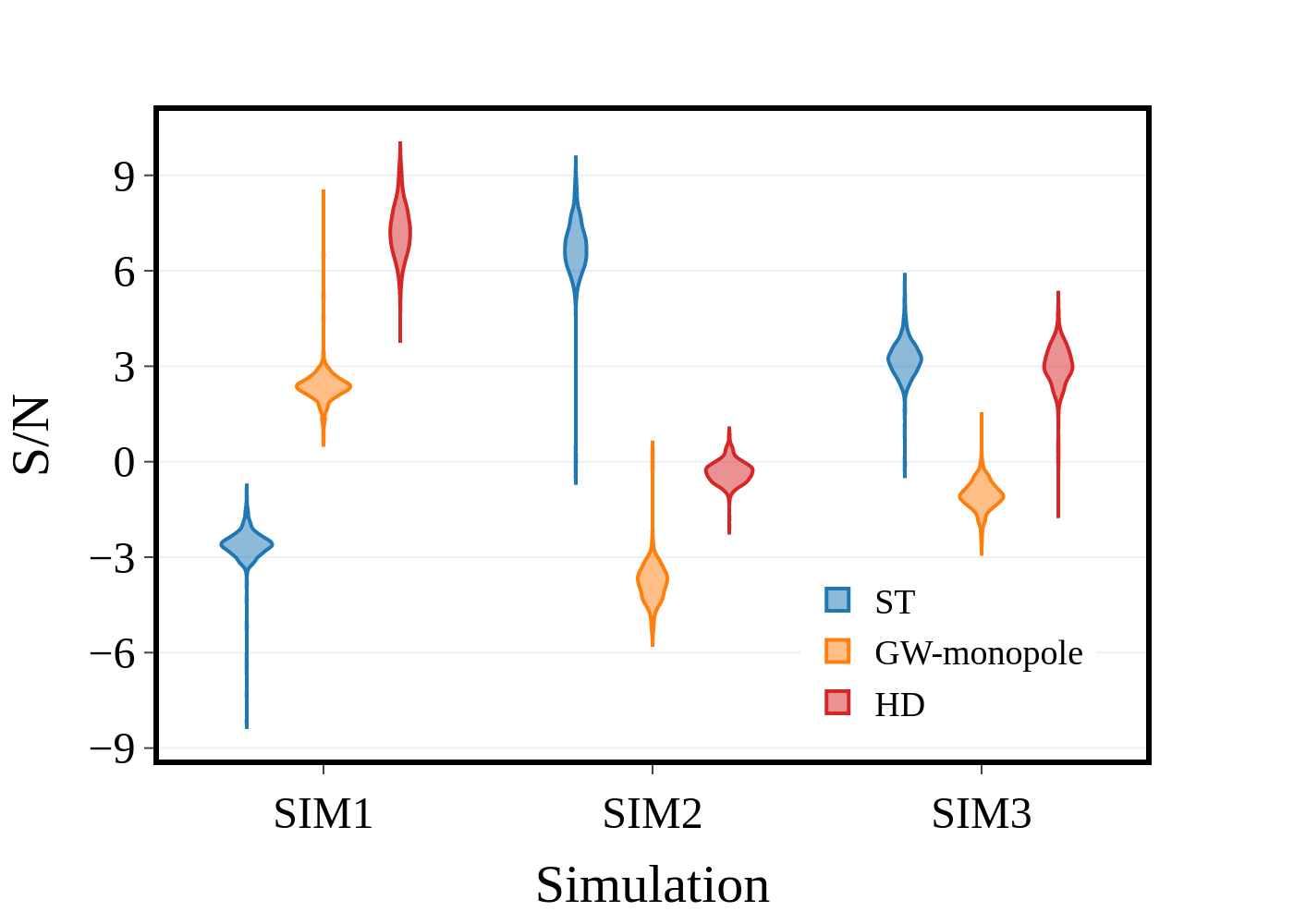}
    \caption{Violin plots showing the S/N distribution of the noise marginalized optimal statistic for one realization of simulated data sets SIM1, SIM2, and SIM3 (see the main text for a description). The data are searched for three different correlation patterns simultaneously: ST (blue), HD (red), and GW-like monopole (orange). The S/N distribution for each simulated data set is obtained from the calculation of the noise-marginalized optimal statistic evaluated 1000 times. Significant improvements in the estimation of the S/N (compared to Fig.~\ref{SN_comp}) is made by changing the noise marginalized optimal statistic to search for ST, HD, and GW-like monopole simultaneously: i) signal to noise is no longer over estimated, and ii) the overlap between S/N values of ST and GW-like monopole correlations is significantly reduced.}
    \label{sn_simult}
\end{figure}

Separating ST from GW-like monopole correlations introduces new challenges to the usual detection procedure as outlined in \S\ref{sec:sims}. These two correlations seem to overlap significantly to the point where they can be used interchangeably in our Bayesian and frequentist analyses. One way to mitigate this problem is to search for such modes simultaneously as opposed to separately, which has been the default procedure thus far for the noise marginalized optimal statistic technique to compute signal-to-noise values. There are two benefits to searching for degenerate correlations simultaneously: i) we avoid the over-estimation of the S/N of mode, ii) we achieve a significant reduction in the overlap of S/N distributions for the various correlation patterns. For instance, searching for ST and GW-like correlation patterns simultaneously in SIM2 results in reduction of the high S/N value of $17$ to $7$ for ST and $17$ to $-4$ for GW-like monopole (see Fig.~\ref{sn_simult}). The addition of this new feature to the noise marginalized optimal statistic technique will be explored in depth in a separate paper. 
\section{Bayesian Methods} \label{app:methods}

We used Markov chain Monte Carlo (MCMC) methods to stochastically sample the joint posterior of our model parameter spaces, and use Monte Carlo integration to deduce marginalized distributions, where $\int f(\theta)p(\theta|d)d\theta \approx \langle f(\theta_i)\rangle$ for the integral of an arbitrary function $f(\theta)$ over the posterior $p(\theta|d)$ of which the samples $\{\theta_i\}$ are randomly drawn.
Where necessary, we estimated the uncertainty on the marginalized posterior value to be the Monte Carlo sampling error of the location $\hat\theta_x$ of the $x$-th quantile:
\begin{equation}
	\frac{\sqrt{x(1-x)/N}}{p(\theta=\hat\theta_x|d)},
\end{equation}
where $N$ is the number of (quasi-)independent samples in our MCMC chain.

As described in \citetalias{abb+18b}, we employ two techniques for model selection based on the relationship between the competing models. For \textit{nested} models that compare the additional presence of a signal to that of noise alone, we used the \textit{Savage-Dickey approximation} \citep{d71}. This requires adequate sampling coverage of low amplitude posterior regions in order to compute the Savage-Dickey density ratio, which corresponds to the prior to posterior density at zero amplitude: Bayes factor $ = p(A=0) / p(A=0|d)$. In practice this means that the method is only useful for moderate model odds contrasts, and while this was used extensively in \citetalias{abb+18b}, the strength of the recovered signal in this paper exceeds the reliability of the Savage-Dickey approximation without additional sampling strategies to explore the low amplitude posterior region. For \textit{disjoint} models, models that are not easily distinguished parametrically, and indeed all model selection in this paper, we used the \textit{product-space method} \citep{cc95,g01,hee15,tvs20}. This recasts model selection as a parameter estimation problem, introducing a model indexing variable that is sampled along with the parameters of the competing models, and which controls which model likelihood is active at each MCMC iteration. The ratio of samples spent in each bin of the model indexing variable returns the posterior odds ratio between models. The efficiency of model transitions is controlled by our prior model probabilities, which we usually set to be equal. However, one can improve the odds ratio computation by performing a pilot run, whose odds ratio estimate can be used to re-weight the models in a follow-up run. This will ensure more equitable chain visitation to each model, after which the model index posterior is re-weighted back to the true model contrast. 

\section{Software} \label{app:software}

We used the software packages \texttt{enterprise} \citep{enterprise} and \texttt{enterprise\_extensions} \citep{enterprise_extensions} to perform the Bayesian and frequentist searches. These packages implement the signal models, likelihood, and priors.
We used the software package \texttt{PTMCMCSampler} \citep{ptmcmc} to perform the MCMC for the Bayesian searches. 
\bibliography{12p5-altpol}
  
\end{document}

%% file: 12p5-altpol-authors.tex
\author{Zaven Arzoumanian}
\affiliation{X-Ray Astrophysics Laboratory, NASA Goddard Space Flight Center, Code 662, Greenbelt, MD 20771, USA}
\author{Paul T. Baker}
\affiliation{Department of Physics and Astronomy, Widener University, One University Place, Chester, PA 19013, USA}
\author{Harsha Blumer}
\affiliation{Department of Physics and Astronomy, West Virginia University, P.O. Box 6315, Morgantown, WV 26506, USA}
\affiliation{Center for Gravitational Waves and Cosmology, West Virginia University, Chestnut Ridge Research Building, Morgantown, WV 26505, USA}
\author{Bence B\'{e}csy}
\affiliation{Department of Physics, Montana State University, Bozeman, MT 59717, USA}
\author{Adam Brazier}
\affiliation{Cornell Center for Astrophysics and Planetary Science and Department of Astronomy, Cornell University, Ithaca, NY 14853, USA}
\affiliation{Cornell Center for Advanced Computing, Cornell University, Ithaca, NY 14853, USA}
\author{Paul R. Brook}
\affiliation{Department of Physics and Astronomy, West Virginia University, P.O. Box 6315, Morgantown, WV 26506, USA}
\affiliation{Center for Gravitational Waves and Cosmology, West Virginia University, Chestnut Ridge Research Building, Morgantown, WV 26505, USA}
\author{Sarah Burke-Spolaor}
\affiliation{Department of Physics and Astronomy, West Virginia University, P.O. Box 6315, Morgantown, WV 26506, USA}
\affiliation{Center for Gravitational Waves and Cosmology, West Virginia University, Chestnut Ridge Research Building, Morgantown, WV 26505, USA}
\affiliation{CIFAR Azrieli Global Scholars program, CIFAR, Toronto, Canada}
\author{Maria Charisi}
\affiliation{Department of Physics and Astronomy, Vanderbilt University, 2301 Vanderbilt Place, Nashville, TN 37235, USA}
\author{Shami Chatterjee}
\affiliation{Cornell Center for Astrophysics and Planetary Science and Department of Astronomy, Cornell University, Ithaca, NY 14853, USA}
\author{Siyuan Chen}
\affiliation{Station de Radioastronomie de Nancay, Observatoire de Paris, Universite PSL, CNRS, Universite d'Orleans, 18330 Nancay, France}
\affiliation{FEMTO-ST Institut de recherche, Department of Time and Frequency, UBFC and CNRS, ENSMM, 25030 Besancon, France}
\affiliation{Laboratoire de Physique et Chimie de l'Environment et de l'Espace, LPC2E UMR7328, Universite d'Orleans, CNRS, 45071 Orleans, France}
\author{James M. Cordes}
\affiliation{Cornell Center for Astrophysics and Planetary Science and Department of Astronomy, Cornell University, Ithaca, NY 14853, USA}
\author{Neil J. Cornish}
\affiliation{Department of Physics, Montana State University, Bozeman, MT 59717, USA}
\author{Fronefield Crawford}
\affiliation{Department of Physics and Astronomy, Franklin \& Marshall College, P.O. Box 3003, Lancaster, PA 17604, USA}
\author{H. Thankful Cromartie}
\affiliation{Cornell Center for Astrophysics and Planetary Science and Department of Astronomy, Cornell University, Ithaca, NY 14853, USA}
\author{Megan E. DeCesar}
\altaffiliation{NANOGrav Physics Frontiers Center Postdoctoral Fellow}
\affiliation{Department of Physics, Lafayette College, Easton, PA 18042, USA}
\affiliation{George Mason University, Fairfax, VA 22030, resident at U.S. Naval Research Laboratory, Washington, D.C. 20375, USA}
\author{Dallas M. DeGan}
\affiliation{Department of Physics, Oregon State University, Corvallis, OR 97331, USA}
\author{Paul B. Demorest}
\affiliation{National Radio Astronomy Observatory, 1003 Lopezville Rd., Socorro, NM 87801, USA}
\author{Timothy Dolch}
\affiliation{Department of Physics, Hillsdale College, 33 E. College Street, Hillsdale, MI 49242, USA}
\affiliation{Eureka Scientific, Inc. 2452 Delmer Street, Suite 100, Oakland, CA 94602-3017, USA}
\author{Brendan Drachler}
\affiliation{School of Physics and Astronomy, Rochester Institute of Technology, Rochester, NY 14623, USA}
\affiliation{Laboratory for Multiwavelength Astrophysics, Rochester Institute of Technology, Rochester, NY 14623, USA}
\author{Justin A. Ellis}
\affiliation{Infinia ML, 202 Rigsbee Avenue, Durham NC, 27701, USA}
\author{Elizabeth C. Ferrara}
\affiliation{Department of Astronomy, University of Maryland, College Park, MD 20742, USA}
\affiliation{Center for Research and Exploration in Space Science and Technology, NASA/GSFC, Greenbelt, MD 20771, USA}
\affiliation{NASA Goddard Space Flight Center, Greenbelt, MD 20771, USA}
\author{William Fiore}
\affiliation{Department of Physics and Astronomy, West Virginia University, P.O. Box 6315, Morgantown, WV 26506, USA}
\affiliation{Center for Gravitational Waves and Cosmology, West Virginia University, Chestnut Ridge Research Building, Morgantown, WV 26505, USA}
\author{Emmanuel Fonseca}
\affiliation{Department of Physics, McGill University, 3600  University St., Montreal, QC H3A 2T8, Canada}
\author{Nathan Garver-Daniels}
\affiliation{Department of Physics and Astronomy, West Virginia University, P.O. Box 6315, Morgantown, WV 26506, USA}
\affiliation{Center for Gravitational Waves and Cosmology, West Virginia University, Chestnut Ridge Research Building, Morgantown, WV 26505, USA}
\author{Peter A. Gentile}
\affiliation{Department of Physics and Astronomy, West Virginia University, P.O. Box 6315, Morgantown, WV 26506, USA}
\affiliation{Center for Gravitational Waves and Cosmology, West Virginia University, Chestnut Ridge Research Building, Morgantown, WV 26505, USA}
\author{Deborah C. Good}
\affiliation{Department of Physics and Astronomy, University of British Columbia, 6224 Agricultural Road, Vancouver, BC V6T 1Z1, Canada}
\author{Jeffrey S. Hazboun}
\altaffiliation{NANOGrav Physics Frontiers Center Postdoctoral Fellow}
\affiliation{University of Washington Bothell, 18115 Campus Way NE, Bothell, WA 98011, USA}
\author{A. Miguel Holgado}
\affiliation{Department of Astronomy and National Center for Supercomputing Applications, University of Illinois at Urbana-Champaign, Urbana, IL 61801, USA}
\affiliation{McWilliams Center for Cosmology and Department of Physics, Carnegie Mellon University, Pittsburgh PA, 15213, USA}
\author{Kristina Islo}
\affiliation{Center for Gravitation, Cosmology and Astrophysics, Department of Physics, University of Wisconsin-Milwaukee,\\ P.O. Box 413, Milwaukee, WI 53201, USA}
\author{Ross J. Jennings}
\affiliation{Cornell Center for Astrophysics and Planetary Science and Department of Astronomy, Cornell University, Ithaca, NY 14853, USA}
\author{Megan L. Jones}
\affiliation{Center for Gravitation, Cosmology and Astrophysics, Department of Physics, University of Wisconsin-Milwaukee,\\ P.O. Box 413, Milwaukee, WI 53201, USA}
\author{Andrew R. Kaiser}
\affiliation{Department of Physics and Astronomy, West Virginia University, P.O. Box 6315, Morgantown, WV 26506, USA}
\affiliation{Center for Gravitational Waves and Cosmology, West Virginia University, Chestnut Ridge Research Building, Morgantown, WV 26505, USA}
\author{David L. Kaplan}
\affiliation{Center for Gravitation, Cosmology and Astrophysics, Department of Physics, University of Wisconsin-Milwaukee,\\ P.O. Box 413, Milwaukee, WI 53201, USA}
\author{Luke Zoltan Kelley}
\affiliation{Center for Interdisciplinary Exploration and Research in Astrophysics (CIERA), Northwestern University, Evanston, IL 60208, USA}
\author{Joey Shapiro Key}
\affiliation{University of Washington Bothell, 18115 Campus Way NE, Bothell, WA 98011, USA}
\author{Nima Laal}
\affiliation{Department of Physics, Oregon State University, Corvallis, OR 97331, USA}
\author{Michael T. Lam}
\affiliation{School of Physics and Astronomy, Rochester Institute of Technology, Rochester, NY 14623, USA}
\affiliation{Laboratory for Multiwavelength Astrophysics, Rochester Institute of Technology, Rochester, NY 14623, USA}
\author{T. Joseph W. Lazio}
\affiliation{Jet Propulsion Laboratory, California Institute of Technology, 4800 Oak Grove Drive, Pasadena, CA 91109, USA}
\author{Duncan R. Lorimer}
\affiliation{Department of Physics and Astronomy, West Virginia University, P.O. Box 6315, Morgantown, WV 26506, USA}
\affiliation{Center for Gravitational Waves and Cosmology, West Virginia University, Chestnut Ridge Research Building, Morgantown, WV 26505, USA}
\author{Tingting Liu}
\affiliation{Center for Gravitation, Cosmology and Astrophysics, Department of Physics, University of Wisconsin-Milwaukee, P.O. Box 413, Milwaukee, WI 53201, USA}
\author{Jing Luo}
\affiliation{Department of Astronomy \& Astrophysics, University of Toronto, 50 Saint George Street, Toronto, ON M5S 3H4, Canada}
\author{Ryan S. Lynch}
\affiliation{Green Bank Observatory, P.O. Box 2, Green Bank, WV 24944, USA}
\author{Dustin R. Madison}
\altaffiliation{NANOGrav Physics Frontiers Center Postdoctoral Fellow}
\affiliation{Department of Physics and Astronomy, West Virginia University, P.O. Box 6315, Morgantown, WV 26506, USA}
\affiliation{Center for Gravitational Waves and Cosmology, West Virginia University, Chestnut Ridge Research Building, Morgantown, WV 26505, USA}
\author{Alexander McEwen}
\affiliation{Center for Gravitation, Cosmology and Astrophysics, Department of Physics, University of Wisconsin-Milwaukee, P.O. Box 413, Milwaukee, WI 53201, USA}
\author{Maura A. McLaughlin}
\affiliation{Department of Physics and Astronomy, West Virginia University, P.O. Box 6315, Morgantown, WV 26506, USA}
\affiliation{Center for Gravitational Waves and Cosmology, West Virginia University, Chestnut Ridge Research Building, Morgantown, WV 26505, USA}
\author{Chiara M. F. Mingarelli}
\affiliation{Center for Computational Astrophysics, Flatiron Institute, 162 5th Avenue, New York, New York, 10010, USA}
\affiliation{Department of Physics, University of Connecticut, 196 Auditorium Road, U-3046, Storrs, CT 06269-3046, USA}
\author{Cherry Ng}
\affiliation{Dunlap Institute for Astronomy and Astrophysics, University of Toronto, 50 St. George St., Toronto, ON M5S 3H4, Canada}
\author{David J. Nice}
\affiliation{Department of Physics, Lafayette College, Easton, PA 18042, USA}
\author{Ken D. Olum}
\affiliation{Tufts Institute of Cosmology, Department of Physics and Astronomy, Tufts University, 574 Boston Avenue, Medford, MA 02155, USA}
\author{Timothy T. Pennucci}
\altaffiliation{NANOGrav Physics Frontiers Center Postdoctoral Fellow}
\affiliation{National Radio Astronomy Observatory, 520 Edgemont Road, Charlottesville, VA 22903, USA}
\affiliation{Institute of Physics, E\"{o}tv\"{o}s Lor\'{a}nd University, P\'{a}zm\'{a}ny P. s. 1/A, 1117 Budapest, Hungary}
\author{Nihan S. Pol}
\affiliation{Department of Physics and Astronomy, West Virginia University, P.O. Box 6315, Morgantown, WV 26506, USA}
\affiliation{Center for Gravitational Waves and Cosmology, West Virginia University, Chestnut Ridge Research Building, Morgantown, WV 26505, USA}
\affiliation{Department of Physics and Astronomy, Vanderbilt University, 2301 Vanderbilt Place, Nashville, TN 37235, USA}
\author{Scott M. Ransom}
\affiliation{National Radio Astronomy Observatory, 520 Edgemont Road, Charlottesville, VA 22903, USA}
\author{Paul S. Ray}
\affiliation{Space Science Division, Naval Research Laboratory, Washington, DC 20375-5352, USA}
\author{Joseph D. Romano}
\affiliation{Department of Physics and Astronomy, Texas Tech University, Lubbock, TX 79409-1051, USA}

\author{Shashwat C. Sardesai}
\affiliation{Center for Gravitation, Cosmology and Astrophysics, Department of Physics, University of Wisconsin-Milwaukee, P.O. Box 413, Milwaukee, WI 53201, USA}
\author{Brent J. Shapiro-Albert}
\affiliation{Department of Physics and Astronomy, West Virginia University, P.O. Box 6315, Morgantown, WV 26506, USA}
\affiliation{Center for Gravitational Waves and Cosmology, West Virginia University, Chestnut Ridge Research Building, Morgantown, WV 26505, USA}
\author{Xavier Siemens}
\affiliation{Department of Physics, Oregon State University, Corvallis, OR 97331, USA}
\affiliation{Center for Gravitation, Cosmology and Astrophysics, Department of Physics, University of Wisconsin-Milwaukee,\\ P.O. Box 413, Milwaukee, WI 53201, USA}
\author{Joseph Simon}
\affiliation{Jet Propulsion Laboratory, California Institute of Technology, 4800 Oak Grove Drive, Pasadena, CA 91109, USA}
\affiliation{Department of Astrophysical and Planetary Sciences, University of Colorado, Boulder, CO 80309, USA}
\author{Magdalena S. Siwek}
\affiliation{Center for Astrophysics, Harvard University, Cambridge, MA 02138, USA}
\author{Ren\'{e}e Spiewak}
\affiliation{Centre for Astrophysics and Supercomputing, Swinburne University of Technology, P.O. Box 218, Hawthorn, Victoria 3122, Australia}
\author{Ingrid H. Stairs}
\affiliation{Department of Physics and Astronomy, University of British Columbia, 6224 Agricultural Road, Vancouver, BC V6T 1Z1, Canada}
\author{Daniel R. Stinebring}
\affiliation{Department of Physics and Astronomy, Oberlin College, Oberlin, OH 44074, USA}
\author{Kevin Stovall}
\affiliation{National Radio Astronomy Observatory, 1003 Lopezville Rd., Socorro, NM 87801, USA}
\author{Jerry P. Sun}
\affiliation{Department of Physics, Oregon State University, Corvallis, OR 97331, USA}
\author{Joseph K. Swiggum}
\altaffiliation{NANOGrav Physics Frontiers Center Postdoctoral Fellow}
\affiliation{Department of Physics, Lafayette College, Easton, PA 18042, USA}
\author{Stephen R. Taylor}
\affiliation{Department of Physics and Astronomy, Vanderbilt University, 2301 Vanderbilt Place, Nashville, TN 37235, USA}
\author{Jacob E. Turner}
\affiliation{Department of Physics and Astronomy, West Virginia University, P.O. Box 6315, Morgantown, WV 26506, USA}
\affiliation{Center for Gravitational Waves and Cosmology, West Virginia University, Chestnut Ridge Research Building, Morgantown, WV 26505, USA}
\author{Michele Vallisneri}
\affiliation{Jet Propulsion Laboratory, California Institute of Technology, 4800 Oak Grove Drive, Pasadena, CA 91109, USA}
\author{Sarah J. Vigeland}
\affiliation{Center for Gravitation, Cosmology and Astrophysics, Department of Physics, University of Wisconsin-Milwaukee,\\ P.O. Box 413, Milwaukee, WI 53201, USA}
\author{Haley M. Wahl}
\affiliation{Department of Physics and Astronomy, West Virginia University, P.O. Box 6315, Morgantown, WV 26506, USA}
\affiliation{Center for Gravitational Waves and Cosmology, West Virginia University, Chestnut Ridge Research Building, Morgantown, WV 26505, USA}
\author{Caitlin A. Witt}
\affiliation{Department of Physics and Astronomy, West Virginia University, P.O. Box 6315, Morgantown, WV 26506, USA}
\affiliation{Center for Gravitational Waves and Cosmology, West Virginia University, Chestnut Ridge Research Building, Morgantown, WV 26505, USA}

%% file: 12p5-altpol-contrib.tex
An alphabetical-order author list was used for this paper in recognition of the fact that a large, decade timescale project such as NANOGrav is necessarily the result of the work of many people. All authors contributed to the activities of the NANOGrav collaboration leading to the work presented here, and reviewed the manuscript, text, and figures prior to the paper's submission. 
Additional specific contributions to this paper are as follows.
ZA, HB, PRB, HTC, MED, PBD, TD, JAE, RDF, ECF, EF, NG-D, PAG, DCG, MLJ, MTL, DRL, RSL, JL, MAM, CN, DJN, TTP, NSP, SMR, KS, IHS, RS, JKS, RS and SJV developed the 12.5-year data set through a combination of observations, arrival time calculations, data checks and refinements, and timing model development and analysis; additional specific contributions to the data set are summarized in \cite{aab+20}.

NL coordinated the writing of the paper and led the search.
NSP developed the simulated data sets featured in this paper.
SCS and DMG cross replicated the most significant results of this paper.
SRT helped in development of the code for Bayesian runs.
SV helped in development of the code for the optimal statistic runs,
and along with JDR developed the optimal statistic technique for the case of multiple correlations. 
JS provided feedback and guidance on how to resolve issues in the early Bayesian and frequentist analyses. 
NJC, XS, SRT, and SV provided feedback on searches and new analysis techniques.
NJC and XS provided the first insights on the inclusion of GW-like monopole correlations.
NL provided all of the figures and performed all of the analyses featured in this paper.
NL and XS wrote the paper and collected the bibliography.

%% file: 12p5-altpol-acknowledge.tex
This work has been carried out by the NANOGrav collaboration, which is part of the International Pulsar Timing Array. 
We thank the anonymous reviewers for useful suggestions and comments, which improved the quality of the manuscript. 
The NANOGrav project receives support from National Science Foundation (NSF) Physics Frontiers Center award number 1430284 and 2020265. 
The Arecibo Observatory is a facility of the NSF operated under cooperative agreement (\#AST-1744119) by the University of Central Florida (UCF) in alliance with Universidad Ana G. M\'{e}ndez (UAGM) and Yang Enterprises (YEI), Inc. 
The Green Bank Observatory is a facility of the NSF operated under cooperative agreement by Associated Universities, Inc. The National Radio Astronomy Observatory is a facility of the NSF operated under cooperative agreement by Associated Universities, Inc. This work is supported in part by NASA under award number 80GSFC17M0002. We also acknowledge support received from NSF AAG award number 2009468. T.D. and M.T.L. are supported by an NSF Astronomy and Astrophysics grant (AAG) award No. 2009468.
Portions of this work performed at NRL were supported by ONR 6.1 basic research funding.
The work of NL, XS, JPS, and DD was partly supported by the George and Hannah Bolinger Memorial Fund in the College of Science at Oregon State University.

%% file: 12p5-altpol-main.bbl
\providecommand{\noopsort}[1]{}\providecommand{\singleletter}[1]{#1}%
\begin{thebibliography}{}
\expandafter\ifx\csname natexlab\endcsname\relax\def\natexlab#1{#1}\fi
\providecommand{\url}[1]{\href{#1}{#1}}
\providecommand{\dodoi}[1]{doi:~\href{http://doi.org/#1}{\nolinkurl{#1}}}
\providecommand{\doeprint}[1]{\href{http://ascl.net/#1}{\nolinkurl{http://ascl.net/#1}}}
\providecommand{\doarXiv}[1]{\href{https://arxiv.org/abs/#1}{\nolinkurl{https://arxiv.org/abs/#1}}}

\bibitem[{{Abbott} {et~al.}(2018){Abbott}, {Abbott}, {Abbott}, {Acernese},
  {Ackley}, {Adams}, {Adams}, {Addesso}, {Adhikari}, {Adya}, {Affeldt},
  {Afrough}, {Agarwal}, {Agathos}, {Agatsuma}, {Aggarwal}, {Aguiar}, {Aiello},
  {Ain}, {Ajith}, {Allen}, {Allocca}, {Altin}, {Amato}, {Ananyeva}, {Anderson},
  {Anderson}, {Antier}, {Appert}, {Arai}, {Araya}, {Areeda}, {Arnaud}, {Arun},
  {Ascenzi}, {Ashton}, {Ast}, {Aston}, {Astone}, {Aufmuth}, {Aulbert},
  {AultONeal}, {Avila-Alvarez}, {Babak}, {Bacon}, {Bader}, {Bae}, {Baker},
  {Baldaccini}, {Ballardin}, {Ballmer}, {Banagiri}, {Barayoga}, {Barclay},
  {Barish}, {Barker}, {Barone}, {Barr}, {Barsotti}, {Barsuglia}, {Barta},
  {Bartlett}, {Bartos}, {Bassiri}, {Basti}, {Batch}, {Baune}, {Bawaj},
  {Bazzan}, {B{\'e}csy}, {Beer}, {Bejger}, {Belahcene}, {Bell}, {Berger},
  {Bergmann}, {Berry}, {Bersanetti}, {Bertolini}, {Betzwieser}, {Bhagwat},
  {Bhandare}, {Bilenko}, {Billingsley}, {Billman}, {Birch}, {Birney},
  {Birnholtz}, {Biscans}, {Bisht}, {Bitossi}, {Biwer}, {Bizouard}, {Blackburn},
  {Blackman}, {Blair}, {Blair}, {Blair}, {Bloemen}, {Bock}, {Bode}, {Boer},
  {Bogaert}, {Bohe}, {Bondu}, {Bonnand}, {Boom}, {Bork}, {Boschi}, {Bose},
  {Bouffanais}, {Bozzi}, {Bradaschia}, {Brady}, {Braginsky}, {Branchesi},
  {Brau}, {Briant}, {Brillet}, {Brinkmann}, {Brisson}, {Brockill}, {Broida},
  {Brooks}, {Brown}, {Brown}, {Brown}, {Brunett}, {Buchanan}, {Buikema},
  {Bulik}, {Bulten}, {Buonanno}, {Buskulic}, {Buy}, {Byer}, {Cabero},
  {Cadonati}, {Cagnoli}, {Cahillane}, {Calder{\'o}n Bustillo}, {Callister},
  {Calloni}, {Camp}, {Canepa}, {Canizares}, {Cannon}, {Cao}, {Cao}, {Capano},
  {Capocasa}, {Carbognani}, {Caride}, {Carney}, {Casanueva Diaz}, {Casentini},
  {Caudill}, {Cavagli{\`a}}, {Cavalier}, {Cavalieri}, {Cella}, {Cepeda},
  {Cerboni Baiardi}, {Cerretani}, {Cesarini}, {Chamberlin}, {Chan}, {Chao},
  {Charlton}, {Chassande-Mottin}, {Chatterjee}, {Cheeseboro}, {Chen}, {Chen},
  {Cheng}, {Chincarini}, {Chiummo}, {Chmiel}, {Cho}, {Cho}, {Chow},
  {Christensen}, {Chu}, {Chua}, {Chua}, {Chung}, {Chung}, {Ciani}, {Ciolfi},
  {Cirelli}, {Cirone}, {Clara}, {Clark}, {Cleva}, {Cocchieri}, {Coccia},
  {Cohadon}, {Colla}, {Collette}, {Cominsky}, {Constancio}, {Conti}, {Cooper},
  {Corban}, {Corbitt}, {Corley}, {Cornish}, {Corsi}, {Cortese}, {Costa},
  {Coughlin}, {Coughlin}, {Coulon}, {Countryman}, {Couvares}, {Covas}, {Cowan},
  {Coward}, {Cowart}, {Coyne}, {Coyne}, {Creighton}, {Creighton}, {Cripe},
  {Crowder}, {Cullen}, {Cumming}, {Cunningham}, {Cuoco}, {Dal Canton},
  {Danilishin}, {D'Antonio}, {Danzmann}, {Dasgupta}, {Da Silva Costa},
  {Dattilo}, {Dave}, {Davier}, {Davis}, {Daw}, {Day}, {De}, {DeBra},
  {Degallaix}, {De Laurentis}, {Del{\'e}glise}, {Del Pozzo}, {Denker}, {Dent},
  {Dergachev}, {De Rosa}, {DeRosa}, {DeSalvo}, {Devenson}, {Devine},
  {Dhurandhar}, {D{\'\i}az}, {Di Fiore}, {Di Giovanni}, {Di Girolamo}, {Di
  Lieto}, {Di Pace}, {Di Palma}, {Di Renzo}, {Doctor}, {Dolique}, {Donovan},
  {Dooley}, {Doravari}, {Dorrington}, {Douglas}, {Dovale {\'A}lvarez},
  {Downes}, {Drago}, {Drever}, {Driggers}, {Du}, {Ducrot}, {Duncan}, {Dwyer},
  {Edo}, {Edwards}, {Effler}, {Eggenstein}, {Ehrens}, {Eichholz}, {Eikenberry},
  {Eisenstein}, {Essick}, {Etienne}, {Etzel}, {Evans}, {Evans}, {Factourovich},
  {Fafone}, {Fair}, {Fairhurst}, {Fan}, {Farinon}, {Farr}, {Farr},
  {Fauchon-Jones}, {Favata}, {Fays}, {Fehrmann}, {Feicht}, {Fejer},
  {Fernandez-Galiana}, {Ferrante}, {Ferreira}, {Ferrini}, {Fidecaro}, {Fiori},
  {Fiorucci}, {Fisher}, {Flaminio}, {Fletcher}, {Fong}, {Forsyth}, {Forsyth},
  {Fournier}, {Frasca}, {Frasconi}, {Frei}, {Freise}, {Frey}, {Frey}, {Fries},
  {Fritschel}, {Frolov}, {Fulda}, {Fyffe}, {Gabbard}, {Gabel}, {Gadre},
  {Gaebel}, {Gair}, {Gammaitoni}, {Ganija}, {Gaonkar}, {Garufi}, {Gaudio},
  {Gaur}, {Gayathri}, {Gehrels}, {Gemme}, {Genin}, {Gennai}, {George},
  {George}, {Gergely}, {Germain}, {Ghonge}, {Ghosh}, {Ghosh}, {Ghosh},
  {Giaime}, {Giardina}, {Giazotto}, {Gill}, {Glover}, {Goetz}, {Goetz},
  {Gomes}, {Gonz{\'a}lez}, {Gonzalez Castro}, {Gopakumar}, {Gorodetsky},
  {Gossan}, {Gosselin}, {Gouaty}, {Grado}, {Graef}, {Granata}, {Grant}, {Gras},
  {Gray}, {Greco}, {Green}, {Groot}, {Grote}, {Grunewald}, {Gruning}, {Guidi},
  {Guo}, {Gupta}, {Gupta}, {Gushwa}, {Gustafson}, {Gustafson}, {Hall}, {Hall},
  {Hammond}, {Haney}, {Hanke}, {Hanks}, {Hanna}, {Hannuksela}, {Hanson},
  {Hardwick}, {Harms}, {Harry}, {Harry}, {Hart}, {Haster}, {Haughian}, {Healy},
  {Heidmann}, {Heintze}, {Heitmann}, {Hello}, {Hemming}, {Hendry}, {Heng},
  {Hennig}, {Henry}, {Heptonstall}, {Heurs}, {Hild}, {Hoak}, {Hofman}, {Holt},
  {Holz}, {Hopkins}, {Horst}, {Hough}, {Houston}, {Howell}, {Hu}, {Huerta},
  {Huet}, {Hughey}, {Husa}, {Huttner}, {Huynh-Dinh}, {Indik}, {Ingram}, {Inta},
  {Intini}, {Isa}, {Isac}, {Isi}, {Iyer}, {Izumi}, {Jacqmin}, {Jani},
  {Jaranowski}, {Jawahar}, {Jim{\'e}nez-Forteza}, {Johnson}, {Jones}, {Jones},
  {Jonker}, {Ju}, {Junker}, {Kalaghatgi}, {Kalogera}, {Kandhasamy}, {Kang},
  {Kanner}, {Karki}, {Karvinen}, {Kasprzack}, {Katolik}, {Katsavounidis},
  {Katzman}, {Kaufer}, {Kawabe}, {K{\'e}f{\'e}lian}, {Keitel}, {Kemball},
  {Kennedy}, {Kent}, {Key}, {Khalili}, {Khan}, {Khan}, {Khan}, {Khazanov},
  {Kijbunchoo}, {Kim}, {Kim}, {Kim}, {Kim}, {Kim}, {Kimbrell}, {King}, {King},
  {Kirchhoff}, {Kissel}, {Kleybolte}, {Klimenko}, {Koch}, {Koehlenbeck},
  {Koley}, {Kondrashov}, {Kontos}, {Korobko}, {Korth}, {Kowalska}, {Kozak},
  {Kr{\"a}mer}, {Kringel}, {Krishnan}, {Kr{\'o}lak}, {Kuehn}, {Kumar}, {Kumar},
  {Kumar}, {Kuo}, {Kutynia}, {Kwang}, {Lackey}, {Lai}, {Landry}, {Lang},
  {Lange}, {Lantz}, {Lanza}, {Lartaux-Vollard}, {Lasky}, {Laxen}, {Lazzarini},
  {Lazzaro}, {Leaci}, {Leavey}, {Lee}, {Lee}, {Lee}, {Lee}, {Lee}, {Lehmann},
  {Lenon}, {Leonardi}, {Leroy}, {Letendre}, {Levin}, {Li}, {Libson},
  {Littenberg}, {Liu}, {Lo}, {Lockerbie}, {London}, {Lord}, {Lorenzini},
  {Loriette}, {Lormand}, {Losurdo}, {Lough}, {Lousto}, {Lovelace}, {L{\"u}ck},
  {Lumaca}, {Lundgren}, {Lynch}, {Ma}, {Macfoy}, {Machenschalk}, {MacInnis},
  {Macleod}, {Maga{\~n}a Hernandez}, {Maga{\~n}a-Sandoval}, {Maga{\~n}a
  Zertuche}, {Magee}, {Majorana}, {Maksimovic}, {Man}, {Mandic}, {Mangano},
  {Mansell}, {Manske}, {Mantovani}, {Marchesoni}, {Marion}, {M{\'a}rka},
  {M{\'a}rka}, {Markakis}, {Markosyan}, {Maros}, {Martelli}, {Martellini},
  {Martin}, {Martynov}, {Mason}, {Masserot}, {Massinger}, {Masso-Reid},
  {Mastrogiovanni}, {Matas}, {Matichard}, {Matone}, {Mavalvala}, {Mazumder},
  {McCarthy}, {McClelland}, {McCormick}, {McCuller}, {McGuire}, {McIntyre},
  {McIver}, {McManus}, {McRae}, {McWilliams}, {Meacher}, {Meadors}, {Meidam},
  {Mejuto-Villa}, {Melatos}, {Mendell}, {Mercer}, {Merilh}, {Merzougui},
  {Meshkov}, {Messenger}, {Messick}, {Metzdorff}, {Meyers}, {Mezzani}, {Miao},
  {Michel}, {Middleton}, {Mikhailov}, {Milano}, {Miller}, {Miller}, {Miller},
  {Miller}, {Millhouse}, {Minazzoli}, {Minenkov}, {Ming}, {Mishra}, {Mitra},
  {Mitrofanov}, {Mitselmakher}, {Mittleman}, {Moggi}, {Mohan}, {Mohapatra},
  {Montani}, {Moore}, {Moore}, {Moraru}, {Moreno}, {Morriss}, {Mours},
  {Mow-Lowry}, {Mueller}, {Muir}, {Mukherjee}, {Mukherjee}, {Mukherjee},
  {Mukund}, {Mullavey}, {Munch}, {Muniz}, {Murray}, {Napier}, {Nardecchia},
  {Naticchioni}, {Nayak}, {Nelemans}, {Nelson}, {Neri}, {Nery}, {Neunzert},
  {Newport}, {Newton}, {Ng}, {Nguyen}, {Nichols}, {Nielsen}, {Nissanke},
  {Nitz}, {Noack}, {Nocera}, {Nolting}, {Normandin}, {Nuttall}, {Oberling},
  {Ochsner}, {Oelker}, {Ogin}, {Oh}, {Oh}, {Ohme}, {Oliver}, {Oppermann},
  {Oram}, {O'Reilly}, {Ormiston}, {Ortega}, {O'Shaughnessy}, {Ottaway},
  {Overmier}, {Owen}, {Pace}, {Page}, {Page}, {Pai}, {Pai}, {Palamos},
  {Palashov}, {Palomba}, {Pal-Singh}, {Pan}, {Pang}, {Pang}, {Pankow},
  {Pannarale}, {Pant}, {Paoletti}, {Paoli}, {Papa}, {Paris}, {Parker},
  {Pascucci}, {Pasqualetti}, {Passaquieti}, {Passuello}, {Patricelli},
  {Pearlstone}, {Pedraza}, {Pedurand}, {Pekowsky}, {Pele}, {Penn}, {Perez},
  {Perreca}, {Perri}, {Pfeiffer}, {Phelps}, {Piccinni}, {Pichot},
  {Piergiovanni}, {Pierro}, {Pillant}, {Pinard}, {Pinto}, {Pitkin}, {Poggiani},
  {Popolizio}, {Porter}, {Post}, {Powell}, {Prasad}, {Pratt}, {Predoi},
  {Prestegard}, {Prijatelj}, {Principe}, {Privitera}, {Prix}, {Prodi},
  {Prokhorov}, {Puncken}, {Punturo}, {Puppo}, {P{\"u}rrer}, {Qi}, {Qin}, {Qiu},
  {Quetschke}, {Quintero}, {Quitzow-James}, {Raab}, {Rabeling}, {Radkins},
  {Raffai}, {Raja}, {Rajan}, {Rakhmanov}, {Ramirez}, {Rapagnani}, {Raymond},
  {Razzano}, {Read}, {Regimbau}, {Rei}, {Reid}, {Reitze}, {Rew}, {Reyes},
  {Ricci}, {Ricker}, {Rieger}, {Riles}, {Rizzo}, {Robertson}, {Robie},
  {Robinet}, {Rocchi}, {Rolland}, {Rollins}, {Roma}, {Romano}, {Romel},
  {Romie}, {Rosi{\'n}ska}, {Ross}, {Rowan}, {R{\"u}diger}, {Ruggi}, {Ryan},
  {Sachdev}, {Sadecki}, {Sadeghian}, {Sakellariadou}, {Salconi}, {Saleem},
  {Salemi}, {Samajdar}, {Sammut}, {Sampson}, {Sanchez}, {Sandberg}, {Sandeen},
  {Sanders}, {Sassolas}, {Sathyaprakash}, {Saulson}, {Sauter}, {Savage},
  {Sawadsky}, {Schale}, {Scheuer}, {Schmidt}, {Schmidt}, {Schmidt}, {Schnabel},
  {Schofield}, {Sch{\"o}nbeck}, {Schreiber}, {Schuette}, {Schulte}, {Schutz},
  {Schwalbe}, {Scott}, {Scott}, {Seidel}, {Sellers}, {Sengupta}, {Sentenac},
  {Sequino}, {Sergeev}, {Shaddock}, {Shaffer}, {Shah}, {Shahriar}, {Shao},
  {Shapiro}, {Shawhan}, {Sheperd}, {Shoemaker}, {Shoemaker}, {Siellez},
  {Siemens}, {Sieniawska}, {Sigg}, {Silva}, {Singer}, {Singer}, {Singh},
  {Singh}, {Singhal}, {Sintes}, {Slagmolen}, {Smith}, {Smith}, {Smith}, {Son},
  {Sonnenberg}, {Sorazu}, {Sorrentino}, {Souradeep}, {Spencer}, {Srivastava},
  {Staley}, {Steinke}, {Steinlechner}, {Steinlechner}, {Steinmeyer},
  {Stephens}, {Stone}, {Strain}, {Stratta}, {Strigin}, {Sturani}, {Stuver},
  {Summerscales}, {Sun}, {Sunil}, {Sutton}, {Swinkels}, {Szczepa{\'n}czyk},
  {Tacca}, {Talukder}, {Tanner}, {T{\'a}pai}, {Taracchini}, {Taylor}, {Taylor},
  {Theeg}, {Thomas}, {Thomas}, {Thomas}, {Thorne}, {Thorne}, {Thrane},
  {Tiwari}, {Tiwari}, {Tokmakov}, {Toland}, {Tonelli}, {Tornasi}, {Torrie},
  {T{\"o}yr{\"a}}, {Travasso}, {Traylor}, {Trifir{\`o}}, {Trinastic},
  {Tringali}, {Trozzo}, {Tsang}, {Tse}, {Tso}, {Tuyenbayev}, {Ueno}, {Ugolini},
  {Unnikrishnan}, {Urban}, {Usman}, {Vahlbruch}, {Vajente}, {Valdes},
  {Vallisneri}, {van Bakel}, {van Beuzekom}, {van den Brand}, {Van Den Broeck},
  {Vander-Hyde}, {van der Schaaf}, {van Heijningen}, {van Veggel}, {Vardaro},
  {Varma}, {Vass}, {Vas{\'u}th}, {Vecchio}, {Vedovato}, {Veitch}, {Veitch},
  {Venkateswara}, {Venugopalan}, {Verkindt}, {Vetrano}, {Vicer{\'e}}, {Viets},
  {Vinciguerra}, {Vine}, {Vinet}, {Vitale}, {Vo}, {Vocca}, {Vorvick}, {Voss},
  {Vousden}, {Vyatchanin}, {Wade}, {Wade}, {Wade}, {Walet}, {Walker},
  {Wallace}, {Walsh}, {Wang}, {Wang}, {Wang}, {Wang}, {Wang}, {Wang}, {Ward},
  {Warner}, {Was}, {Watchi}, {Weaver}, {Wei}, {Weinert}, {Weinstein}, {Weiss},
  {Wen}, {Wessel}, {We{\ss}els}, {Westphal}, {Wette}, {Whelan}, {Whiting},
  {Whittle}, {Williams}, {Williams}, {Williamson}, {Willis}, {Willke},
  {Wimmer}, {Winkler}, {Wipf}, {Wittel}, {Woan}, {Woehler}, {Wofford}, {Wong},
  {Worden}, {Wright}, {Wu}, {Wu}, {Yam}, {Yamamoto}, {Yancey}, {Yap}, {Yu},
  {Yu}, {Yvert}, {Zadro{\.Z}ny}, {Zanolin}, {Zelenova}, {Zendri}, {Zevin},
  {Zhang}, {Zhang}, {Zhang}, {Zhang}, {Zhao}, {Zhou}, {Zhou}, {Zhu}, {Zhu},
  {Zucker}, {Zweizig}, {Buchner}, {Cognard}, {Corongiu}, {Freire}, {Guillemot},
  {Hobbs}, {Kerr}, {Lyne}, {Possenti}, {Ridolfi}, {Shannon}, {Stappers},
  {Weltevrede}, {LIGO Scientific Collaboration}, \& {Virgo
  Collaboration}}]{LIGO3}
{Abbott}, B.~P., {Abbott}, R., {Abbott}, T.~D., {et~al.} 2018, \prl, 120,
  031104, \dodoi{10.1103/PhysRevLett.120.031104}

\bibitem[{{Abbott} {et~al.}(2019){Abbott}, {Abbott}, {Abbott}, {Abraham},
  {Acernese}, {Ackley}, {Adams}, {Adhikari}, {Adya}, {Affeldt}, {Agathos},
  {Agatsuma}, {Aggarwal}, {Aguiar}, {Aiello}, {Ain}, {Ajith}, {Allen},
  {Allocca}, {Aloy}, {Altin}, {Amato}, {Ananyeva}, {Anderson}, {Anderson},
  {Angelova}, {Antier}, {Appert}, {Arai}, {Araya}, {Areeda}, {Ar{\`e}ne},
  {Arnaud}, {Arun}, {Ascenzi}, {Ashton}, {Aston}, {Astone}, {Aubin}, {Aufmuth},
  {AultONeal}, {Austin}, {Avendano}, {Avila-Alvarez}, {Babak}, {Bacon},
  {Badaracco}, {Bader}, {Bae}, {Baker}, {Baldaccini}, {Ballardin}, {Ballmer},
  {Banagiri}, {Barayoga}, {Barclay}, {Barish}, {Barker}, {Barkett}, {Barnum},
  {Barone}, {Barr}, {Barsotti}, {Barsuglia}, {Barta}, {Bartlett}, {Bartos},
  {Bassiri}, {Basti}, {Bawaj}, {Bayley}, {Bazzan}, {B{\'e}csy}, {Bejger},
  {Belahcene}, {Bell}, {Beniwal}, {Berger}, {Bergmann}, {Bernuzzi}, {Bero},
  {Berry}, {Bersanetti}, {Bertolini}, {Betzwieser}, {Bhandare}, {Bidler},
  {Bilenko}, {Bilgili}, {Billingsley}, {Birch}, {Birney}, {Birnholtz},
  {Biscans}, {Biscoveanu}, {Bisht}, {Bitossi}, {Bizouard}, {Blackburn},
  {Blair}, {Blair}, {Blair}, {Bloemen}, {Bode}, {Boer}, {Boetzel}, {Bogaert},
  {Bondu}, {Bonilla}, {Bonnand}, {Booker}, {Boom}, {Booth}, {Bork}, {Boschi},
  {Bose}, {Bossie}, {Bossilkov}, {Bosveld}, {Bouffanais}, {Bozzi},
  {Bradaschia}, {Brady}, {Bramley}, {Branchesi}, {Brau}, {Breschi}, {Briant},
  {Briggs}, {Brighenti}, {Brillet}, {Brinkmann}, {Brisson}, {Brito},
  {Brockill}, {Brooks}, {Brown}, {Brunett}, {Buikema}, {Bulik}, {Bulten},
  {Buonanno}, {Buskulic}, {Rosell}, {Buy}, {Byer}, {Cabero}, {Cadonati},
  {Cagnoli}, {Cahillane}, {Bustillo}, {Callister}, {Calloni}, {Camp},
  {Campbell}, {Canepa}, {Cannon}, {Cao}, {Cao}, {Capano}, {Capocasa},
  {Carbognani}, {Caride}, {Carney}, {Carullo}, {Diaz}, {Casentini}, {Caudill},
  {Cavagli{\`a}}, {Cavalier}, {Cavalieri}, {Cella}, {Cerd{\'a}-Dur{\'a}n},
  {Cerretani}, {Cesarini}, {Chaibi}, {Chakravarti}, {Chamberlin}, {Chan},
  {Chao}, {Charlton}, {Chase}, {Chassande-Mottin}, {Chatterjee}, {Chaturvedi},
  {Chatziioannou}, {Cheeseboro}, {Chen}, {Chen}, {Chen}, {Cheng}, {Cheong},
  {Chia}, {Chincarini}, {Chiummo}, {Cho}, {Cho}, {Cho}, {Christensen}, {Chu},
  {Chua}, {Chung}, {Chung}, {Ciani}, {Ciobanu}, {Ciolfi}, {Cipriano}, {Cirone},
  {Clara}, {Clark}, {Clearwater}, {Cleva}, {Cocchieri}, {Coccia}, {Cohadon},
  {Cohen}, {Colgan}, {Colleoni}, {Collette}, {Collins}, {Cominsky},
  {Constancio}, {Conti}, {Cooper}, {Corban}, {Corbitt}, {Cordero-Carri{\'o}n},
  {Corley}, {Cornish}, {Corsi}, {Cortese}, {Costa}, {Cotesta}, {Coughlin},
  {Coughlin}, {Coulon}, {Countryman}, {Couvares}, {Covas}, {Cowan}, {Coward},
  {Cowart}, {Coyne}, {Coyne}, {Creighton}, {Creighton}, {Cripe}, {Croquette},
  {Crowder}, {Cullen}, {Cumming}, {Cunningham}, {Cuoco}, {Canton}, {D{\'a}lya},
  {Danilishin}, {D'Antonio}, {Danzmann}, {Dasgupta}, {Costa}, {Datrier},
  {Dattilo}, {Dave}, {Davier}, {Davis}, {Daw}, {DeBra}, {Deenadayalan},
  {Degallaix}, {De Laurentis}, {Del{\'e}glise}, {Del Pozzo}, {DeMarchi},
  {Demos}, {Dent}, {De Pietri}, {Derby}, {De Rosa}, {De Rossi}, {DeSalvo}, {de
  Varona}, {Dhurandhar}, {D{\'\i}az}, {Dietrich}, {Di Fiore}, {Di Giovanni},
  {Di Girolamo}, {Di Lieto}, {Ding}, {Di Pace}, {Di Palma}, {Di Renzo},
  {Dmitriev}, {Doctor}, {Donovan}, {Dooley}, {Doravari}, {Dorrington},
  {Downes}, {Drago}, {Driggers}, {Du}, {Ducoin}, {Dupej}, {Dwyer}, {Easter},
  {Edo}, {Edwards}, {Effler}, {Ehrens}, {Eichholz}, {Eikenberry}, {Eisenmann},
  {Eisenstein}, {Essick}, {Estelles}, {Estevez}, {Etienne}, {Etzel}, {Evans},
  {Evans}, {Fafone}, {Fair}, {Fairhurst}, {Fan}, {Farinon}, {Farr}, {Farr},
  {Fauchon-Jones}, {Favata}, {Fays}, {Fazio}, {Fee}, {Feicht}, {Fejer}, {Feng},
  {Fernandez-Galiana}, {Ferrante}, {Ferreira}, {Ferreira}, {Ferrini},
  {Fidecaro}, {Fiori}, {Fiorucci}, {Fishbach}, {Fisher}, {Fishner},
  {Fitz-Axen}, {Flaminio}, {Fletcher}, {Flynn}, {Fong}, {Font}, {Forsyth},
  {Fournier}, {Frasca}, {Frasconi}, {Frei}, {Freise}, {Frey}, {Frey},
  {Fritschel}, {Frolov}, {Fulda}, {Fyffe}, {Gabbard}, {Gadre}, {Gaebel},
  {Gair}, {Gammaitoni}, {Ganija}, {Gaonkar}, {Garcia},
  {Garc{\'\i}a-Quir{\'o}s}, {Garufi}, {Gateley}, {Gaudio}, {Gaur}, {Gayathri},
  {Gemme}, {Genin}, {Gennai}, {George}, {George}, {Gergely}, {Germain},
  {Ghonge}, {Ghosh}, {Ghosh}, {Ghosh}, {Giacomazzo}, {Giaime}, {Giardina},
  {Giazotto}, {Gill}, {Giordano}, {Glover}, {Godwin}, {Goetz}, {Goetz},
  {Goncharov}, {Gonz{\'a}lez}, {Castro}, {Gopakumar}, {Gorodetsky}, {Gossan},
  {Gosselin}, {Gouaty}, {Grado}, {Graef}, {Granata}, {Grant}, {Gras},
  {Grassia}, {Gray}, {Gray}, {Greco}, {Green}, {Green}, {Gretarsson}, {Groot},
  {Grote}, {Grunewald}, {Gruning}, {Guidi}, {Gulati}, {Guo}, {Gupta}, {Gupta},
  {Gustafson}, {Gustafson}, {Haegel}, {Halim}, {Hall}, {Hall}, {Hamilton},
  {Hammond}, {Haney}, {Hanke}, {Hanks}, {Hanna}, {Hannam}, {Hannuksela},
  {Hanson}, {Hardwick}, {Haris}, {Harms}, {Harry}, {Harry}, {Haster},
  {Haughian}, {Hayes}, {Healy}, {Heidmann}, {Heintze}, {Heitmann}, {Hello},
  {Hemming}, {Hendry}, {Heng}, {Hennig}, {Heptonstall}, {Vivanco}, {Heurs},
  {Hild}, {Hinderer}, {Hoak}, {Hochheim}, {Hofman}, {Holgado}, {Holland},
  {Holt}, {Holz}, {Hopkins}, {Horst}, {Hough}, {Howell}, {Hoy}, {Hreibi},
  {Huerta}, {Huet}, {Hughey}, {Hulko}, {Husa}, {Huttner}, {Huynh-Dinh},
  {Idzkowski}, {Iess}, {Ingram}, {Inta}, {Intini}, {Irwin}, {Isa}, {Isac},
  {Isi}, {Iyer}, {Izumi}, {Jacqmin}, {Jadhav}, {Jani}, {Janthalur},
  {Jaranowski}, {Jenkins}, {Jiang}, {Johnson}, {Johnson-McDaniel}, {Jones},
  {Jones}, {Jones}, {Jonker}, {Ju}, {Junker}, {Kalaghatgi}, {Kalogera},
  {Kamai}, {Kandhasamy}, {Kang}, {Kanner}, {Kapadia}, {Karki}, {Karvinen},
  {Kashyap}, {Kasprzack}, {Katsanevas}, {Katsavounidis}, {Katzman}, {Kaufer},
  {Kawabe}, {Keerthana}, {K{\'e}f{\'e}lian}, {Keitel}, {Kennedy}, {Key},
  {Khalili}, {Khan}, {Khan}, {Khan}, {Khan}, {Khazanov}, {Khursheed},
  {Kijbunchoo}, {Kim}, {Kim}, {Kim}, {Kim}, {Kim}, {Kim}, {Kimball}, {King},
  {King}, {Kinley-Hanlon}, {Kirchhoff}, {Kissel}, {Kleybolte}, {Klika},
  {Klimenko}, {Knowles}, {Koch}, {Koehlenbeck}, {Koekoek}, {Koley},
  {Kondrashov}, {Kontos}, {Koper}, {Korobko}, {Korth}, {Kowalska}, {Kozak},
  {Kringel}, {Krishnendu}, {Kr{\'o}lak}, {Kuehn}, {Kumar}, {Kumar}, {Kumar},
  {Kumar}, {Kuo}, {Kutynia}, {Kwang}, {Lackey}, {Lai}, {Lam}, {Landry}, {Lane},
  {Lang}, {Lange}, {Lantz}, {Lanza}, {Lartaux-Vollard}, {Lasky}, {Laxen},
  {Lazzarini}, {Lazzaro}, {Leaci}, {Leavey}, {Lecoeuche}, {Lee}, {Lee}, {Lee},
  {Lee}, {Lee}, {Lee}, {Lehmann}, {Lenon}, {Leroy}, {Letendre}, {Levin}, {Li},
  {Li}, {Li}, {Li}, {Lin}, {Linde}, {Linker}, {Littenberg}, {Liu}, {Liu}, {Lo},
  {Lockerbie}, {London}, {Longo}, {Lorenzini}, {Loriette}, {Lormand},
  {Losurdo}, {Lough}, {Lousto}, {Lovelace}, {Lower}, {L{\"u}ck}, {Lumaca},
  {Lundgren}, {Lynch}, {Ma}, {Macas}, {Macfoy}, {MacInnis}, {Macleod},
  {Macquet}, {Maga{\~n}a-Sandoval}, {Zertuche}, {Magee}, {Majorana},
  {Maksimovic}, {Malik}, {Man}, {Mandic}, {Mangano}, {Mansell}, {Manske},
  {Mantovani}, {Marchesoni}, {Marion}, {M{\'a}rka}, {M{\'a}rka}, {Markakis},
  {Markosyan}, {Markowitz}, {Maros}, {Marquina}, {Marsat}, {Martelli},
  {Martin}, {Martin}, {Martynov}, {Mason}, {Massera}, {Masserot}, {Massinger},
  {Masso-Reid}, {Mastrogiovanni}, {Matas}, {Matichard}, {Matone}, {Mavalvala},
  {Mazumder}, {McCann}, {McCarthy}, {McClelland}, {McCormick}, {McCuller},
  {McGuire}, {McIver}, {McManus}, {McRae}, {McWilliams}, {Meacher}, {Meadors},
  {Mehmet}, {Mehta}, {Meidam}, {Melatos}, {Mendell}, {Mercer}, {Mereni},
  {Merilh}, {Merzougui}, {Meshkov}, {Messenger}, {Messick}, {Metzdorff},
  {Meyers}, {Miao}, {Michel}, {Middleton}, {Mikhailov}, {Milano}, {Miller},
  {Miller}, {Millhouse}, {Mills}, {Milovich-Goff}, {Minazzoli}, {Minenkov},
  {Mishkin}, {Mishra}, {Mistry}, {Mitra}, {Mitrofanov}, {Mitselmakher},
  {Mittleman}, {Mo}, {Moffa}, {Mogushi}, {Mohapatra}, {Montani}, {Moore},
  {Moraru}, {Moreno}, {Morisaki}, {Mours}, {Mow-Lowry}, {Mukherjee},
  {Mukherjee}, {Mukherjee}, {Mukund}, {Mullavey}, {Munch}, {Mu{\~n}iz},
  {Muratore}, {Murray}, {Nagar}, {Nardecchia}, {Naticchioni}, {Nayak},
  {Neilson}, {Nelemans}, {Nelson}, {Nery}, {Neunzert}, {Ng}, {Ng}, {Nguyen},
  {Nichols}, {Nielsen}, {Nissanke}, {Nitz}, {Nocera}, {North}, {Nuttall},
  {Obergaulinger}, {Oberling}, {O'Brien}, {O'Dea}, {Ogin}, {Oh}, {Oh}, {Ohme},
  {Ohta}, {Okada}, {Oliver}, {Oppermann}, {Oram}, {O'Reilly}, {Ormiston},
  {Ortega}, {O'Shaughnessy}, {Ossokine}, {Ottaway}, {Overmier}, {Owen}, {Pace},
  {Pagano}, {Page}, {Pai}, {Pai}, {Palamos}, {Palashov}, {Palomba},
  {Pal-Singh}, {Pan}, {Pang}, {Pang}, {Pankow}, {Pannarale}, {Pant},
  {Paoletti}, {Paoli}, {Parida}, {Parker}, {Pascucci}, {Pasqualetti},
  {Passaquieti}, {Passuello}, {Patil}, {Patricelli}, {Pearlstone}, {Pedersen},
  {Pedraza}, {Pedurand}, {Pele}, {Penn}, {Perez}, {Perreca}, {Pfeiffer},
  {Phelps}, {Phukon}, {Piccinni}, {Pichot}, {Piergiovanni}, {Pillant},
  {Pinard}, {Pirello}, {Pitkin}, {Poggiani}, {Pong}, {Ponrathnam}, {Popolizio},
  {Porter}, {Powell}, {Prajapati}, {Prasad}, {Prasai}, {Prasanna}, {Pratten},
  {Prestegard}, {Privitera}, {Prodi}, {Prokhorov}, {Puncken}, {Punturo},
  {Puppo}, {P{\"u}rrer}, {Qi}, {Quetschke}, {Quinonez}, {Quintero},
  {Quitzow-James}, {Raab}, {Radkins}, {Radulescu}, {Raffai}, {Raja}, {Rajan},
  {Rajbhandari}, {Rakhmanov}, {Ramirez}, {Ramos-Buades}, {Rana}, {Rao},
  {Rapagnani}, {Raymond}, {Razzano}, {Read}, {Regimbau}, {Rei}, {Reid},
  {Reitze}, {Ren}, {Ricci}, {Richardson}, {Richardson}, {Ricker}, {Riles},
  {Rizzo}, {Robertson}, {Robie}, {Robinet}, {Rocchi}, {Rolland}, {Rollins},
  {Roma}, {Romanelli}, {Romano}, {Romel}, {Romie}, {Rose}, {Rosi{\'n}ska},
  {Rosofsky}, {Ross}, {Rowan}, {R{\"u}diger}, {Ruggi}, {Rutins}, {Ryan},
  {Sachdev}, {Sadecki}, {Sakellariadou}, {Salconi}, {Saleem}, {Samajdar},
  {Sammut}, {Sanchez}, {Sanchez}, {Sanchis-Gual}, {Sandberg}, {Sanders},
  {Santiago}, {Sarin}, {Sassolas}, {Sathyaprakash}, {Saulson}, {Sauter},
  {Savage}, {Schale}, {Scheel}, {Scheuer}, {Schmidt}, {Schnabel}, {Schofield},
  {Sch{\"o}nbeck}, {Schreiber}, {Schulte}, {Schutz}, {Schwalbe}, {Scott},
  {Scott}, {Seidel}, {Sellers}, {Sengupta}, {Sennett}, {Sentenac}, {Sequino},
  {Sergeev}, {Setyawati}, {Shaddock}, {Shaffer}, {Shahriar}, {Shaner}, {Shao},
  {Sharma}, {Shawhan}, {Shen}, {Shink}, {Shoemaker}, {Shoemaker},
  {ShyamSundar}, {Siellez}, {Sieniawska}, {Sigg}, {Silva}, {Singer}, {Singh},
  {Singhal}, {Sintes}, {Sitmukhambetov}, {Skliris}, {Slagmolen},
  {Slaven-Blair}, {Smith}, {Smith}, {Somala}, {Son}, {Sorazu}, {Sorrentino},
  {Souradeep}, {Sowell}, {Spencer}, {Srivastava}, {Srivastava}, {Staats},
  {Stachie}, {Standke}, {Steer}, {Steinke}, {Steinlechner}, {Steinlechner},
  {Steinmeyer}, {Stevenson}, {Stocks}, {Stone}, {Stops}, {Strain}, {Stratta},
  {Strigin}, {Strunk}, {Sturani}, {Stuver}, {Sudhir}, {Summerscales}, {Sun},
  {Sunil}, {Suresh}, {Sutton}, {Swinkels}, {Szczepa{\'n}czyk}, {Tacca}, {Tait},
  {Talbot}, {Talukder}, {Tanner}, {T{\'a}pai}, {Taracchini}, {Tasson},
  {Taylor}, {Thies}, {Thomas}, {Thomas}, {Thondapu}, {Thorne}, {Thrane},
  {Tiwari}, {Tiwari}, {Tiwari}, {Toland}, {Tonelli}, {Tornasi},
  {Torres-Forn{\'e}}, {Torrie}, {T{\"o}yr{\"a}}, {Travasso}, {Traylor},
  {Tringali}, {Trovato}, {Trozzo}, {Trudeau}, {Tsang}, {Tse}, {Tso}, {Tsukada},
  {Tsuna}, {Tuyenbayev}, {Ueno}, {Ugolini}, {Unnikrishnan}, {Urban}, {Usman},
  {Vahlbruch}, {Vajente}, {Valdes}, {van Bakel}, {van Beuzekom}, {van den
  Brand}, {Van Den Broeck}, {Vander-Hyde}, {van Heijningen}, {van der Schaaf},
  {van Veggel}, {Vardaro}, {Varma}, {Vass}, {Vas{\'u}th}, {Vecchio},
  {Vedovato}, {Veitch}, {Veitch}, {Venkateswara}, {Venugopalan}, {Verkindt},
  {Vetrano}, {Vicer{\'e}}, {Viets}, {Vine}, {Vinet}, {Vitale}, {Vo}, {Vocca},
  {Vorvick}, {Vyatchanin}, {Wade}, {Wade}, {Wade}, {Wald}, {Walet}, {Walker},
  {Wallace}, {Walsh}, {Wang}, {Wang}, {Wang}, {Wang}, {Wang}, {Ward}, {Warden},
  {Warner}, {Was}, {Watchi}, {Weaver}, {Wei}, {Weinert}, {Weinstein}, {Weiss},
  {Wellmann}, {Wen}, {Wessel}, {We{\ss}els}, {Westhouse}, {Wette}, {Whelan},
  {Whiting}, {Whittle}, {Wilken}, {Williams}, {Williamson}, {Willis}, {Willke},
  {Wimmer}, {Winkler}, {Wipf}, {Wittel}, {Woan}, {Woehler}, {Wofford},
  {Worden}, {Wright}, {Wu}, {Wysocki}, {Xiao}, {Yamamoto}, {Yancey}, {Yang},
  {Yap}, {Yazback}, {Yeeles}, {Yu}, {Yu}, {Yuen}, {Yvert}, {Zadro{\.Z}ny},
  {Zanolin}, {Zelenova}, {Zendri}, {Zevin}, {Zhang}, {Zhang}, {Zhang}, {Zhao},
  {Zhou}, {Zhou}, {Zhu}, {Zimmerman}, {Zucker}, {Zweizig}, {LIGO Scientific
  Collaboration}, \& {Virgo Collaboration}}]{LIGO2}
---. 2019, \prd, 100, 104036, \dodoi{10.1103/PhysRevD.100.104036}

\bibitem[{{Abbott} {et~al.}(2021){Abbott}, {Abbott}, {Abraham}, {Acernese},
  {Ackley}, {Adams}, {Adams}, {Adhikari}, {Adya}, {Affeldt}, {Agarwal},
  {Agathos}, {Agatsuma}, {Aggarwal}, {Aguiar}, {Aiello}, {Ain}, {Akutsu},
  {Aleman}, {Allen}, {Allocca}, {Altin}, {Amato}, {Anand}, {Ananyeva},
  {Anderson}, {Anderson}, {Ando}, {Angelova}, {Ansoldi}, {Antelis}, {Antier},
  {Appert}, {Arai}, {Arai}, {Arai}, {Araki}, {Araya}, {Araya}, {Areeda},
  {Ar{\`e}ne}, {Aritomi}, {Arnaud}, {Aronson}, {Asada}, {Asali}, {Ashton},
  {Aso}, {Aston}, {Astone}, {Aubin}, {Aufmuth}, {Aultoneal}, {Austin}, {Babak},
  {Badaracco}, {Bader}, {Bae}, {Bae}, {Baer}, {Bagnasco}, {Bai}, {Baiotti},
  {Baird}, {Bajpai}, {Ball}, {Ballardin}, {Ballmer}, {Bals}, {Balsamo},
  {Baltus}, {Banagiri}, {Bankar}, {Bankar}, {Barayoga}, {Barbieri}, {Barish},
  {Barker}, {Barneo}, {Barnum}, {Barone}, {Barr}, {Barsotti}, {Barsuglia},
  {Barta}, {Bartlett}, {Barton}, {Bartos}, {Bassiri}, {Basti}, {Bawaj},
  {Bayley}, {Baylor}, {Bazzan}, {B{\'e}csy}, {Bedakihale}, {Bejger},
  {Belahcene}, {Benedetto}, {Beniwal}, {Benjamin}, {Bennett}, {Bentley},
  {Benyaala}, {Bergamin}, {Berger}, {Bernuzzi}, {Bersanetti}, {Bertolini},
  {Betzwieser}, {Bhandare}, {Bhandari}, {Bhattacharjee}, {Bhaumik}, {Bidler},
  {Bilenko}, {Billingsley}, {Birney}, {Birnholtz}, {Biscans}, {Bischi},
  {Biscoveanu}, {Bisht}, {Biswas}, {Bitossi}, {Bizouard}, {Blackburn},
  {Blackman}, {Blair}, {Blair}, {Blair}, {Bobba}, {Bode}, {Boer}, {Bogaert},
  {Boldrini}, {Bondu}, {Bonilla}, {Bonnand}, {Booker}, {Boom}, {Bork},
  {Boschi}, {Bose}, {Bose}, {Bossilkov}, {Boudart}, {Bouffanais}, {Bozzi},
  {Bradaschia}, {Brady}, {Bramley}, {Branch}, {Branchesi}, {Brau}, {Breschi},
  {Briant}, {Briggs}, {Brillet}, {Brinkmann}, {Brockill}, {Brooks}, {Brooks},
  {Brown}, {Brunett}, {Bruno}, {Bruntz}, {Bryant}, {Buikema}, {Bulik},
  {Bulten}, {Buonanno}, {Buscicchio}, {Buskulic}, {Byer}, {Cadonati}, {Caesar},
  {Cagnoli}, {Cahillane}, {Cain}, {Bustillo}, {Callaghan}, {Callister},
  {Calloni}, {Camp}, {Canepa}, {Cannavacciuolo}, {Cannon}, {Cao}, {Cao}, {Cao},
  {Capocasa}, {Capote}, {Carapella}, {Carbognani}, {Carlin}, {Carney},
  {Carpinelli}, {Carullo}, {Carver}, {Diaz}, {Casentini}, {Castaldi},
  {Caudill}, {Cavagli{\`a}}, {Cavalier}, {Cavalieri}, {Cella},
  {Cerd{\'a}-Dur{\'a}n}, {Cesarini}, {Chaibi}, {Chakravarti}, {Champion},
  {Chan}, {Chan}, {Chan}, {Chan}, {Chandra}, {Chanial}, {Chao}, {Charlton},
  {Chase}, {Chassande-Mottin}, {Chatterjee}, {Chaturvedi}, {Chen}, {Chen},
  {Chen}, {Chen}, {Chen}, {Chen}, {Chen}, {Chen}, {Chen}, {Cheng}, {Cheong},
  {Cheung}, {Chia}, {Chiadini}, {Chiang}, {Chierici}, {Chincarini}, {Chiofalo},
  {Chiummo}, {Cho}, {Cho}, {Choate}, {Choudhary}, {Choudhary}, {Christensen},
  {Chu}, {Chu}, {Chu}, {Chua}, {Chung}, {Ciani}, {Ciecielag}, {Cie{\'s}lar},
  {Cifaldi}, {Ciobanu}, {Ciolfi}, {Cipriano}, {Cirone}, {Clara}, {Clark},
  {Clark}, {Clarke}, {Clearwater}, {Clesse}, {Cleva}, {Coccia}, {Cohadon},
  {Cohen}, {Cohen}, {Colleoni}, {Collette}, {Colpi}, {Compton}, {Constancio},
  {Conti}, {Cooper}, {Corban}, {Corbitt}, {Cordero-Carri{\'o}n}, {Corezzi},
  {Corley}, {Cornish}, {Corre}, {Corsi}, {Cortese}, {Costa}, {Cotesta},
  {Coughlin}, {Coughlin}, {Coulon}, {Countryman}, {Cousins}, {Couvares},
  {Covas}, {Coward}, {Cowart}, {Coyne}, {Coyne}, {Creighton}, {Creighton},
  {Criswell}, {Croquette}, {Crowder}, {Cudell}, {Cullen}, {Cumming},
  {Cummings}, {Cuoco}, {Cury{\l}o}, {Canton}, {D{\'a}lya}, {Dana},
  {Daneshgaranbajastani}, {D'Angelo}, {Danilishin}, {D'Antonio}, {Danzmann},
  {Darsow-Fromm}, {Dasgupta}, {Datrier}, {Dattilo}, {Dave}, {Davier}, {Davies},
  {Davis}, {Daw}, {Dean}, {Deenadayalan}, {Degallaix}, {de Laurentis},
  {Del{\'e}glise}, {Del Favero}, {de Lillo}, {de Lillo}, {Del Pozzo},
  {Demarchi}, {de Matteis}, {D'Emilio}, {Demos}, {Dent}, {Depasse}, {de
  Pietri}, {De Rosa}, {de Rossi}, {Desalvo}, {de Simone}, {Dhurandhar},
  {D{\'\i}az}, {Diaz-Ortiz}, {Didio}, {Dietrich}, {di Fiore}, {di Fronzo}, {di
  Giorgio}, {di Giovanni}, {di Girolamo}, {di Lieto}, {Ding}, {di Pace}, {di
  Palma}, {di Renzo}, {Divakarla}, {Dmitriev}, {Doctor}, {D'Onofrio},
  {Donovan}, {Dooley}, {Doravari}, {Dorrington}, {Drago}, {Driggers}, {Drori},
  {Du}, {Ducoin}, {Dupej}, {Durante}, {D'Urso}, {Duverne}, {Dvorkin}, {Dwyer},
  {Easter}, {Ebersold}, {Eddolls}, {Edelman}, {Edo}, {Edy}, {Effler}, {Eguchi},
  {Eichholz}, {Eikenberry}, {Eisenmann}, {Eisenstein}, {Ejlli}, {Enomoto},
  {Errico}, {Essick}, {Estell{\'e}s}, {Estevez}, {Etienne}, {Etzel}, {Evans},
  {Evans}, {Ewing}, {Fafone}, {Fair}, {Fairhurst}, {Fan}, {Farah}, {Farinon},
  {Farr}, {Farr}, {Farrow}, {Fauchon-Jones}, {Favata}, {Fays}, {Fazio},
  {Feicht}, {Fejer}, {Feng}, {Fenyvesi}, {Ferguson}, {Fernandez-Galiana},
  {Ferrante}, {Ferreira}, {Fidecaro}, {Figura}, {Fiori}, {Fishbach}, {Fisher},
  {Fishner}, {Fittipaldi}, {Fiumara}, {Flaminio}, {Floden}, {Flynn}, {Fong},
  {Font}, {Fornal}, {Forsyth}, {Franke}, {Frasca}, {Frasconi}, {Frederick},
  {Frei}, {Freise}, {Frey}, {Fritschel}, {Frolov}, {Fronz{\'e}}, {Fujii},
  {Fujikawa}, {Fukunaga}, {Fukushima}, {Fulda}, {Fyffe}, {Gabbard}, {Gadre},
  {Gaebel}, {Gair}, {Gais}, {Galaudage}, {Gamba}, {Ganapathy}, {Ganguly},
  {Gao}, {Gaonkar}, {Garaventa}, {Garc{\'\i}a-N{\'u}{\~n}ez},
  {Garc{\'\i}a-Quir{\'o}s}, {Garufi}, {Gateley}, {Gaudio}, {Gayathri}, {Ge},
  {Gemme}, {Gennai}, {George}, {Gergely}, {Gewecke}, {Ghonge}, {Ghosh},
  {Ghosh}, {Ghosh}, {Ghosh}, {Ghosh}, {Giacomazzo}, {Giacoppo}, {Giaime},
  {Giardina}, {Gibson}, {Gier}, {Giesler}, {Giri}, {Gissi}, {Glanzer},
  {Gleckl}, {Godwin}, {Goetz}, {Goetz}, {Gohlke}, {Goncharov}, {Gonz{\'a}lez},
  {Gopakumar}, {Gosselin}, {Gouaty}, {Grace}, {Grado}, {Granata}, {Granata},
  {Grant}, {Gras}, {Grassia}, {Gray}, {Gray}, {Greco}, {Green}, {Green},
  {Gretarsson}, {Gretarsson}, {Griffith}, {Griffiths}, {Griggs}, {Grignani},
  {Grimaldi}, {Grimes}, {Grimm}, {Grote}, {Grunewald}, {Gruning}, {Guerrero},
  {Guidi}, {Guimaraes}, {Guix{\'e}}, {Gulati}, {Guo}, {Guo}, {Gupta}, {Gupta},
  {Gupta}, {Gustafson}, {Gustafson}, {Guzman}, {Ha}, {Haegel}, {Hagiwara},
  {Haino}, {Halim}, {Hall}, {Hamilton}, {Hammond}, {Han}, {Haney}, {Hanks},
  {Hanna}, {Hannam}, {Hannuksela}, {Hansen}, {Hansen}, {Hanson}, {Harder},
  {Hardwick}, {Haris}, {Harms}, {Harry}, {Harry}, {Hartwig}, {Hasegawa},
  {Haskell}, {Hasskew}, {Haster}, {Hattori}, {Haughian}, {Hayakawa}, {Hayama},
  {Hayes}, {Healy}, {Heidmann}, {Heintze}, {Heinze}, {Heinzel}, {Heitmann},
  {Hellman}, {Hello}, {Helmling-Cornell}, {Hemming}, {Hendry}, {Heng},
  {Hennes}, {Hennig}, {Hennig}, {Vivanco}, {Heurs}, {Hild}, {Hill}, {Himemoto},
  {Hines}, {Hiranuma}, {Hirata}, {Hirose}, {Hochheim}, {Hofman}, {Hohmann},
  {Holgado}, {Holland}, {Hollows}, {Holmes}, {Holt}, {Holz}, {Hong}, {Hopkins},
  {Hough}, {Howell}, {Hoy}, {Hoyland}, {Hreibi}, {Hsieh}, {Hsu}, {Huang},
  {Huang}, {Huang}, {Huang}, {Huang}, {Huang}, {H{\"u}bner}, {Huddart},
  {Huerta}, {Hughey}, {Hui}, {Hui}, {Husa}, {Huttner}, {Huxford}, {Huynh-Dinh},
  {Ide}, {Idzkowski}, {Iess}, {Ikenoue}, {Imam}, {Inayoshi}, {Inchauspe},
  {Ingram}, {Inoue}, {Intini}, {Ioka}, {Isi}, {Isleif}, {Ito}, {Itoh}, {Iyer},
  {Izumi}, {Jaberianhamedan}, {Jacqmin}, {Jadhav}, {Jadhav}, {James}, {Jan},
  {Jani}, {Janssens}, {Janthalur}, {Jaranowski}, {Jariwala}, {Jaume},
  {Jenkins}, {Jeon}, {Jeunon}, {Jia}, {Jiang}, {Jin}, {Johns}, {Jones},
  {Jones}, {Jones}, {Jones}, {Jones}, {Jonker}, {Ju}, {Jung}, {Jung}, {Junker},
  {Kaihotsu}, {Kajita}, {Kakizaki}, {Kalaghatgi}, {Kalogera}, {Kamai},
  {Kamiizumi}, {Kanda}, {Kandhasamy}, {Kang}, {Kanner}, {Kao}, {Kapadia},
  {Kapasi}, {Karathanasis}, {Karki}, {Kashyap}, {Kasprzack}, {Kastaun},
  {Katsanevas}, {Katsavounidis}, {Katzman}, {Kaur}, {Kawabe}, {Kawaguchi},
  {Kawai}, {Kawasaki}, {K{\'e}f{\'e}lian}, {Keitel}, {Key}, {Khadka},
  {Khalili}, {Khan}, {Khan}, {Khazanov}, {Khetan}, {Khursheed}, {Kijbunchoo},
  {Kim}, {Kim}, {Kim}, {Kim}, {Kim}, {Kim}, {Kimball}, {Kimura}, {King},
  {Kinley-Hanlon}, {Kirchhoff}, {Kissel}, {Kita}, {Kitazawa}, {Kleybolte},
  {Klimenko}, {Knee}, {Knowles}, {Knyazev}, {Koch}, {Koekoek}, {Kojima},
  {Kokeyama}, {Koley}, {Kolitsidou}, {Kolstein}, {Komori}, {Kondrashov},
  {Kong}, {Kontos}, {Koper}, {Korobko}, {Kotake}, {Kovalam}, {Kozak},
  {Kozakai}, {Kozu}, {Kringel}, {Krishnendu}, {Kr{\'o}lak}, {Kuehn}, {Kuei},
  {Kumar}, {Kumar}, {Kumar}, {Kumar}, {Kume}, {Kuns}, {Kuo}, {Kuo}, {Kuromiya},
  {Kuroyanagi}, {Kusayanagi}, {Kwak}, {Kwang}, {Laghi}, {Lalande}, {Lam},
  {Lamberts}, {Landry}, {Lane}, {Lang}, {Lange}, {Lantz}, {La Rosa},
  {Lartaux-Vollard}, {Lasky}, {Laxen}, {Lazzarini}, {Lazzaro}, {Leaci},
  {Leavey}, {Lecoeuche}, {Lee}, {Lee}, {Lee}, {Lee}, {Lee}, {Lee}, {Lehmann},
  {Lema{\^\i}tre}, {Leon}, {Leonardi}, {Leroy}, {Letendre}, {Levin}, {Leviton},
  {Li}, {Li}, {Li}, {Li}, {Li}, {Li}, {Lin}, {Lin}, {Lin}, {Lin}, {Lin},
  {Linde}, {Linker}, {Linley}, {Littenberg}, {Liu}, {Liu}, {Liu}, {Liu},
  {Llorens-Monteagudo}, {Lo}, {Lockwood}, {Lollie}, {London}, {Longo}, {Lopez},
  {Lorenzini}, {Loriette}, {Lormand}, {Losurdo}, {Lough}, {Lousto}, {Lovelace},
  {L{\"u}ck}, {Lumaca}, {Lundgren}, {Luo}, {Macas}, {Macinnis}, {MacLeod},
  {MacMillan}, {Macquet}, {Hernandez}, {Maga{\~n}a-Sandoval}, {Magazz{\`u}},
  {Magee}, {Maggiore}, {Majorana}, {Maksimovic}, {Maliakal}, {Malik}, {Man},
  {Mandic}, {Mangano}, {Mango}, {Mansell}, {Manske}, {Mantovani}, {Mapelli},
  {Marchesoni}, {Marchio}, {Marion}, {Mark}, {M{\'a}rka}, {M{\'a}rka},
  {Markakis}, {Markosyan}, {Markowitz}, {Maros}, {Marquina}, {Marsat},
  {Martelli}, {Martin}, {Martin}, {Martinez}, {Martinez}, {Martinovic},
  {Martynov}, {Marx}, {Masalehdan}, {Mason}, {Massera}, {Masserot},
  {Massinger}, {Masso-Reid}, {Mastrogiovanni}, {Matas}, {Mateu-Lucena},
  {Matichard}, {Matiushechkina}, {Mavalvala}, {McCann}, {McCarthy},
  {McClelland}, {McClincy}, {McCormick}, {McCuller}, {McGhee}, {McGuire},
  {McIsaac}, {McIver}, {McManus}, {McRae}, {McWilliams}, {Meacher}, {Mehmet},
  {Mehta}, {Melatos}, {Melchor}, {Mendell}, {Menendez-Vazquez}, {Menoni},
  {Mercer}, {Mereni}, {Merfeld}, {Merilh}, {Merritt}, {Merzougui}, {Meshkov},
  {Messenger}, {Messick}, {Meyers}, {Meylahn}, {Mhaske}, {Miani}, {Miao},
  {Michaloliakos}, {Michel}, {Michimura}, {Middleton}, {Milano}, {Miller},
  {Millhouse}, {Mills}, {Milotti}, {Milovich-Goff}, {Minazzoli}, {Minenkov},
  {Mio}, {Mir}, {Mishkin}, {Mishra}, {Mishra}, {Mistry}, {Mitra}, {Mitrofanov},
  {Mitselmakher}, {Mittleman}, {Miyakawa}, {Miyamoto}, {Miyazaki}, {Miyo},
  {Miyoki}, {Mo}, {Mogushi}, {Mohapatra}, {Mohite}, {Molina}, {Molina-Ruiz},
  {Mondin}, {Montani}, {Moore}, {Moraru}, {Morawski}, {More}, {Moreno},
  {Moreno}, {Mori}, {Morisaki}, {Moriwaki}, {Mours}, {Mow-Lowry}, {Mozzon},
  {Muciaccia}, {Mukherjee}, {Mukherjee}, {Mukherjee}, {Mukherjee}, {Mukund},
  {Mullavey}, {Munch}, {Mu{\~n}iz}, {Murray}, {Musenich}, {Nadji}, {Nagano},
  {Nagano}, {Nagar}, {Nakamura}, {Nakano}, {Nakano}, {Nakashima}, {Nakayama},
  {Nardecchia}, {Narikawa}, {Naticchioni}, {Nayak}, {Nayak}, {Negishi}, {Neil},
  {Neilson}, {Nelemans}, {Nelson}, {Nery}, {Neunzert}, {Ng}, {Ng}, {Nguyen},
  {Nguyen}, {Nguyen}, {Quynh}, {Ni}, {Nichols}, {Nishizawa}, {Nissanke},
  {Nocera}, {Noh}, {Norman}, {North}, {Nozaki}, {Nuttall}, {Oberling},
  {O'Brien}, {Obuchi}, {O'Dell}, {Ogaki}, {Oganesyan}, {Oh}, {Oh}, {Oh},
  {Ohashi}, {Ohishi}, {Ohkawa}, {Ohme}, {Ohta}, {Okada}, {Okutani}, {Okutomi},
  {Olivetto}, {Oohara}, {Ooi}, {Oram}, {O'Reilly}, {Ormiston}, {Ormsby},
  {Ortega}, {O'Shaughnessy}, {O'Shea}, {Oshino}, {Ossokine}, {Osthelder},
  {Otabe}, {Ottaway}, {Overmier}, {Pace}, {Pagano}, {Page}, {Pagliaroli},
  {Pai}, {Pai}, {Palamos}, {Palashov}, {Palomba}, {Pan}, {Panda}, {Pang},
  {Pang}, {Pankow}, {Pannarale}, {Pant}, {Paoletti}, {Paoli}, {Paolone},
  {Parisi}, {Park}, {Parker}, {Pascucci}, {Pasqualetti}, {Passaquieti},
  {Passuello}, {Patel}, {Patricelli}, {Payne}, {Pechsiri}, {Pedraza},
  {Pegoraro}, {Pele}, {Arellano}, {Penn}, {Perego}, {Pereira}, {Pereira},
  {Perez}, {P{\'e}rigois}, {Perreca}, {Perri{\`e}s}, {Petermann}, {Petterson},
  {Pfeiffer}, {Pham}, {Phukon}, {Piccinni}, {Pichot}, {Piendibene},
  {Piergiovanni}, {Pierini}, {Pierro}, {Pillant}, {Pilo}, {Pinard}, {Pinto},
  {Piotrzkowski}, {Piotrzkowski}, {Pirello}, {Pitkin}, {Placidi}, {Plastino},
  {Pluchar}, {Poggiani}, {Polini}, {Pong}, {Ponrathnam}, {Popolizio}, {Porter},
  {Powell}, {Pracchia}, {Pradier}, {Prajapati}, {Prasai}, {Prasanna},
  {Pratten}, {Prestegard}, {Principe}, {Prodi}, {Prokhorov}, {Prosposito},
  {Prudenzi}, {Puecher}, {Punturo}, {Puosi}, {Puppo}, {P{\"u}rrer}, {Qi},
  {Quetschke}, {Quinonez}, {Quitzow-James}, {Raab}, {Raaijmakers}, {Radkins},
  {Radulesco}, {Raffai}, {Rail}, {Raja}, {Rajan}, {Ramirez}, {Ramirez},
  {Ramos-Buades}, {Rana}, {Rapagnani}, {Rapol}, {Ratto}, {Raymond}, {Raza},
  {Razzano}, {Read}, {Rees}, {Regimbau}, {Rei}, {Reid}, {Reitze}, {Relton},
  {Rettegno}, {Ricci}, {Richardson}, {Richardson}, {Richardson}, {Ricker},
  {Riemenschneider}, {Riles}, {Rizzo}, {Robertson}, {Robie}, {Robinet},
  {Rocchi}, {Rocha}, {Rodriguez}, {Rodriguez-Soto}, {Rolland}, {Rollins},
  {Roma}, {Romanelli}, {Romano}, {Romano}, {Romel}, {Romero}, {Romero-Shaw},
  {Romie}, {Rose}, {Rosi{\'n}ska}, {Rosofsky}, {Ross}, {Rowan}, {Rowlinson},
  {Roy}, {Roy}, {Rozza}, {Ruggi}, {Ryan}, {Sachdev}, {Sadecki}, {Sadiq},
  {Sago}, {Saito}, {Saito}, {Sakai}, {Sakai}, {Sakellariadou}, {Sakuno},
  {Salafia}, {Salconi}, {Saleem}, {Salemi}, {Samajdar}, {Sanchez}, {Sanchez},
  {Sanchez}, {Sanchis-Gual}, {Sanders}, {Sanuy}, {Saravanan}, {Sarin},
  {Sassolas}, {Satari}, {Sato}, {Sato}, {Sauter}, {Savage}, {Savant}, {Sawada},
  {Sawant}, {Sawant}, {Sayah}, {Schaetzl}, {Scheel}, {Scheuer},
  {Schindler-Tyka}, {Schmidt}, {Schnabel}, {Schneewind}, {Schofield},
  {Sch{\"o}nbeck}, {Schulte}, {Schutz}, {Schwartz}, {Scott}, {Scott},
  {Seglar-Arroyo}, {Seidel}, {Sekiguchi}, {Sekiguchi}, {Sellers}, {Sengupta},
  {Sennett}, {Sentenac}, {Seo}, {Sequino}, {Sergeev}, {Setyawati}, {Shaffer},
  {Shahriar}, {Shams}, {Shao}, {Sharifi}, {Sharma}, {Sharma}, {Shawhan},
  {Shcheblanov}, {Shen}, {Shibagaki}, {Shikauchi}, {Shimizu}, {Shimoda},
  {Shimode}, {Shink}, {Shinkai}, {Shishido}, {Shoda}, {Shoemaker}, {Shoemaker},
  {Shukla}, {Shyamsundar}, {Sieniawska}, {Sigg}, {Singer}, {Singh}, {Singh},
  {Singha}, {Sintes}, {Sipala}, {Skliris}, {Slagmolen}, {Slaven-Blair},
  {Smetana}, {Smith}, {Smith}, {Somala}, {Somiya}, {Son}, {Soni}, {Soni},
  {Sorazu}, {Sordini}, {Sorrentino}, {Sorrentino}, {Sotani}, {Soulard},
  {Souradeep}, {Sowell}, {Spagnuolo}, {Spencer}, {Spera}, {Srivastava},
  {Srivastava}, {Staats}, {Stachie}, {Steer}, {Steinlechner}, {Steinlechner},
  {Stops}, {Stover}, {Strain}, {Strang}, {Stratta}, {Strunk}, {Sturani},
  {Stuver}, {S{\"u}dbeck}, {Sudhagar}, {Sudhir}, {Sugimoto}, {Suh},
  {Summerscales}, {Sun}, {Sun}, {Sunil}, {Sur}, {Suresh}, {Sutton}, {Suzuki},
  {Suzuki}, {Swinkels}, {Szczepa{\'n}czyk}, {Szewczyk}, {Tacca}, {Tagoshi},
  {Tait}, {Takahashi}, {Takahashi}, {Takamori}, {Takano}, {Takeda}, {Takeda},
  {Talbot}, {Tanaka}, {Tanaka}, {Tanaka}, {Tanaka}, {Tanaka}, {Tanasijczuk},
  {Tanioka}, {Tanner}, {Tao}, {Tapia}, {Martin}, {Martin}, {Tasson}, {Telada},
  {Tenorio}, {Terkowski}, {Test}, {Thirugnanasambandam}, {Thomas}, {Thomas},
  {Thompson}, {Thondapu}, {Thorne}, {Thrane}, {Tiwari}, {Tiwari}, {Tiwari},
  {Toland}, {Tolley}, {Tomaru}, {Tomigami}, {Tomura}, {Tonelli},
  {Torres-Forn{\'e}}, {Torrie}, {E Melo}, {T{\"o}yr{\"a}}, {Trapananti},
  {Travasso}, {Traylor}, {Tringali}, {Tripathee}, {Troiano}, {Trovato},
  {Trozzo}, {Trudeau}, {Tsai}, {Tsai}, {Tsang}, {Tsang}, {Tsao}, {Tse}, {Tso},
  {Tsubono}, {Tsuchida}, {Tsukada}, {Tsuna}, {Tsutsui}, {Tsuzuki}, {Turconi},
  {Tuyenbayev}, {Ubhi}, {Uchikata}, {Uchiyama}, {Udall}, {Ueda}, {Uehara},
  {Ueno}, {Ueshima}, {Ugolini}, {Unnikrishnan}, {Uraguchi}, {Urban}, {Ushiba},
  {Usman}, {Utina}, {Vahlbruch}, {Vajente}, {Vajpeyi}, {Valdes}, {Valentini},
  {Valsan}, {van Bakel}, {van Beuzekom}, {van den Brand}, {van den Broeck},
  {van Remortel}, {Vander-Hyde}, {van der Schaaf}, {van Heijningen}, {van
  Putten}, {Vardaro}, {Vargas}, {Varma}, {Vas{\'u}th}, {Vecchio}, {Vedovato},
  {Veitch}, {Veitch}, {Venkateswara}, {Venneberg}, {Venugopalan}, {Verkindt},
  {Verma}, {Veske}, {Vetrano}, {Vicer{\'e}}, {Viets}, {Villa-Ortega}, {Vinet},
  {Vitale}, {Vo}, {Vocca}, {von Reis}, {von Wrangel}, {Vorvick}, {Vyatchanin},
  {Wade}, {Wade}, {Wagner}, {Walet}, {Walker}, {Wallace}, {Wallace}, {Walsh},
  {Wang}, {Wang}, {Wang}, {Ward}, {Warner}, {Was}, {Washimi}, {Washington},
  {Watchi}, {Weaver}, {Wei}, {Weinert}, {Weinstein}, {Weiss}, {Weller},
  {Wellmann}, {Wen}, {We{\ss}els}, {Westhouse}, {Wette}, {Whelan}, {White},
  {Whiting}, {Whittle}, {Wilken}, {Williams}, {Williams}, {Williamson},
  {Willis}, {Willke}, {Wilson}, {Winkler}, {Wipf}, {Wlodarczyk}, {Woan},
  {Woehler}, {Wofford}, {Wong}, {Wu}, {Wu}, {Wu}, {Wu}, {Wysocki}, {Xiao},
  {Xu}, {Yamada}, {Yamamoto}, {Yamamoto}, {Yamamoto}, {Yamamoto}, {Yamashita},
  {Yamazaki}, {Yang}, {Yang}, {Yang}, {Yang}, {Yang}, {Yap}, {Yeeles},
  {Yelikar}, {Ying}, {Yokogawa}, {Yokoyama}, {Yokozawa}, {Yoon}, {Yoshioka},
  {Yu}, {Yu}, {Yuzurihara}, {Zadro{\.z}ny}, {Zanolin}, {Zeidler}, {Zelenova},
  {Zendri}, {Zevin}, {Zhan}, {Zhang}, {Zhang}, {Zhang}, {Zhang}, {Zhang},
  {Zhao}, {Zhao}, {Zhao}, {Zhao}, {Zhou}, {Zhu}, {Zhu}, {Zucker}, {Zweizig},
  {Ligo Scientific Collaboration}, {VIRGO Collaboration}, \& {Kagra
  Collaboration}}]{LIGO1}
{Abbott}, R., {Abbott}, T.~D., {Abraham}, S., {et~al.} 2021, \prd, 104, 022004,
  \dodoi{10.1103/PhysRevD.104.022004}

\bibitem[{{Addazi} {et~al.}(2020){Addazi}, {Cai}, {Gan}, {Marciano}, \&
  {Zeng}}]{PT2}
{Addazi}, A., {Cai}, Y.-F., {Gan}, Q., {Marciano}, A., \& {Zeng}, K. 2020,
  arXiv e-prints, arXiv:2009.10327.
\newblock \doarXiv{2009.10327}

\bibitem[{{Aggarwal} {et~al.}(2019){Aggarwal}, {Arzoumanian}, {Baker},
  {Brazier}, {Brinson}, {Brook}, {Burke-Spolaor}, {Chatterjee}, {Cordes},
  {Cornish}, {Crawford}, {Crowter}, {Cromartie}, {DeCesar}, {Demorest},
  {Dolch}, {Ellis}, {Ferdman}, {Ferrara}, {Fonseca}, {Garver-Daniels},
  {Gentile}, {Hazboun}, {Holgado}, {Huerta}, {Islo}, {Jennings}, {Jones},
  {Jones}, {Kaiser}, {Kaplan}, {Kelley}, {Key}, {Lam}, {Lazio}, {Levin},
  {Lorimer}, {Luo}, {Lynch}, {Madison}, {McLaughlin}, {McWilliams},
  {Mingarelli}, {Ng}, {Nice}, {Pennucci}, {Pol}, {Ransom}, {Ray}, {Siemens},
  {Simon}, {Spiewak}, {Stairs}, {Stinebring}, {Stovall}, {Swiggum}, {Taylor},
  {Turner}, {Vallisneri}, {van Haasteren}, {Vigeland}, {Witt}, {Zhu}, \&
  {NANOGrav Collaboration}}]{NG11}
{Aggarwal}, K., {Arzoumanian}, Z., {Baker}, P.~T., {et~al.} 2019, \apj, 880,
  116, \dodoi{10.3847/1538-4357/ab2236}

\bibitem[{{Alam} {et~al.}(2020){Alam}, {Arzoumanian}, {Baker}, {Blumer},
  {Bohler}, {Brazier}, {Brook}, {Burke-Spolaor}, {Caballero}, {Camuccio},
  {Chamberlain}, {Chatterjee}, {Cordes}, {Cornish}, {Crawford}, {Cromartie},
  {DeCesar}, {Demorest}, {Dolch}, {Ellis}, {Ferdman}, {Ferrara}, {Fiore},
  {Fonseca}, {Garcia}, {Garver-Daniels}, {Gentile}, {Good}, {Gusdorff},
  {Halmrast}, {Hazboun}, {Islo}, {Jennings}, {Jessup}, {Jones}, {Kaiser},
  {Kaplan}, {Kelley}, {Shapiro Key}, {Lam}, {Lazio}, {Lorimer}, {Luo}, {Lynch},
  {Madison}, {Maraccini}, {McLaughlin}, {Mingarelli}, {Ng}, {Nguyen}, {Nice},
  {Pennucci}, {Pol}, {Ramette}, {Ransom}, {Ray}, {Shapiro-Albert}, {Siemens},
  {Simon}, {Spiewak}, {Stairs}, {Stinebring}, {Stovall}, {Swiggum}, {Taylor},
  {Tripepi}, {Vallisneri}, {Vigeland}, {Witt}, \& {Zhu}}]{aab+20}
{Alam}, M.~F., {Arzoumanian}, Z., {Baker}, P.~T., {et~al.} 2020, arXiv
  e-prints, arXiv:2005.06490.
\newblock \doarXiv{2005.06490}

\bibitem[{{Allen}(1988)}]{PhysRevD.37.2078}
{Allen}, B. 1988, \prd, 37, 2078, \dodoi{10.1103/PhysRevD.37.2078}

\bibitem[{{Anholm} {et~al.}(2009){Anholm}, {Ballmer}, {Creighton}, {Price}, \&
  {Siemens}}]{abc+09}
{Anholm}, M., {Ballmer}, S., {Creighton}, J.~D.~E., {Price}, L.~R., \&
  {Siemens}, X. 2009, \prd, 79, 084030, \dodoi{10.1103/PhysRevD.79.084030}

\bibitem[{Arzoumanian {et~al.}(2016)Arzoumanian, Brazier, Burke-Spolaor,
  Chamberlin, Chatterjee, Christy, Cordes, Cornish, Crowter, Demorest, Deng,
  Dolch, Ellis, Ferdman, Fonseca, Garver-Daniels, Gonzalez, Jenet, Jones,
  Jones, Kaspi, Koop, Lam, Lazio, Levin, Lommen, Lorimer, Luo, Lynch, Madison,
  Mclaughlin, Mcwilliams, Mingarelli, Nice, Palliyaguru, Pennucci, Ransom,
  Sampson, Sanidas, Sesana, Siemens, Simon, Stairs, Stinebring, Stovall,
  Swiggum, Taylor, Vallisneri, van Haasteren, Wang, Zhu, \&
  Collaboration}]{abb+16}
Arzoumanian, Z., Brazier, A., Burke-Spolaor, S., {et~al.} 2016, \apj, 821, 13

\bibitem[{{Arzoumanian} {et~al.}(2018){Arzoumanian}, {Baker}, {Brazier},
  {Burke-Spolaor}, {Chamberlin}, {Chatterjee}, {Christy}, {Cordes}, {Cornish},
  {Crawford}, {Thankful Cromartie}, {Crowter}, {DeCesar}, {Demorest}, {Dolch},
  {Ellis}, {Ferdman}, {Ferrara}, {Folkner}, {Fonseca}, {Garver-Daniels},
  {Gentile}, {Haas}, {Hazboun}, {Huerta}, {Islo}, {Jones}, {Jones}, {Kaplan},
  {Kaspi}, {Lam}, {Lazio}, {Levin}, {Lommen}, {Lorimer}, {Luo}, {Lynch},
  {Madison}, {McLaughlin}, {McWilliams}, {Mingarelli}, {Ng}, {Nice}, {Park},
  {Pennucci}, {Pol}, {Ransom}, {Ray}, {Rasskazov}, {Siemens}, {Simon},
  {Spiewak}, {Stairs}, {Stinebring}, {Stovall}, {Swiggum}, {Taylor},
  {Vallisneri}, {van Haasteren}, {Vigeland}, {Zhu}, \& {NANOGrav
  Collaboration}}]{abb+18b}
{Arzoumanian}, Z., {Baker}, P.~T., {Brazier}, A., {et~al.} 2018, \apj, 859, 47,
  \dodoi{10.3847/1538-4357/aabd3b}

\bibitem[{{Arzoumanian} {et~al.}(2020){Arzoumanian}, {Baker}, {Blumer},
  {B{\'e}csy}, {Brazier}, {Brook}, {Burke-Spolaor}, {Chatterjee}, {Chen},
  {Cordes}, {Cornish}, {Crawford}, {Cromartie}, {Decesar}, {Demorest}, {Dolch},
  {Ellis}, {Ferrara}, {Fiore}, {Fonseca}, {Garver-Daniels}, {Gentile}, {Good},
  {Hazboun}, {Holgado}, {Islo}, {Jennings}, {Jones}, {Kaiser}, {Kaplan},
  {Kelley}, {Key}, {Laal}, {Lam}, {Lazio}, {Lorimer}, {Luo}, {Lynch},
  {Madison}, {McLaughlin}, {Mingarelli}, {Ng}, {Nice}, {Pennucci}, {Pol},
  {Ransom}, {Ray}, {Shapiro-Albert}, {Siemens}, {Simon}, {Spiewak}, {Stairs},
  {Stinebring}, {Stovall}, {Sun}, {Swiggum}, {Taylor}, {Turner}, {Vallisneri},
  {Vigeland}, {Witt}, \& {Nanograv Collaboration}}]{NG}
{Arzoumanian}, Z., {Baker}, P.~T., {Blumer}, H., {et~al.} 2020, \apjl, 905,
  L34, \dodoi{10.3847/2041-8213/abd401}

\bibitem[{{Arzoumanian} {et~al.}(2021){Arzoumanian}, {Baker}, {Blumer},
  {B{\'e}csy}, {Brazier}, {Brook}, {Burke-Spolaor}, {Charisi}, {Chatterjee},
  {Chen}, {Cordes}, {Cornish}, {Crawford}, {Cromartie}, {DeCesar}, {Demorest},
  {Dolch}, {Ellis}, {Ferrara}, {Fiore}, {Fonseca}, {Garver-Daniels}, {Gentile},
  {Good}, {Hazboun}, {Holgado}, {Islo}, {Jennings}, {Jones}, {Kaiser},
  {Kaplan}, {Kelley}, {Shapiro Key}, {Laal}, {Lam}, {Lazio}, {Lee}, {Lorimer},
  {Luo}, {Lynch}, {Madison}, {McLaughlin}, {Mingarelli}, {Mitridate}, {Ng},
  {Nice}, {Pennucci}, {Pol}, {Ransom}, {Ray}, {Shapiro-Albert}, {Siemens},
  {Simon}, {Spiewak}, {Stairs}, {Stinebring}, {Stovall}, {Sun}, {Swiggum},
  {Taylor}, {Turner}, {Vallisneri}, {Vigeland}, {Witt}, \& {Zurek}}]{PT1}
---. 2021, arXiv e-prints, arXiv:2104.13930.
\newblock \doarXiv{2104.13930}

\bibitem[{{Ashoorioon} {et~al.}(2021){Ashoorioon}, {Rostami}, \&
  {Firouzjaee}}]{INF1}
{Ashoorioon}, A., {Rostami}, A., \& {Firouzjaee}, J.~T. 2021, \prd, 103,
  123512, \dodoi{10.1103/PhysRevD.103.123512}

\bibitem[{{Bailes} {et~al.}(2018){Bailes}, {Barr}, {Bhat}, {Brink}, {Buchner},
  {Burgay}, {Camilo}, {Champion}, {Hessels}, {Janssen}, {Jameson}, {Johnston},
  {Karastergiou}, {Karuppusamy}, {Kaspi}, {Keith}, {Kramer}, {McLaughlin},
  {Moodley}, {Oslowski}, {Possenti}, {Ransom}, {Rasio}, {Sievers}, {Serylak},
  {Stappers}, {Stairs}, {Theureau}, {van Straten}, {Weltevrede}, \&
  {Wex}}]{MeerTime}
{Bailes}, M., {Barr}, E., {Bhat}, N.~D.~R., {et~al.} 2018, arXiv e-prints,
  arXiv:1803.07424.
\newblock \doarXiv{1803.07424}

\bibitem[{B\'ecsy \& Cornish(2021)}]{Becsy:2020utk}
B\'ecsy, B., \& Cornish, N.~J. 2021, Class. Quant. Grav., 38, 095012,
  \dodoi{10.1088/1361-6382/abf1c6}

\bibitem[{{Berezinsky} {et~al.}(2004){Berezinsky}, {Hnatyk}, \&
  {Vilenkin}}]{Berezinsky:2000vn}
{Berezinsky}, V., {Hnatyk}, B., \& {Vilenkin}, A. 2004, Baltic Astronomy, 13,
  289

\bibitem[{{Blanco-Pillado} {et~al.}(2018){Blanco-Pillado}, {Olum}, \&
  {Siemens}}]{2018PhLB..778..392B}
{Blanco-Pillado}, J.~J., {Olum}, K.~D., \& {Siemens}, X. 2018, Physics Letters
  B, 778, 392, \dodoi{10.1016/j.physletb.2018.01.050}

\bibitem[{{Blanco-Pillado} {et~al.}(2021){Blanco-Pillado}, {Olum}, \&
  {Wachter}}]{CS4}
{Blanco-Pillado}, J.~J., {Olum}, K.~D., \& {Wachter}, J.~M. 2021, \prd, 103,
  103512, \dodoi{10.1103/PhysRevD.103.103512}

\bibitem[{{Bonetti} {et~al.}(2018){Bonetti}, {Sesana}, {Barausse}, \&
  {Haardt}}]{2018MNRAS.477.2599B}
{Bonetti}, M., {Sesana}, A., {Barausse}, E., \& {Haardt}, F. 2018, \mnras, 477,
  2599, \dodoi{10.1093/mnras/sty874}

\bibitem[{{Borah} {et~al.}(2021){Borah}, {Dasgupta}, \& {Kang}}]{PT4}
{Borah}, D., {Dasgupta}, A., \& {Kang}, S.~K. 2021, arXiv e-prints,
  arXiv:2105.01007.
\newblock \doarXiv{2105.01007}

\bibitem[{{Brandenburg} {et~al.}(2021){Brandenburg}, {Clarke}, {He}, \&
  {Kahniashvili}}]{PT6}
{Brandenburg}, A., {Clarke}, E., {He}, Y., \& {Kahniashvili}, T. 2021, arXiv
  e-prints, arXiv:2102.12428.
\newblock \doarXiv{2102.12428}

\bibitem[{{Caprini} {et~al.}(2010){Caprini}, {Durrer}, \& {Siemens}}]{ccd+10}
{Caprini}, C., {Durrer}, R., \& {Siemens}, X. 2010, \prd, 82, 063511,
  \dodoi{10.1103/PhysRevD.82.063511}

\bibitem[{Carlin \& Chib(1995)}]{cc95}
Carlin, B.~P., \& Chib, S. 1995, Journal of the Royal Statistical Society.
  Series B (Methodological), 57, 473.
\newblock \url{http://www.jstor.org/stable/2346151}

\bibitem[{{Chakrabortty} {et~al.}(2021){Chakrabortty}, {Lazarides}, {Maji}, \&
  {Shafi}}]{CS8}
{Chakrabortty}, J., {Lazarides}, G., {Maji}, R., \& {Shafi}, Q. 2021, Journal
  of High Energy Physics, 2021, 114, \dodoi{10.1007/JHEP02(2021)114}

\bibitem[{{Chamberlin} \& {Siemens}(2012{\natexlab{a}})}]{ss12}
{Chamberlin}, S.~J., \& {Siemens}, X. 2012{\natexlab{a}}, \prd, 85, 082001,
  \dodoi{10.1103/PhysRevD.85.082001}

\bibitem[{{Chamberlin} \& {Siemens}(2012{\natexlab{b}})}]{SX}
---. 2012{\natexlab{b}}, \prd, 85, 082001, \dodoi{10.1103/PhysRevD.85.082001}

\bibitem[{{Chang} \& {Cui}(2021)}]{CS1}
{Chang}, C.-F., \& {Cui}, Y. 2021, arXiv e-prints, arXiv:2106.09746.
\newblock \doarXiv{2106.09746}

\bibitem[{Chen {et~al.}(2021{\natexlab{a}})Chen, Wu, \& Huang}]{Chen:2021ncc}
Chen, Z.-C., Wu, Y.-M., \& Huang, Q.-G. 2021{\natexlab{a}}.
\newblock \doarXiv{2109.00296}

\bibitem[{Chen {et~al.}(2021{\natexlab{b}})Chen, Yuan, \& Huang}]{Chen:2021wdo}
Chen, Z.-C., Yuan, C., \& Huang, Q.-G. 2021{\natexlab{b}}.
\newblock \doarXiv{2101.06869}

\bibitem[{{Chen} {et~al.}(2021){Chen}, {Yuan}, \& {Huang}}]{Chen}
{Chen}, Z.-C., {Yuan}, C., \& {Huang}, Q.-G. 2021, arXiv e-prints,
  arXiv:2101.06869.
\newblock \doarXiv{2101.06869}

\bibitem[{{Chiang} \& {Lu}(2021)}]{CS6}
{Chiang}, C.-W., \& {Lu}, B.-Q. 2021, \jcap, 2021, 049,
  \dodoi{10.1088/1475-7516/2021/05/049}

\bibitem[{{Cordes} \& {Jenet}(2012)}]{Cordes:2012zz}
{Cordes}, J.~M., \& {Jenet}, F.~A. 2012, \apj, 752, 54,
  \dodoi{10.1088/0004-637X/752/1/54}

\bibitem[{{Cornish} {et~al.}(2018{\natexlab{a}}){Cornish}, {O'Beirne},
  {Taylor}, \& {Yunes}}]{2018PhRvL.120r1101C}
{Cornish}, N.~J., {O'Beirne}, L., {Taylor}, S.~R., \& {Yunes}, N.
  2018{\natexlab{a}}, Physical Review Letters, 120, 181101,
  \dodoi{10.1103/PhysRevLett.120.181101}

\bibitem[{{Cornish} {et~al.}(2018{\natexlab{b}}){Cornish}, {O'Beirne},
  {Taylor}, \& {Yunes}}]{NanoFirstAlt}
---. 2018{\natexlab{b}}, \prl, 120, 181101,
  \dodoi{10.1103/PhysRevLett.120.181101}

\bibitem[{{Damour} \& {Vilenkin}(2000)}]{Damour:2000wa}
{Damour}, T., \& {Vilenkin}, A. 2000, Physical Review Letters, 85, 3761,
  \dodoi{10.1103/PhysRevLett.85.3761}

\bibitem[{{Damour} \& {Vilenkin}(2001)}]{Damour:2001bk}
---. 2001, \prd, 64, 064008, \dodoi{10.1103/PhysRevD.64.064008}

\bibitem[{{Damour} \& {Vilenkin}(2005)}]{Damour:2004kw}
---. 2005, \prd, 71, 063510, \dodoi{10.1103/PhysRevD.71.063510}

\bibitem[{{Demorest} {et~al.}(2013){Demorest}, {Ferdman}, {Gonzalez}, {Nice},
  {Ransom}, {Stairs}, {Arzoumanian}, {Brazier}, {Burke-Spolaor}, {Chamberlin},
  {Cordes}, {Ellis}, {Finn}, {Freire}, {Giampanis}, {Jenet}, {Kaspi}, {Lazio},
  {Lommen}, {McLaughlin}, {Palliyaguru}, {Perrodin}, {Shannon}, {Siemens},
  {Stinebring}, {Swiggum}, \& {Zhu}}]{dfg+13}
{Demorest}, P.~B., {Ferdman}, R.~D., {Gonzalez}, M.~E., {et~al.} 2013, \apj,
  762, 94, \dodoi{10.1088/0004-637X/762/2/94}

\bibitem[{{Desvignes} {et~al.}(2016){Desvignes}, {Caballero}, {Lentati},
  {Verbiest}, {Champion}, {Stappers}, {Janssen}, {Lazarus}, {Os{\l}owski},
  {Babak}, {Bassa}, {Brem}, {Burgay}, {Cognard}, {Gair}, {Graikou},
  {Guillemot}, {Hessels}, {Jessner}, {Jordan}, {Karuppusamy}, {Kramer},
  {Lassus}, {Lazaridis}, {Lee}, {Liu}, {Lyne}, {McKee}, {Mingarelli},
  {Perrodin}, {Petiteau}, {Possenti}, {Purver}, {Rosado}, {Sanidas}, {Sesana},
  {Shaifullah}, {Smits}, {Taylor}, {Theureau}, {Tiburzi}, {van Haasteren}, \&
  {Vecchio}}]{dcl+16}
{Desvignes}, G., {Caballero}, R.~N., {Lentati}, L., {et~al.} 2016, \mnras, 458,
  3341, \dodoi{10.1093/mnras/stw483}

\bibitem[{{Detweiler}(1979)}]{Detw}
{Detweiler}, S. 1979, \apj, 234, 1100, \dodoi{10.1086/157593}

\bibitem[{{Di Bari} {et~al.}(2021){Di Bari}, {Marfatia}, \& {Zhou}}]{PT3}
{Di Bari}, P., {Marfatia}, D., \& {Zhou}, Y.-L. 2021, arXiv e-prints,
  arXiv:2106.00025.
\newblock \doarXiv{2106.00025}

\bibitem[{Dickey(1971)}]{d71}
Dickey, J.~M. 1971, The Annals of Mathematical Statistics, 42, 204.
\newblock \url{http://www.jstor.org/stable/2958475}

\bibitem[{{Dvorkin} \& {Barausse}(2017)}]{2017MNRAS.470.4547D}
{Dvorkin}, I., \& {Barausse}, E. 2017, \mnras, 470, 4547,
  \dodoi{10.1093/mnras/stx1454}

\bibitem[{{Eardley} {et~al.}(1973{\natexlab{a}}){Eardley}, {Lee}, \&
  {Lightman}}]{Eardley:1974nw}
{Eardley}, D.~M., {Lee}, D.~L., \& {Lightman}, A.~P. 1973{\natexlab{a}}, \prd,
  8, 3308, \dodoi{10.1103/PhysRevD.8.3308}

\bibitem[{{Eardley} {et~al.}(1973{\natexlab{b}}){Eardley}, {Lee}, \&
  {Lightman}}]{Eard}
---. 1973{\natexlab{b}}, \prd, 8, 3308, \dodoi{10.1103/PhysRevD.8.3308}

\bibitem[{{Eardley} {et~al.}(1973{\natexlab{c}}){Eardley}, {Lee}, {Lightman},
  {Wagoner}, \& {Will}}]{Eardley:1973br}
{Eardley}, D.~M., {Lee}, D.~L., {Lightman}, A.~P., {Wagoner}, R.~V., \& {Will},
  C.~M. 1973{\natexlab{c}}, Physical Review Letters, 30, 884,
  \dodoi{10.1103/PhysRevLett.30.884}

\bibitem[{{Ellis} \& {Lewicki}(2021)}]{CS9}
{Ellis}, J., \& {Lewicki}, M. 2021, \prl, 126, 041304,
  \dodoi{10.1103/PhysRevLett.126.041304}

\bibitem[{Ellis \& van Haasteren(2017)}]{ptmcmc}
Ellis, J., \& van Haasteren, R. 2017, jellis18/PTMCMCSampler: Official Release,
  \dodoi{10.5281/zenodo.1037579}

\bibitem[{{Ellis} {et~al.}(2020){Ellis}, {Vallisneri}, {Taylor}, \&
  {Baker}}]{ENTP}
{Ellis}, J.~A., {Vallisneri}, M., {Taylor}, S.~R., \& {Baker}, P.~T. 2020,
  {ENTERPRISE: Enhanced Numerical Toolbox Enabling a Robust PulsaR Inference
  SuitE}, v3.0.0,  Zenodo, \dodoi{10.5281/zenodo.4059815}

\bibitem[{Ellis {et~al.}(2020)Ellis, Vallisneri, Taylor, \& Baker}]{enterprise}
Ellis, J.~A., Vallisneri, M., Taylor, S.~R., \& Baker, P.~T. 2020, ENTERPRISE:
  Enhanced Numerical Toolbox Enabling a Robust PulsaR Inference SuitE, Zenodo,
  \dodoi{10.5281/zenodo.4059815}

\bibitem[{{Enoki} {et~al.}(2004){Enoki}, {Inoue}, {Nagashima}, \&
  {Sugiyama}}]{ein+04}
{Enoki}, M., {Inoue}, K.~T., {Nagashima}, M., \& {Sugiyama}, N. 2004, \apj,
  615, 19, \dodoi{10.1086/424475}

\bibitem[{{Gair} {et~al.}(2015){Gair}, {Romano}, \&
  {Taylor}}]{2015PhRvD..92j2003G}
{Gair}, J.~R., {Romano}, J.~D., \& {Taylor}, S.~R. 2015, \prd, 92, 102003,
  \dodoi{10.1103/PhysRevD.92.102003}

\bibitem[{{Ghayour} {et~al.}(2021){Ghayour}, {Khodagholizadeh}, {Afkani},
  {Torkamani}, \& {Vahedi}}]{CS10}
{Ghayour}, B., {Khodagholizadeh}, J., {Afkani}, M., {Torkamani}, M.~R., \&
  {Vahedi}, A. 2021, International Journal of Modern Physics D, 30, 2150023,
  \dodoi{10.1142/S0218271821500231}

\bibitem[{Godsill(2001)}]{g01}
Godsill, S.~J. 2001, Journal of Computational and Graphical Statistics, 10,
  230.
\newblock \url{http://www.jstor.org/stable/1391010}

\bibitem[{Goncharov {et~al.}(2021)}]{Goncharov:2021oub}
Goncharov, B., {et~al.} 2021, \dodoi{10.3847/2041-8213/ac17f4}

\bibitem[{{Gorghetto} {et~al.}(2021){Gorghetto}, {Hardy}, \&
  {Nicolaescu}}]{CS2}
{Gorghetto}, M., {Hardy}, E., \& {Nicolaescu}, H. 2021, \jcap, 2021, 034,
  \dodoi{10.1088/1475-7516/2021/06/034}

\bibitem[{{Hazboun} {et~al.}(2020{\natexlab{a}}){Hazboun}, {Simon}, {Siemens},
  \& {Romano}}]{ModelDep}
{Hazboun}, J.~S., {Simon}, J., {Siemens}, X., \& {Romano}, J.~D.
  2020{\natexlab{a}}, \apjl, 905, L6, \dodoi{10.3847/2041-8213/abca92}

\bibitem[{{Hazboun} {et~al.}(2020{\natexlab{b}}){Hazboun}, {Simon}, {Taylor},
  {Lam}, {Vigeland}, {Islo}, {Key}, {Arzoumanian}, {Baker}, {Brazier}, {Brook},
  {Burke-Spolaor}, {Chatterjee}, {Cordes}, {Cornish}, {Crawford}, {Crowter},
  {Cromartie}, {DeCesar}, {Demorest}, {Dolch}, {Ellis}, {Ferdman}, {Ferrara},
  {Fonseca}, {Garver-Daniels}, {Gentile}, {Good}, {Holgado}, {Huerta},
  {Jennings}, {Jones}, {Jones}, {Kaiser}, {Kaplan}, {Kelley}, {Lazio}, {Levin},
  {Lommen}, {Lorimer}, {Luo}, {Lynch}, {Madison}, {McLaughlin}, {McWilliams},
  {Mingarelli}, {Ng}, {Nice}, {Pennucci}, {Pol}, {Ransom}, {Ray}, {Siemens},
  {Spiewak}, {Stairs}, {Stinebring}, {Stovall}, {Swiggum}, {Turner},
  {Vallisneri}, {van Haasteren}, {Witt}, \& {Zhu}}]{11YEAR_SLICES}
{Hazboun}, J.~S., {Simon}, J., {Taylor}, S.~R., {et~al.} 2020{\natexlab{b}},
  \apj, 890, 108, \dodoi{10.3847/1538-4357/ab68db}

\bibitem[{Hee {et~al.}(2015)Hee, Handley, Hobson, \& Lasenby}]{hee15}
Hee, S., Handley, W.~J., Hobson, M.~P., \& Lasenby, A.~N. 2015, Monthly Notices
  of the Royal Astronomical Society, 455, 2461, \dodoi{10.1093/mnras/stv2217}

\bibitem[{{Hellings} \& {Downs}(1983{\natexlab{a}})}]{hd83}
{Hellings}, R.~W., \& {Downs}, G.~S. 1983{\natexlab{a}}, \apjl, 265, L39,
  \dodoi{10.1086/183954}

\bibitem[{{Hellings} \& {Downs}(1983{\natexlab{b}})}]{HD}
---. 1983{\natexlab{b}}, \apjl, 265, L39, \dodoi{10.1086/183954}

\bibitem[{{Hobbs} \& {Edwards}(2012)}]{tempo2}
{Hobbs}, G., \& {Edwards}, R. 2012, {Tempo2: Pulsar Timing Package}.
\newblock \doeprint{1210.015}

\bibitem[{{Hunter}(2007)}]{matplotlib}
{Hunter}, J.~D. 2007, Computing in Science and Engineering, 9, 90,
  \dodoi{10.1109/MCSE.2007.55}

\bibitem[{Inc.(2015)}]{plotly}
Inc., P.~T. 2015, Collaborative data science,  Montreal, QC: Plotly
  Technologies Inc.
\newblock \url{https://plot.ly}

\bibitem[{{Islo} {et~al.}(2019){Islo}, {Simon}, {Burke-Spolaor}, \&
  {Siemens}}]{Islo:2019qht}
{Islo}, K., {Simon}, J., {Burke-Spolaor}, S., \& {Siemens}, X. 2019, arXiv
  e-prints, arXiv:1906.11936.
\newblock \doarXiv{1906.11936}

\bibitem[{{Jaffe} \& {Backer}(2003)}]{jb03}
{Jaffe}, A.~H., \& {Backer}, D.~C. 2003, \apj, 583, 616, \dodoi{10.1086/345443}

\bibitem[{{Joshi} {et~al.}(2018){Joshi}, {Arumugasamy}, {Bagchi},
  {Bandyopadhyay}, {Basu}, {Dhand a Batra}, {Bethapudi}, {Choudhary}, {De},
  {Dey}, {Gopakumar}, {Gupta}, {Krishnakumar}, {Maan}, {Manoharan}, {Naidu},
  {Nandi}, {Pathak}, {Surnis}, \& {Susobhanan}}]{InPTA}
{Joshi}, B.~C., {Arumugasamy}, P., {Bagchi}, M., {et~al.} 2018, Journal of
  Astrophysics and Astronomy, 39, 51, \dodoi{10.1007/s12036-018-9549-y}

\bibitem[{Kelley {et~al.}(2016)Kelley, Blecha, \& Hernquist}]{Kelley_2016}
Kelley, L.~Z., Blecha, L., \& Hernquist, L. 2016, Monthly Notices of the Royal
  Astronomical Society, 464, 3131–3157, \dodoi{10.1093/mnras/stw2452}

\bibitem[{{Kelley} {et~al.}(2017){Kelley}, {Blecha}, {Hernquist}, {Sesana}, \&
  {Taylor}}]{2017MNRAS.471.4508K}
{Kelley}, L.~Z., {Blecha}, L., {Hernquist}, L., {Sesana}, A., \& {Taylor},
  S.~R. 2017, \mnras, 471, 4508, \dodoi{10.1093/mnras/stx1638}

\bibitem[{{Kelley} {et~al.}(2018){Kelley}, {Blecha}, {Hernquist}, {Sesana}, \&
  {Taylor}}]{2018MNRAS.477..964K}
---. 2018, \mnras, 477, 964, \dodoi{10.1093/mnras/sty689}

\bibitem[{{Kerr} {et~al.}(2020){Kerr}, {Reardon}, {Hobbs}, {Shannon},
  {Manchester}, {Dai}, {Russell}, {Zhang}, {van Straten}, {Os{\l}owski},
  {Parthasarathy}, {Spiewak}, {Bailes}, {Bhat}, {Cameron}, {Coles}, {Dempsey},
  {Deng}, {Goncharov}, {Kaczmarek}, {Keith}, {Lasky}, {Lower}, {Preisig},
  {Sarkissian}, {Toomey}, {Wang}, {Wang}, {Zhang}, \& {Zhu}}]{krh+20}
{Kerr}, M., {Reardon}, D.~J., {Hobbs}, G., {et~al.} 2020, arXiv e-prints,
  arXiv:2003.09780.
\newblock \doarXiv{2003.09780}

\bibitem[{{Lazarides} {et~al.}(2021{\natexlab{a}}){Lazarides}, {Maji}, \&
  {Shafi}}]{CS7}
{Lazarides}, G., {Maji}, R., \& {Shafi}, Q. 2021{\natexlab{a}}, arXiv e-prints,
  arXiv:2104.02016.
\newblock \doarXiv{2104.02016}

\bibitem[{{Lazarides} {et~al.}(2021{\natexlab{b}}){Lazarides}, {Maji}, \&
  {Shafi}}]{INF5}
---. 2021{\natexlab{b}}, arXiv e-prints, arXiv:2104.02016.
\newblock \doarXiv{2104.02016}

\bibitem[{{Lee}(2016)}]{CPTA}
{Lee}, K.~J. 2016, in Astronomical Society of the Pacific Conference Series,
  Vol. 502, Frontiers in Radio Astronomy and FAST Early Sciences Symposium
  2015, ed. L.~{Qain} \& D.~{Li}, 19

\bibitem[{{Lee} {et~al.}(2008{\natexlab{a}}){Lee}, {Jenet}, \& {Price}}]{ljp08}
{Lee}, K.~J., {Jenet}, F.~A., \& {Price}, R.~H. 2008{\natexlab{a}}, \apj, 685,
  1304, \dodoi{10.1086/591080}

\bibitem[{{Lee} {et~al.}(2008{\natexlab{b}}){Lee}, {Jenet}, \& {Price}}]{VLORF}
---. 2008{\natexlab{b}}, \apj, 685, 1304, \dodoi{10.1086/591080}

\bibitem[{{Lentati} {et~al.}(2015){Lentati}, {Taylor}, {Mingarelli}, {Sesana},
  {Sanidas}, {Vecchio}, {Caballero}, {Lee}, {van Haasteren}, {Babak}, {Bassa},
  {Brem}, {Burgay}, {Champion}, {Cognard}, {Desvignes}, {Gair}, {Guillemot},
  {Hessels}, {Janssen}, {Karuppusamy}, {Kramer}, {Lassus}, {Lazarus}, {Liu},
  {Os{\l}owski}, {Perrodin}, {Petiteau}, {Possenti}, {Purver}, {Rosado},
  {Smits}, {Stappers}, {Theureau}, {Tiburzi}, \& {Verbiest}}]{ltm+15}
{Lentati}, L., {Taylor}, S.~R., {Mingarelli}, C.~M.~F., {et~al.} 2015, \mnras,
  453, 2576, \dodoi{10.1093/mnras/stv1538}

\bibitem[{{Li} {et~al.}(2021){Li}, {Ye}, \& {Piao}}]{INF3}
{Li}, H.-H., {Ye}, G., \& {Piao}, Y.-S. 2021, Physics Letters B, 816, 136211,
  \dodoi{10.1016/j.physletb.2021.136211}

\bibitem[{{Lin}(2021)}]{CS5}
{Lin}, C.-M. 2021, \jcap, 2021, 056, \dodoi{10.1088/1475-7516/2021/05/056}

\bibitem[{{Lommen} \& {Backer}(2001)}]{lb01}
{Lommen}, A.~N., \& {Backer}, D.~C. 2001, \apj, 562, 297,
  \dodoi{10.1086/323491}

\bibitem[{{Madison} {et~al.}(2017){Madison}, {Chernoff}, \&
  {Cordes}}]{2017PhRvD..96l3016M}
{Madison}, D.~R., {Chernoff}, D.~F., \& {Cordes}, J.~M. 2017, \prd, 96, 123016,
  \dodoi{10.1103/PhysRevD.96.123016}

\bibitem[{{McWilliams} {et~al.}(2012){McWilliams}, {Ostriker}, \&
  {Pretorius}}]{McWilliams:2012jj}
{McWilliams}, S.~T., {Ostriker}, J.~P., \& {Pretorius}, F. 2012, arXiv
  e-prints, arXiv:1211.4590.
\newblock \doarXiv{1211.4590}

\bibitem[{{Mingarelli} {et~al.}(2012){Mingarelli}, {Grover}, {Sidery}, {Smith},
  \& {Vecchio}}]{Mingarelli:2012hh}
{Mingarelli}, C.~M.~F., {Grover}, K., {Sidery}, T., {Smith}, R.~J.~E., \&
  {Vecchio}, A. 2012, Physical Review Letters, 109, 081104,
  \dodoi{10.1103/PhysRevLett.109.081104}

\bibitem[{{Mingarelli} {et~al.}(2017){Mingarelli}, {Lazio}, {Sesana}, {Greene},
  {Ellis}, {Ma}, {Croft}, {Burke-Spolaor}, \& {Taylor}}]{2017NatAs...1..886M}
{Mingarelli}, C.~M.~F., {Lazio}, T.~J.~W., {Sesana}, A., {et~al.} 2017, Nature
  Astronomy, 1, 886, \dodoi{10.1038/s41550-017-0299-6}

\bibitem[{{Nakai} {et~al.}(2021){Nakai}, {Suzuki}, {Takahashi}, \&
  {Yamada}}]{PT5}
{Nakai}, Y., {Suzuki}, M., {Takahashi}, F., \& {Yamada}, M. 2021, Physics
  Letters B, 816, 136238, \dodoi{10.1016/j.physletb.2021.136238}

\bibitem[{{Neronov} {et~al.}(2021){Neronov}, {Pol}, {Caprini}, \&
  {Semikoz}}]{PT7}
{Neronov}, A., {Pol}, A.~R., {Caprini}, C., \& {Semikoz}, D. 2021, \prd, 103,
  L041302, \dodoi{10.1103/PhysRevD.103.L041302}

\bibitem[{{Newman} \& {Penrose}(1962)}]{NP-not}
{Newman}, E., \& {Penrose}, R. 1962, Journal of Mathematical Physics, 3, 566,
  \dodoi{10.1063/1.1724257}

\bibitem[{{Ng}(2018)}]{CHIMEPulsar}
{Ng}, C. 2018, in IAU Symposium, Vol. 337, Pulsar Astrophysics the Next Fifty
  Years, ed. P.~{Weltevrede}, B.~B.~P. {Perera}, L.~L. {Preston}, \&
  S.~{Sanidas}, 179--182, \dodoi{10.1017/S1743921317010638}

\bibitem[{{O'Beirne} {et~al.}(2019){O'Beirne}, {Cornish}, {Vigeland}, \&
  {Taylor}}]{2019PhRvD..99l4039O}
{O'Beirne}, L., {Cornish}, N.~J., {Vigeland}, S.~J., \& {Taylor}, S.~R. 2019,
  \prd, 99, 124039, \dodoi{10.1103/PhysRevD.99.124039}

\bibitem[{{{\"O}lmez} {et~al.}(2010){{\"O}lmez}, {Mandic}, \&
  {Siemens}}]{Olmez:2010bi}
{{\"O}lmez}, S., {Mandic}, V., \& {Siemens}, X. 2010, \prd, 81, 104028,
  \dodoi{10.1103/PhysRevD.81.104028}

\bibitem[{{Perera} {et~al.}(2019){Perera}, {DeCesar}, {Demorest}, {Kerr},
  {Lentati}, {Nice}, {Os{\l}owski}, {Ransom}, {Keith}, {Arzoumanian}, {Bailes},
  {Baker}, {Bassa}, {Bhat}, {Brazier}, {Burgay}, {Burke-Spolaor}, {Caballero},
  {Champion}, {Chatterjee}, {Chen}, {Cognard}, {Cordes}, {Crowter}, {Dai},
  {Desvignes}, {Dolch}, {Ferdman}, {Ferrara}, {Fonseca}, {Goldstein},
  {Graikou}, {Guillemot}, {Hazboun}, {Hobbs}, {Hu}, {Islo}, {Janssen},
  {Karuppusamy}, {Kramer}, {Lam}, {Lee}, {Liu}, {Luo}, {Lyne}, {Manchester},
  {McKee}, {McLaughlin}, {Mingarelli}, {Parthasarathy}, {Pennucci}, {Perrodin},
  {Possenti}, {Reardon}, {Russell}, {Sanidas}, {Sesana}, {Shaifullah},
  {Shannon}, {Siemens}, {Simon}, {Spiewak}, {Stairs}, {Stappers}, {Swiggum},
  {Taylor}, {Theureau}, {Tiburzi}, {Vallisneri}, {Vecchio}, {Wang}, {Zhang},
  {Zhang}, {Zhu}, \& {Zhu}}]{pdd+19}
{Perera}, B.~B.~P., {DeCesar}, M.~E., {Demorest}, P.~B., {et~al.} 2019, \mnras,
  490, 4666, \dodoi{10.1093/mnras/stz2857}

\bibitem[{{Pol} {et~al.}(2021){Pol}, {Taylor}, {Kelley}, {Vigeland}, {Simon},
  {Chen}, {Arzoumanian}, {Baker}, {B{\'e}csy}, {Brazier}, {Brook},
  {Burke-Spolaor}, {Chatterjee}, {Cordes}, {Cornish}, {Crawford}, {Thankful
  Cromartie}, {Decesar}, {Demorest}, {Dolch}, {Ferrara}, {Fiore}, {Fonseca},
  {Garver-Daniels}, {Good}, {Hazboun}, {Jennings}, {Jones}, {Kaiser}, {Kaplan},
  {Shapiro Key}, {Lam}, {Lazio}, {Luo}, {Lynch}, {Madison}, {McEwen},
  {McLaughlin}, {Mingarelli}, {Ng}, {Nice}, {Pennucci}, {Ransom}, {Ray},
  {Shapiro-Albert}, {Siemens}, {Stairs}, {Stinebring}, {Swiggum}, {Vallisneri},
  {Wahl}, {Witt}, \& {Nanograv Collaboration}}]{Astroforcast}
{Pol}, N.~S., {Taylor}, S.~R., {Kelley}, L.~Z., {et~al.} 2021, \apjl, 911, L34,
  \dodoi{10.3847/2041-8213/abf2c9}

\bibitem[{{Poletti}(2021)}]{INF4}
{Poletti}, D. 2021, \jcap, 2021, 052, \dodoi{10.1088/1475-7516/2021/05/052}

\bibitem[{{Qin} {et~al.}(2021){Qin}, {Boddy}, \& {Kamionkowski}}]{MG}
{Qin}, W., {Boddy}, K.~K., \& {Kamionkowski}, M. 2021, \prd, 103, 024045,
  \dodoi{10.1103/PhysRevD.103.024045}

\bibitem[{{Ransom} {et~al.}(2019){Ransom}, {Brazier}, {Chatterjee}, {Cohen},
  {Cordes}, {DeCesar}, {Demorest}, {Hazboun}, {Lam}, {Lynch}, {McLaughlin},
  {Ransom}, {Siemens}, {Taylor}, \& {Vigeland}}]{Brazier:2019mmu}
{Ransom}, S., {Brazier}, A., {Chatterjee}, S., {et~al.} 2019, in Bulletin of
  the American Astronomical Society, Vol.~51, 195.
\newblock \doarXiv{1908.05356}

\bibitem[{{Ravi} {et~al.}(2012){Ravi}, {Wyithe}, {Hobbs}, {Shannon},
  {Manchester}, {Yardley}, \& {Keith}}]{rwh+12}
{Ravi}, V., {Wyithe}, J.~S.~B., {Hobbs}, G., {et~al.} 2012, \apj, 761, 84,
  \dodoi{10.1088/0004-637X/761/2/84}

\bibitem[{Ravi {et~al.}(2015)Ravi, Wyithe, Shannon, \& Hobbs}]{Ravi_2015}
Ravi, V., Wyithe, J. S.~B., Shannon, R.~M., \& Hobbs, G. 2015, Monthly Notices
  of the Royal Astronomical Society, 447, 2772–2783,
  \dodoi{10.1093/mnras/stu2659}

\bibitem[{{Roedig} \& {Sesana}(2012)}]{rs11}
{Roedig}, C., \& {Sesana}, A. 2012, Journal of Physics Conference Series, 363,
  012035, \dodoi{10.1088/1742-6596/363/1/012035}

\bibitem[{{Romano} \& {Cornish}(2017)}]{Laplace}
{Romano}, J.~D., \& {Cornish}, N.~J. 2017, Living Reviews in Relativity, 20, 2,
  \dodoi{10.1007/s41114-017-0004-1}

\bibitem[{Rosado {et~al.}(2015)Rosado, Sesana, \& Gair}]{Rosado_2015}
Rosado, P.~A., Sesana, A., \& Gair, J. 2015, Monthly Notices of the Royal
  Astronomical Society, 451, 2417–2433, \dodoi{10.1093/mnras/stv1098}

\bibitem[{Ryu {et~al.}(2018)Ryu, Perna, Haiman, Ostriker, \&
  Stone}]{Ryu:2018yhv}
Ryu, T., Perna, R., Haiman, Z., Ostriker, J.~P., \& Stone, N.~C. 2018, Mon.
  Not. Roy. Astron. Soc., 473, 3410, \dodoi{10.1093/mnras/stx2524}

\bibitem[{{Sampson} {et~al.}(2015){Sampson}, {Cornish}, \&
  {McWilliams}}]{kappa_param}
{Sampson}, L., {Cornish}, N.~J., \& {McWilliams}, S.~T. 2015, \prd, 91, 084055,
  \dodoi{10.1103/PhysRevD.91.084055}

\bibitem[{{Sanidas} {et~al.}(2013){Sanidas}, {Battye}, \&
  {Stappers}}]{Sanidas:2012tf}
{Sanidas}, S.~A., {Battye}, R.~A., \& {Stappers}, B.~W. 2013, \apj, 764, 108,
  \dodoi{10.1088/0004-637X/764/1/108}

\bibitem[{{Sazhin}(1978)}]{Shazin}
{Sazhin}, M.~V. 1978, \sovast, 22, 36

\bibitem[{{Schutz} \& {Ma}(2016)}]{2016MNRAS.459.1737S}
{Schutz}, K., \& {Ma}, C.-P. 2016, \mnras, 459, 1737,
  \dodoi{10.1093/mnras/stw768}

\bibitem[{{Sesana}(2013)}]{s13}
{Sesana}, A. 2013, \mnras, 433, L1, \dodoi{10.1093/mnrasl/slt034}

\bibitem[{Sesana {et~al.}(2016)Sesana, Shankar, Bernardi, \&
  Sheth}]{Sesana:2016yky}
Sesana, A., Shankar, F., Bernardi, M., \& Sheth, R.~K. 2016, Mon. Not. Roy.
  Astron. Soc., 463, L6, \dodoi{10.1093/mnrasl/slw139}

\bibitem[{{Sesana} \& {Vecchio}(2010)}]{sv10}
{Sesana}, A., \& {Vecchio}, A. 2010, \prd, 81, 104008,
  \dodoi{10.1103/PhysRevD.81.104008}

\bibitem[{{Sesana} {et~al.}(2008){Sesana}, {Vecchio}, \& {Colacino}}]{svc08}
{Sesana}, A., {Vecchio}, A., \& {Colacino}, C.~N. 2008, \mnras, 390, 192,
  \dodoi{10.1111/j.1365-2966.2008.13682.x}

\bibitem[{{Sesana} {et~al.}(2009){Sesana}, {Vecchio}, \& {Volonteri}}]{svv09}
{Sesana}, A., {Vecchio}, A., \& {Volonteri}, M. 2009, \mnras, 394, 2255,
  \dodoi{10.1111/j.1365-2966.2009.14499.x}

\bibitem[{{Shannon} {et~al.}(2013){Shannon}, {Ravi}, {Coles}, {Hobbs}, {Keith},
  {Manchester}, {Wyithe}, {Bailes}, {Bhat}, {Burke-Spolaor}, {Khoo}, {Levin},
  {Oslowski}, {Sarkissian}, {van Straten}, {Verbiest}, \& {Want}}]{src+13}
{Shannon}, R.~M., {Ravi}, V., {Coles}, W.~A., {et~al.} 2013, Science, 342, 334.
\newblock \doarXiv{1310.4569}

\bibitem[{{Shannon} {et~al.}(2015){Shannon}, {Ravi}, {Lentati}, {Lasky},
  {Hobbs}, {Kerr}, {Manchester}, {Coles}, {Levin}, {Bailes}, {Bhat},
  {Burke-Spolaor}, {Dai}, {Keith}, {Os{\l}owski}, {Reardon}, {van Straten},
  {Toomey}, {Wang}, {Wen}, {Wyithe}, \& {Zhu}}]{srl+15}
{Shannon}, R.~M., {Ravi}, V., {Lentati}, L.~T., {et~al.} 2015, Science, 349,
  1522, \dodoi{10.1126/science.aab1910}

\bibitem[{{Sharma}(2021)}]{INF7}
{Sharma}, R. 2021, arXiv e-prints, arXiv:2102.09358.
\newblock \doarXiv{2102.09358}

\bibitem[{{Siemens} {et~al.}(2006){Siemens}, {Creighton}, {Maor}, {Majumder},
  {Cannon}, \& {Read}}]{Siemens:2006vk}
{Siemens}, X., {Creighton}, J., {Maor}, I., {et~al.} 2006, \prd, 73, 105001,
  \dodoi{10.1103/PhysRevD.73.105001}

\bibitem[{{Siemens} {et~al.}(2013){Siemens}, {Ellis}, {Jenet}, \&
  {Romano}}]{Siemens:2013zla}
{Siemens}, X., {Ellis}, J., {Jenet}, F., \& {Romano}, J.~D. 2013, Classical and
  Quantum Gravity, 30, 224015, \dodoi{10.1088/0264-9381/30/22/224015}

\bibitem[{{Siemens} {et~al.}(2007){Siemens}, {Mandic}, \&
  {Creighton}}]{Siemens:2006yp}
{Siemens}, X., {Mandic}, V., \& {Creighton}, J. 2007, Physical Review Letters,
  98, 111101, \dodoi{10.1103/PhysRevLett.98.111101}

\bibitem[{{Starobinski{\v \i}}(1979)}]{sa79}
{Starobinski{\v \i}}, A.~A. 1979, Soviet Journal of Experimental and
  Theoretical Physics Letters, 30, 682

\bibitem[{{Taylor} {et~al.}(2020{\natexlab{a}}){Taylor}, {Baker}, {Hazboun},
  {Simon}, \& {Vigeland}}]{enterprise_extensions}
{Taylor}, S.~R., {Baker}, P.~T., {Hazboun}, J.~S., {Simon}, J.~J., \&
  {Vigeland}, S.~J. 2020{\natexlab{a}}, enterprise extensions.
\newblock \url{https://github.com/nanograv/enterprise_extensions}

\bibitem[{{Taylor} {et~al.}(2016){Taylor}, {Vallisneri}, {Ellis}, {Mingarelli},
  {Lazio}, \& {van Haasteren}}]{2016ApJ...819L...6T}
{Taylor}, S.~R., {Vallisneri}, M., {Ellis}, J.~A., {et~al.} 2016, \apjl, 819,
  L6, \dodoi{10.3847/2041-8205/819/1/L6}

\bibitem[{{Taylor} {et~al.}(2020{\natexlab{b}}){Taylor}, {van Haasteren}, \&
  {Sesana}}]{tvs20}
{Taylor}, S.~R., {van Haasteren}, R., \& {Sesana}, A. 2020{\natexlab{b}}, \prd,
  102, 084039, \dodoi{10.1103/PhysRevD.102.084039}

\bibitem[{{Vagnozzi}(2021)}]{INF6}
{Vagnozzi}, S. 2021, \mnras, 502, L11, \dodoi{10.1093/mnrasl/slaa203}

\bibitem[{{Vallisneri}(2020)}]{libstempo}
{Vallisneri}, M. 2020, {libstempo: Python wrapper for Tempo2}.
\newblock \doeprint{2002.017}

\bibitem[{{Vallisneri} {et~al.}(2020){Vallisneri}, {Taylor}, {Simon},
  {Folkner}, {Park}, {Cutler}, {Ellis}, {Lazio}, {Vigeland}, {Aggarwal},
  {Arzoumanian}, {Baker}, {Brazier}, {Brook}, {Burke-Spolaor}, {Chatterjee},
  {Cordes}, {Cornish}, {Crawford}, {Cromartie}, {Crowter}, {DeCesar},
  {Demorest}, {Dolch}, {Ferdman}, {Ferrara}, {Fonseca}, {Garver-Daniels},
  {Gentile}, {Good}, {Hazboun}, {Holgado}, {Huerta}, {Islo}, {Jennings},
  {Jones}, {Jones}, {Kaplan}, {Kelley}, {Key}, {Lam}, {Levin}, {Lorimer},
  {Luo}, {Lynch}, {Madison}, {McLaughlin}, {McWilliams}, {Mingarelli}, {Ng},
  {Nice}, {Pennucci}, {Pol}, {Ransom}, {Ray}, {Siemens}, {Spiewak}, {Stairs},
  {Stinebring}, {Stovall}, {Swiggum}, {van Haasteren}, {Witt}, \&
  {Zhu}}]{BayesEphem}
{Vallisneri}, M., {Taylor}, S.~R., {Simon}, J., {et~al.} 2020, \apj, 893, 112,
  \dodoi{10.3847/1538-4357/ab7b67}

\bibitem[{{van Haasteren} \& {Levin}(2010)}]{vl10}
{van Haasteren}, R., \& {Levin}, Y. 2010, \mnras, 401, 2372,
  \dodoi{10.1111/j.1365-2966.2009.15885.x}

\bibitem[{{van Haasteren} {et~al.}(2011){van Haasteren}, {Levin}, {Janssen},
  {Lazaridis}, {Kramer}, {Stappers}, {Desvignes}, {Purver}, {Lyne}, {Ferdman},
  {Jessner}, {Cognard}, {Theureau}, {D'Amico}, {Possenti}, {Burgay},
  {Corongiu}, {Hessels}, {Smits}, \& {Verbiest}}]{vhj+11}
{van Haasteren}, R., {Levin}, Y., {Janssen}, G.~H., {et~al.} 2011, \mnras, 414,
  3117, \dodoi{10.1111/j.1365-2966.2011.18613.x}

\bibitem[{{Verbiest} {et~al.}(2016){Verbiest}, {Lentati}, {Hobbs}, {van
  Haasteren}, {Demorest}, {Janssen}, {Wang}, {Desvignes}, {Caballero}, {Keith},
  {Champion}, {Arzoumanian}, {Babak}, {Bassa}, {Bhat}, {Brazier}, {Brem},
  {Burgay}, {Burke-Spolaor}, {Chamberlin}, {Chatterjee}, {Christy}, {Cognard},
  {Cordes}, {Dai}, {Dolch}, {Ellis}, {Ferdman}, {Fonseca}, {Gair},
  {Garver-Daniels}, {Gentile}, {Gonzalez}, {Graikou}, {Guillemot}, {Hessels},
  {Jones}, {Karuppusamy}, {Kerr}, {Kramer}, {Lam}, {Lasky}, {Lassus},
  {Lazarus}, {Lazio}, {Lee}, {Levin}, {Liu}, {Lynch}, {Lyne}, {Mckee},
  {McLaughlin}, {McWilliams}, {Madison}, {Manchester}, {Mingarelli}, {Nice},
  {Os{\l}owski}, {Palliyaguru}, {Pennucci}, {Perera}, {Perrodin}, {Possenti},
  {Petiteau}, {Ransom}, {Reardon}, {Rosado}, {Sanidas}, {Sesana}, {Shaifullah},
  {Shannon}, {Siemens}, {Simon}, {Smits}, {Spiewak}, {Stairs}, {Stappers},
  {Stinebring}, {Stovall}, {Swiggum}, {Taylor}, {Theureau}, {Tiburzi},
  {Toomey}, {Vallisneri}, {van Straten}, {Vecchio}, {Wang}, {Wen}, {You},
  {Zhu}, \& {Zhu}}]{vlh+16}
{Verbiest}, J.~P.~W., {Lentati}, L., {Hobbs}, G., {et~al.} 2016, \mnras, 458,
  1267, \dodoi{10.1093/mnras/stw347}

\bibitem[{{Vigeland} {et~al.}(2018){Vigeland}, {Islo}, {Taylor}, \&
  {Ellis}}]{OPT11}
{Vigeland}, S.~J., {Islo}, K., {Taylor}, S.~R., \& {Ellis}, J.~A. 2018, \prd,
  98, 044003, \dodoi{10.1103/PhysRevD.98.044003}

\bibitem[{{Volonteri} {et~al.}(2003){Volonteri}, {Haardt}, \& {Madau}}]{vhm03}
{Volonteri}, M., {Haardt}, F., \& {Madau}, P. 2003, \apj, 582, 559,
  \dodoi{10.1086/344675}

\bibitem[{{Will}(1977)}]{GeneralEQ}
{Will}, C.~M. 1977, \apj, 214, 826, \dodoi{10.1086/155313}

\bibitem[{{Will}(1993)}]{Will}
---. 1993, {Theory and Experiment in Gravitational Physics}

\bibitem[{Witten(1984)}]{PhysRevD.30.272}
Witten, E. 1984, Phys. Rev. D, 30, 272, \dodoi{10.1103/PhysRevD.30.272}

\bibitem[{{Wu} {et~al.}(2021){Wu}, {Gong}, \& {Li}}]{CS3}
{Wu}, L., {Gong}, Y., \& {Li}, T. 2021, arXiv e-prints, arXiv:2105.07694.
\newblock \doarXiv{2105.07694}

\bibitem[{Wu {et~al.}(2021)Wu, Chen, \& Huang}]{Wu:2021kmd}
Wu, Y.-M., Chen, Z.-C., \& Huang, Q.-G. 2021.
\newblock \doarXiv{2108.10518}

\bibitem[{{Wyithe} \& {Loeb}(2003)}]{Wyithe:2002ep}
{Wyithe}, J.~S.~B., \& {Loeb}, A. 2003, \apj, 590, 691, \dodoi{10.1086/375187}

\bibitem[{{Yi} \& {Zhu}(2021)}]{INF2}
{Yi}, Z., \& {Zhu}, Z.-H. 2021, arXiv e-prints, arXiv:2105.01943.
\newblock \doarXiv{2105.01943}

\bibitem[{{Yunes} \& {Siemens}(2013)}]{Yunes:2013dva}
{Yunes}, N., \& {Siemens}, X. 2013, Living Reviews in Relativity, 16, 9,
  \dodoi{10.12942/lrr-2013-9}

\end{thebibliography}
